\documentclass[fleqn,usenatbib,useAMS]{mnras}

\usepackage{graphicx}	
\usepackage{amsmath}	
\usepackage{amssymb}	
\usepackage{multicol}        
\usepackage{bm}		
\usepackage{pdflscape}	

\usepackage{subcaption}

\usepackage{enumitem}

\usepackage{booktabs}
\usepackage{multirow}

\usepackage{stfloats}

\usepackage[T1]{fontenc}
\usepackage{ae,aecompl}

\usepackage{newtxtext,newtxmath}

	\title[State-space PTA]{State-space algorithm for detecting the nanohertz gravitational wave background}

\author[Kimpson]{Tom Kimpson$^{1,2}$\thanks{Contact e-mail: \href{tom.kimpson@unimelb.edu.au}{tom.kimpson@unimelb.edu.au}}, Andrew Melatos$^{1,2}$, Joseph O'Leary$^{1,2}$, Julian B. Carlin$^{1,2}$, Robin J. Evans$^{3}$, \newauthor William Moran$^{3}$, Tong Cheunchitra$^{1,2}$, Wenhao Dong$^{1,2}$, Liam Dunn$^{1,2}$, Julian Greentree$^{3}$, Nicholas J. O'Neill$^{1,2}$, \newauthor Sofia Suvorova$^{3}$, Kok Hong Thong$^{1,2}$, Andrés F. Vargas$^{1,2}$%
	\\
	$^{1}$School of Physics, University of Melbourne, Parkville, VIC 3010, Australia \\
	$^{2}$OzGrav, University of Melbourne, Parkville, VIC 3010, Australia \\
	$^{3}$Department of Electrical and Electronic Engineering, University of Melbourne, Parkville, VIC 3010, Australia }
\date{Last updated \today}

\pubyear{2024}

\begin{document}
\label{firstpage}
\pagerange{\pageref{firstpage}--\pageref{lastpage}}
\maketitle

\begin{abstract}	
	The stochastic gravitational wave background (SGWB) can be observed in the nanohertz band using a pulsar timing array (PTA). Here a computationally efficient state-space framework is developed for analysing SGWB data, in which the stochastic gravitational wave strain at Earth is tracked with a non-linear Kalman filter and separated simultaneously from intrinsic, achromatic pulsar spin wandering. The filter is combined with a nested sampler to estimate the parameters of the model, and to calculate a Bayes factor for selecting between models with and without a SGWB. The procedure extends previous state-space formulations of PTA data analysis applied to individually resolvable binary black hole sources. The performance of the new algorithm is tested on synthetic data from the first International PTA Mock Data Challenge. It is shown that the algorithm distinguishes a SGWB from pure noise for $A_{\rm gw} \geq 3 \times 10^{-14}$, where $A_{\rm gw}$ denotes the standard normalization factor for a power spectral density with power-law exponent $-13/3$. Additional, systematic validation tests are also performed with synthetic data generated independently by adjusting the injected parameters to cover astrophysically plausible ranges. Full posterior distributions are recovered and tested for accuracy. The state-space procedure is memory-light and evaluates the likelihood for a standard-sized PTA dataset in $\lesssim 10^{-1}$ s without optimization on a standard central processing unit.
\end{abstract}




\begingroup
\let\clearpage\relax
\endgroup
\newpage

\section{Introduction}\label{sec:intro}
Pulsar timing array \citep[PTA;][]{ Tiburzi2018, 2021hgwa.bookE...4V} experiments find congruent evidence for a stochastic gravitational wave (GW) background \citep{2023ApJ...951L...8A,2023arXiv230616214A,2023ApJ...951L...6R,2023RAA....23g5024X} by measuring the cross-correlation in the timing residuals between pairs of pulsars as a function of their angular separation – the Hellings-Downs curve \citep{Hellings,2023arXiv230805847R}. The stochastic background is theorised to result from the incoherent superposition of multiple GW sources from inspiralling supermassive black hole binaries  \citep[SMBHBs;][]{Rajagopal1995,Jaffe_2003, Wyithe2003,Sesana2013,McWilliams_2014,Ravi2015MNRAS.447.2772R,Burke2019, Skyes2022}, as well as more exotic sources such as cosmic strings \citep{PhysRevD.108.103511,KITAJIMA2023138213} or phase transitions in the early universe \citep{FUJIKURA2023138203,PhysRevD.108.055018,PhysRevD.109.023522}. \newline 

SMBHBs that emit GWs with sufficiently large amplitudes may be resolvable individually with PTAs \citep{Jenet2004,Sesana2010,Yardley2010,Babak2012,2013CQGra..30v4004E,Zhu2014PPTA,Zhu10, Babak2016,Zhupulsarterms,2023arXiv230616226A,Arzoumanian2023}. State-space algorithms are a promising and complementary approach to standard PTA cross-correlation analyses for GWs from individual SMBHBs \citep{KimpsonPTA1,KimpsonPTA2}. PTA state-space algorithms use a Kalman filter \citep{Kalman1,Meyers2021,Melatos2023} to track the intrinsic rotational state of the pulsars in the array. The optimal estimate of the state-space evolution provided by the Kalman filter is combined with a Bayesian nested sampler \citep{Skilling,Ashton2022} to infer the time invariant parameters of a single-source GW and calculate the Bayesian evidence for models with and without the GW present. State-space algorithms disentangle GW-induced modulations in every PTA pulsar's spin frequency from other fluctuations by tracking the actual, measured, time-ordered, random realisation of the intrinsic, achromatic timing noise in every PTA pulsar \citep[e.g.][]{Shannon2010,Lasky2015,Caballero2016,Goncharov2021} by harnessing the adaptive gain of the Kalman filter \citep{Kalman1,zarchan2000fundamentals}. In contrast, traditional PTA algorithms average over the ensemble of admissible timing noise realisations when inferring the noise power spectral density (PSD). \newline 

The goal of this paper is to develop an independent framework for detecting a stochastic GW background, based on a state-space algorithm, which can be applied as a mutual cross-check in tandem with traditional analyses, based on the Hellings-Downs cross-correlation. The independent framework is a generalization of an existing state-space algorithm for detecting individual SMBHBs, which has been validated through systematic Monte Carlo testing with synthetic data \citep{KimpsonPTA1,KimpsonPTA2}. The paper is organised as follows. In Section \ref{sec:state_space} we briefly review state-space representations of noisy, dynamical systems in general. In Section \ref{sec:state_space_formulation} we  show how PTA data analysis maps readily onto the state-space structure, and introduce a description of the stochastic GW background as a Gauss-Markov process, specifically an Ornstein-Uhlenbeck process. In Section \ref{sec:kfns} we review how state-space algorithms for PTA data analysis estimate the parameters of the inference model and calculate the model evidence. In Section \ref{sec:representative_analysis_mdc} we employ synthetic data from the International Pulsar Timing Array (IPTA) Mock Data Challenge (MDC) to validate the new algorithm. We quantify its sensitivity and the accuracy with which it infers the parameters of the GW background. In Section \ref{sec:computation_costs} we briefly discuss the computational cost of the method and compare with traditional PTA analyses. Conclusions are drawn in Section \ref{sec:conclusion}. Throughout the paper we adopt natural units, with $c = G = \hbar =1$, and the metric signature $(-,+,+,+)$

\section{State-space representation} \label{sec:state_space}
\begin{figure}
	\includegraphics[width=\columnwidth, height =0.67\columnwidth]{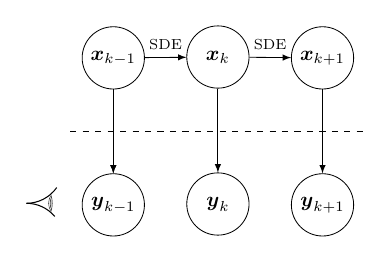} 	
	\caption{Diagram of the general state-space structure used in this paper. A discrete, temporal sequence of latent (hidden), state variables, $\boldsymbol{x}(t_k) = \boldsymbol{x}_k$, evolve according to a first-order stochastic differential equation (SDE). At each timestep, the latent state is related to the measurement recorded by the detector, $\boldsymbol{y}_k$, through a noisy measurement process. State-space algorithms, such as the Kalman filter, solve the inverse problem of estimating $p(\boldsymbol{x}_k | \boldsymbol{y}_{1:k})$.}
	\label{fig:state_space_diagram}
\end{figure}
In this section we outline the general form of the state-space framework used in this paper and its underlying assumptions. In Section \ref{sec:state_space_formulation} we explain how the general framework maps onto the specific case of PTA data analysis and the search for a stochastic GW background. \newline 

The state-space representation describes a sequence of hidden or latent states, $\boldsymbol{x}(t)$, which are not directly observed, and are related to a set of noisy measurements, $\boldsymbol{y}(t)$. The hidden states form a Markov sequence and evolve according to a first-order stochastic differential equation 
\begin{equation}
\dot{\boldsymbol{x}}(t) = \boldsymbol{f}\left[t,\boldsymbol{x}(t)\right] + \boldsymbol{w}(t) \, ,\label{eq:state_evolution}
\end{equation}
where $\boldsymbol{f}$ is an arbitrary function which governs the deterministic dynamics, and $\boldsymbol{w}(t)$ is a stochastic process. Importantly, the ``noise'' in $\boldsymbol{w}(t)$ is physical; it is intrinsic to the dynamical system and does not include measurement errors, for example. The hidden states are related to the measurements recorded at a detector via a measurement equation, viz.
\begin{equation}
	\boldsymbol{y}(t) = \boldsymbol{h}\left[t,\boldsymbol{x}(t)\right] + \boldsymbol{v}(t)\, . \label{eq:measurement_diagram}
\end{equation}
In Equation \eqref{eq:measurement_diagram}, $\boldsymbol{h}$ is an arbitrary deterministic function, and $\boldsymbol{v}(t)$ is a stochastic process associated with the measurement procedure which includes measurement errors and is distinct from the stochastic process which acts on the states, $\boldsymbol{w}(t)$. Figure \ref{fig:state_space_diagram} presents a diagram of the state-space structure for a discrete subsequence of states and measurements, where the subscript $k$ indexes the timestep, viz. $\boldsymbol{x}_k = \boldsymbol{x}(t_k)$. One moves horizontally from $\boldsymbol{x}_{k-1}$ through $\boldsymbol{x}_{k}$ to $\boldsymbol{x}_{k+1}$ via Equation \eqref{eq:state_evolution}. One moves vertically from $\boldsymbol{x}_{k}$ to $\boldsymbol{y}_{k}$ via Equation \eqref{eq:measurement_diagram}. \newline

Typically, one wants to solve the inverse problem of estimating the discrete sequence of $T$ hidden states $\boldsymbol{x}_{1:T} = \{\boldsymbol{x}_{1},\dots,\boldsymbol{x}_{T}  \}$ given the discrete sequence of measurements $\boldsymbol{y}_{1:T}=\{\boldsymbol{y}_{1},\dots,\boldsymbol{y}_{T}  \}$. In the Bayesian sense, one wishes to compute the joint posterior $p(\boldsymbol{x}_{1:T} | \boldsymbol{y}_{1:T})$. However, computing the joint posterior becomes computationally intractable, as $T$ increases, and the dimensionality of the posterior increases in tandem. Instead, state-space algorithms such as the Kalman filter \citep{Kalman1}, seek to compute at each timestep $k$ the marginalised posterior $p(\boldsymbol{x}_{k} | \boldsymbol{y}_{1:k})$. One can show that the marginalised posterior at $k$ satisfies \citep{Sarkka_2013,9581918}
\begin{equation}
	p(\boldsymbol{x}_{k} | \boldsymbol{y}_{1:k}) \propto p(\boldsymbol{y}_{k} | \boldsymbol{x}_{1:k}) \int d\boldsymbol{x}_{k-1} p(\boldsymbol{x}_{k} | \boldsymbol{x}_{k-1}) p(\boldsymbol{x}_{k-1} | \boldsymbol{y}_{1:k-1})  \, .
\end{equation}
Equation \eqref{eq:state_evolution} determines $p(\boldsymbol{x}_{k} | \boldsymbol{x}_{k-1})$. Equation \eqref{eq:measurement_diagram} determines $p(\boldsymbol{y}_{k} | \boldsymbol{x}_{1:k})$. The Kalman filter is the optimal estimator of $p({\boldsymbol{x}}_k | {\boldsymbol{y}}_{1:k})$, in the sense that it minimizes the mean square error between the estimated and true state sequences, if ${\boldsymbol{f}}$ and ${\boldsymbol{h}}$ are linear in ${\boldsymbol{x}}(t)$, irrespective of the statistics of ${\boldsymbol{w}}(t)$ and ${\boldsymbol{v}}(t)$. It is also optimal, when ${\boldsymbol{f}}$ and ${\boldsymbol{h}}$ are nonlinear, as long as the statistics of ${\boldsymbol{w}}(t)$ and ${\boldsymbol{v}}(t)$ are Gaussian. Importantly, and counterintuitively, the Kalman filter distinguishes the physical noise ${\boldsymbol{w}}(t)$ from the measurement noise ${\boldsymbol{v}}(t)$, even if both noises share the same statistics. This is a key advantage in PTA data analysis, where the challenge is to distinguish stochastic spin wandering intrinsic to the pulsar, which enters through Equation \eqref{eq:state_evolution}, from stochastic modulations of the pulse arrival times caused by the stochastic GW background, which enter through Equation \eqref{eq:measurement_diagram}. This key point is discussed further below.

\section{State-space formulation of PTA analysis}\label{sec:state_space_formulation}
In this section we formulate PTA detection of the stochastic GW background as a state-space problem. There are $N$ pulsars in the array, labelled $1\leq n\leq N$, whose independent rotational states define an $N$-dimensional state space. The $n$-th pulsar’s spin frequency $f_{\rm p}^{(n)}(t)$, as measured in the local, freely-falling rest frame of the pulsar’s centre of mass, evolves according to a stochastic differential equation of motion. In this paper, we copy an approach adopted successfully elsewhere \citep{Meyers2021,Vargas, KimpsonPTA1, KimpsonPTA2} and assume on phenomenological grounds that $f_{\rm p}(t)$ obeys a mean-reverting Ornstein-Uhlenbeck process. A short review of the model is presented in Section \ref{sec:spin_evolution}; details regarding its justification may be found in the foregoing references. The spin frequency $f_{\rm p}^{(n)}(t)$ is related to the pulse frequency measured by an observer at Earth, $f_{\rm m}^{(n)}(t)$, via a measurement equation, which involves the time-varying random amplitude $a^{(n)}(t)$ of the stochastic GW background at the location of the $n$-th pulsar as well as the instrumental noise in the PTA. The measurement equation relating $f^{(n)}_{\rm p}(t)$ to $f^{(n)}_{\rm m}(t)$ via $a^{(n)}(t)$ is presented in Section \ref{sec:gw_background_modulation}, supplemented by stochastic differential equations describing the approximate evolution of $a^{(n)}(t)$ for $1\leq n \leq N$, which are derived rigorously via a generalized Hellings-Downs argument in Appendix \ref{appendix:justify_OU_process_for_background}. The dynamical equations of motion in Section \ref{sec:spin_evolution}, and the measurement equations in Section \ref{sec:gw_background_modulation}, match exactly the mathematical state-space structure presented in Section \ref{sec:state_space}. That is, given a temporal sequence of measurements $f_{\rm m}^{(n)}(t_k)$ labelled by the integer $k$, one can use a non-linear Kalman filter \citep{Kalman1,zarchan2000fundamentals} to estimate the most likely sequence of hidden states $f_{\rm p}^{(n)}(t_k)$ and $a^{(n)}(t_k)$, as well as the most likely values for the model parameters in the problem, e.g.\ the root-mean-square amplitude of the stochastic GW background. We summarise how the PTA problem maps onto the state-space structure in Section \ref{sec:state_space_strcuture}. We summarise the parameters of the model in Section  \ref{sec:summary_of_static_parameters}  and set out the connection to the Kalman filter in Section \ref{sec:kfns} and Appendix \ref{sec:kalman}. \newline 

To maintain consistency with previous work \citep{Meyers2021,KimpsonPTA1,KimpsonPTA2}, and in order to validate the approach in its simplest form, the analysis in this paper operates on $f_{\rm m}^{(n)}(t_k)$ instead of a sequence of pulse times of arrival (TOAs). When applying the algorithm to real astronomical data it will be necessary to modify it to accept pulse TOAs, a subtle problem which is solved in a forthcoming paper.

\subsection{Spin evolution}\label{sec:spin_evolution}
In this paper, we assume that the rest frame spin frequency of the $n$-th pulsar evolves according to a mean-reverting Ornstein-Uhlenbeck process, described by a Langevin equation with a time-dependent drift term \citep{Vargas},
\begin{equation}
	\frac{df_{\rm p}^{(n)}}{dt} = -\gamma^{(n)}	 [f_{\rm p}^{(n)} - f_{\rm em}^{(n)} (t)] + \dot{f}_{\rm em}^{(n)}(t) +\xi^{(n)}(t) \ , 
	\label{eq:frequency_evolution}
\end{equation}
where $f_{\rm em}^{(n)}(t)$ represents the deterministic evolution  (e.g.\ due to electromagnetic braking; see below), an overdot denotes a derivative with respect to $t$, $\gamma^{(n)}$ is a damping constant whose reciprocal specifies the mean-reversion timescale, and $\xi^{(n)}(t)$ is a white noise stochastic process which satisfies
\begin{align}
	\langle \xi^{(n)}(t) \rangle &= 0 \ , \\
	\langle \xi^{(n)}(t) \xi^{(n')}(t') \rangle &= [\sigma^{(n)}]^2 \delta_{n,n'} \delta(t - t') \ .	\label{eq:xieqn_new}
\end{align}
In Equation \eqref{eq:xieqn_new}, $[\sigma^{(n)}]^2$ parameterises the noise amplitude and produces characteristic root mean square fluctuations $\approx \sigma^{(n)} / [\gamma^{(n)}]^{1/2}$ in $f_{\rm p}^{(n)}(t)$ \citep{gardiner2009stochastic}. The factor $\delta_{n,n'}$ in the right-hand side of Equation \eqref{eq:xieqn_new} ensures that the non-GW-induced spin fluctuations in distinct pulsars $n\neq n'$ are uncorrelated, as expected physically. The deterministic evolution $f_{\rm em}^{(n)}$ is attributed to magnetic dipole braking for the sake of definiteness, with braking index $n_{\rm em}=3$ \citep{1969ApJ...157..869G}. PTAs are typically composed of millisecond pulsars (MSPs), for which the quadratic correction due to $n_{\rm em}$ in $f_{\rm p}^{(n)}(t)$ is negligible over the observation time $T_{\rm obs} \sim 10 \, {\rm yr}$. Consequently, $f_{\rm em}^{(n)}(t)$ can be approximated accurately by 
\begin{equation}
	f_{\rm em}^{(n)}(t) = f_{\rm em}^{(n)}(t_1) + \dot{f}_{\rm em}^{(n)}(t_1)t \ , \label{eq:spinevol}
\end{equation} 
where $t_1$ labels the time of the first TOA. \newline

Equations \eqref{eq:frequency_evolution}--\eqref{eq:spinevol} imply that the PSD of the fluctuations in $f_{\rm p}^{(n)}(t)$ scales $\propto f^{-2}$ in the regime $f \gtrsim \gamma^{(n)} \sim 10^{-13} \, {\rm s^{-1}}$, where $f$ denotes the Fourier frequency, and is flat with $f$ for $f \lesssim \gamma^{(n)}$. Hence the PSD of the fluctuations in the rotational phase $\phi_{\rm p}^{(n)}(t) = \int_0^{t} dt' \, f_{\rm p}^{(n)}(t')$ scales $\propto f^{-4}$ for $f\gtrsim \gamma^{(n)}$ and $\propto f^{-2}$ for $f \lesssim \gamma^{(n)}$. These scalings are consistent with data from some PTA MSPs \citep{Caballero2016,Goncharov2021,Parthasarathy2021} but not others, e.g.\ PSR J1909$-$3744 \citep{Melatos2014}. We do not tune the exact form of the implied phase residual PSD while performing the validation tests in this paper for two reasons. First, when analysing real astronomical data, it is straightforward to modify Equations \eqref{eq:frequency_evolution}--\eqref{eq:spinevol} to reproduce any desired PSD, e.g.\ by adding colour to the white noise $\xi^{(n)}(t)$ \citep{allen2004signal,tenoudji2016analog} or by augmenting the equations of motion and injecting white noise into higher derivatives such as $\ddot{f}_{\rm p}^{(n)}$ \citep{Meyers2021,Myers2021MNRAS.502.3113M,Antonelli2023}. The latter approach has been applied successfully to interpret measurements of anomalous pulsar braking indices, e.g.\ PSR J0942$-$5552. \citep{Vargas}. Second, Equations \eqref{eq:frequency_evolution}--\eqref{eq:spinevol} are a good approximation in the PTA context, because timing observations are relatively frequent compared to the mean reversion time-scale for PTA MSPs, with TOAs separated typically by intervals $\lesssim 10^7 \, {\rm s} \ll [\gamma^{(n)}]^{-1}$. The white-noise approximation for $\xi^{(n)}(t)$ in Equations \eqref{eq:frequency_evolution}--\eqref{eq:spinevol}  has enjoyed practical success when analysing real astronomical data in two related but  independent applications: detecting rotational glitches with a hidden Markov model \citep{Melatos2020ApJ...896...78M,Lower2021MNRAS.508.3251L,2021MNRAS.504.3399D,Dunn2022,Dunn2023MNRAS.522.5469D} and estimating parameters for the two-component crust-superfluid model in PSR J1359$-$6038 \citep{Myers2021MNRAS.502.3113M,Meyers2021,2024MNRAS.tmp..891O}.

\subsection{Modulation of pulsar frequency by GW background}\label{sec:gw_background_modulation}

In the presence of a stochastic GW background, $f_{\rm p}^{(n)}(t)$ is related to  $f_{\rm m}^{(n)}(t)$ via a measurement equation \citep{Maggiore},
\begin{equation}
	f_{\rm m}^{(n)}(t) = f_{\rm p}^{(n)} [t-d^{(n)}] g^{(n)}(t) +  \varepsilon^{(n)}(t)\ ,
	\label{eq:measurement}
\end{equation}
where $d^{(n)}$ labels the distance to the $n$-th pulsar, $f_{\rm p}^{(n)}$ is evaluated at the retarded time $t-d^{(n)}$, and $\varepsilon^{(n)}(t)$ is a Gaussian measurement noise variate which satisfies 
\begin{align}
	\langle \varepsilon^{(n)}(t) \rangle &= 0 \ , \\
	\langle \varepsilon^{(n)}(t) \varepsilon^{(n')}(t') \rangle &= [ \sigma_{\rm m}^{(n)}]^2  \delta_{n,n'} \delta(t - t') \ .	\label{eq:vareps}
\end{align}
In Equation \eqref{eq:vareps}, $[ \sigma_{\rm m}^{(n)}]^2$ is the variance of measurement noise at the telescope for the $n$-th pulsar. The factor $\delta_{n,n'}$ indicates that the noise is uncorrelated between pulsars. Although adequate for a first pass at the problem, this approximation may not always hold in reality, e.g.\ if TOAs are measured for a group of pulsars on the same day, and the measurement noise $\varepsilon^{(n)}(t)$ is correlated temporally over one day.\newline 

The measurement function is
\begin{equation}
	g^{(n)}(t) = 1 - a^{(n)}(t) \, ,
	\label{eq:gfunc}
\end{equation} 
where $a^{(n)}(t)$ quantifies the modulation imprinted by the stochastic GW background on the $n$-th pulsar. It can be expressed as the linear combination of the modulations induced by $M$ discrete SMBHB sources \footnote{A related but different analysis applies to other backgrounds, such as from cosmic strings \citep{PhysRevD.108.103511,KITAJIMA2023138213,PhysRevD.109.023522}, which are not a superposition of discrete stellar sources. The analysis of non-SMBHB backgrounds is postponed to future work.}, viz.
\begin{equation}
a^{(n)}(t) = \sum_{m=1}^{M} z^{(n,m)}(t) \ , \label{eq:afunc}
\end{equation} 
where $z^{(n,m)}(t)$ is the redshift induced by the $m$-th SMBHB source on the $n$-th pulsar and is given by \citep[e.g.][]{Sesana2010,Perrodin2018,Agazie2023ApJ...951L..50A,2024A&A...690A.118E,KimpsonPTA1} 
\begin{equation}
z^{(n,m)}(t) = \sum_{A=+,\times} F_A^{(n,m)} \Delta h_A^{(n,m)}(t) \, . \label{eq:z_of_t}
\end{equation}
In Equation \eqref{eq:z_of_t}, $F_A^{(n,m)}$ is the antenna beam pattern
\begin{equation}
	F_A^{(n,m)} = \frac{1}{2} \frac{[q^{(n)}]^i [q^{(n)}]^j}{1 + [q^{(n)}]^k [n^{(m)}]_k} [e_{ij}^{A}]^{(m)} \, ,
\end{equation}
where $[q^{(n)}]^i$ labels the $i$-th (contravariant) coordinate of the $n$-th pulsar's position vector, $[n^{(m)}]_i$ labels the $i$-th (covariant) coordinate of the $m$-th SMBHB's position vector, and $[e_{ij}^{A}]^{(m)}$ (with $A = +,\times$) are the polarisation tensors of the $m$-th GW. In Equation \eqref{eq:z_of_t}, $\Delta h_A^{(n,m)}(t)$ is the difference between the $A$-component of the $m$-th GW, evaluated at time $t_{\rm p}^{(n)}$ at the $n$-th pulsar and at the observer on Earth viz.
\begin{equation}
\Delta h_A^{(n,m)}(t) = h_{A}^{(m)}\left[t_{\rm p}^{(n)}\right] - h_{A}^{(m)}(t) \, . \label{eq:delta_h}
\end{equation}

 In reality, when $M$ is large, $a^{(n)}(t)$ is a complicated and ``wiggly'' but still deterministic time series, whose time invariant parameters can be estimated individually in principle by Bayesian inference from the measured data. In practice, however, it turns out that the $7M$ source parameters (see Appendix \ref{appendix:justify_OU_process_for_background}) are not identifiable; they cannot be inferred uniquely from the data. Identifiability is a fundamental issue in state-space estimation problems across a wide range of electrical engineering applications \citep{e5be7c83a0d24500826f6e1b414d1733,WALTER1996125,DOBRE20122740,GUILLAUME2019418,casella2021statistical}. Rigorous mathematical tests have been devised to separate weakly and strongly identifiable parameters from those that are not identifiable at all, supplemented by empirical experiments \citep{BELLU200752,KARLSSON2012941,10.1093/bioinformatics/btad065,10.1093/bioinformatics/bty1069}. Unsurprisingly, the $7M$ parameters  in Equations \eqref{eq:afunc}--\eqref{eq:delta_h} are not identifiable in practice in the PTA context, where one has $M \gtrsim 10^{4}$ typically \citep{2009MNRAS.394.2255S,10.1111/j.1365-2966.2008.13682.x}, compared to only $\sim 2.5 \times 10^{4}$ TOAs (e.g. 50 pulsars observed with a weekly cadence for 10 years). \footnote{The parameters are identifiable to high accuracy for $M=1$ using a Kalman filter, as demonstrated by \cite{KimpsonPTA1} and \cite{KimpsonPTA2}, as well as by traditional methods \citep{Jenet2004,Zhu2014PPTA,Babak2016,Arzoumanian2023,2023arXiv230616226A}} \newline  
 
 Despite the identifiability challenge, one can still make progress, by approximating the ``wiggly'' time series $a^{(n)}(t)$ as a random variable, whose statistics are determined by the ensemble of $M$ sources. In this paper, we approximate the evolution of $a^{(n)}(t)$ using a Langevin equation\footnote{In the same spirit, the traditional Hellings-Downs analysis seeks to measure an ensemble-averaged quantity, by cross-correlating phase residuals of pulsar pairs, instead of resolving the background into individual sources and inferring their properties individually.},
\begin{eqnarray}
	\frac{d a^{(n)}}{dt} = -\gamma_{\rm a} a^{(n)} + \Xi_{\rm a}^{(n)}(t) \, , \label{eq:ornstein_for_at}
\end{eqnarray}
with damping constant $\gamma_{\rm a}$ and a white noise stochastic process $\Xi_{\rm a}^{(n)}(t)$ with ensemble statistics
\begin{align}
	\langle \Xi^{(n)}_{\rm a}(t) \rangle &= 0 \ , 	\label{eq:xieqn1} \\
	\langle \Xi^{(n)}_{\rm a}(t) \, \Xi^{(n')}_{\rm a}(t') \rangle &= \left[\sigma^{(n,n')}_{\rm a}\right]^2 \delta(t - t') \ .	\label{eq:xieqn2}
\end{align}
Equations \eqref{eq:ornstein_for_at}--\eqref{eq:xieqn2} are approximate but not arbitrary. They are derived from first principles in Appendix \ref{appendix:justify_OU_process_for_background} by evaluating $\langle a^{(n)}(t) a^{(n')}(t') \rangle$ averaged over a SMBHB source ensemble for the specific modulation in Equations \eqref{eq:afunc}--\eqref{eq:delta_h}. In Equation \eqref{eq:xieqn2}, the noise covariance satisfies 
\begin{eqnarray}
	 \left[\sigma^{(n,n')}_{\rm a}\right]^2 = \frac{\langle h^2\rangle}{6} \gamma_{\rm a} \, \Gamma \left[ \theta^{(n,n')} \right] \, , \label{eq:sigma_a_expression}
\end{eqnarray}
where $\langle h^2 \rangle$ is the mean square GW strain from the $M$ summed sources at the Earth's position, $\theta^{(n,n')}$ is the angle between the $n$-th and $n'$-th pulsars, and one has
	\begin{eqnarray}
		\Gamma\left[\theta^{(n,n')} \right] =  \frac{3}{2} x_{n n'} \ln x_{n n'}  -\frac{x_{n n'} }{4}+\frac{1}{2} + \frac{1}{2} \delta_{n n'}\label{eq:correlation} \, ,
	\end{eqnarray}
with $x_{nn'} = \left[1 - \cos \theta^{(n,n')}\right]/2$. Equation \eqref{eq:correlation} is related closely to the traditional Hellings-Downs cross-correlation formula \citep{Hellings}, cf. Equation (5) in \cite{2023ApJ...951L...6R} and Equation (4) in \cite{2023ApJ...951L...8A}. In Equation \eqref{eq:ornstein_for_at} we have
\begin{eqnarray}
\gamma_{\rm a} \sim \Omega_{\rm min} \label{eq:gamma_a_min}
\end{eqnarray}
where $\Omega_{\rm min}$ is the lower bound on the probability distribution of $\Omega$ over the SMBHB population; see Appendix \ref{app:OU_rep} and Appendix \ref{sec:generate_a}. Note that $\gamma_{\rm a}$ is shared by all pulsars, i.e. it does not carry an $(n)$ superscript. Equations \eqref{eq:ornstein_for_at}--\eqref{eq:correlation} approximate formally the modulations imprinted on PTA TOAs by the stochastic GW background as an Ornstein-Uhlnebeck process, following the derivation in Appendix \ref{app:OU_rep}. \newline 

We emphasise that Equations \eqref{eq:ornstein_for_at}--\eqref{eq:correlation} represent an inference model, specified so as to reproduce accurately the statistics of a stochastic GW  background from SMBHBs, and to circumvent the non-identifiability of the $7M$ SMBHB source parameters. The procedure for generating synthetic data with which to validate the inference model does not use Equations \eqref{eq:ornstein_for_at}--\eqref{eq:correlation}, as explained in Section \ref{sec:representative_analysis_mdc}. Moreover, there exists a well-defined connection between the Ornstein-Uhlenbeck process in Equations \eqref{eq:ornstein_for_at}--\eqref{eq:correlation} and the PSD (e.g.\ broken power law) used in traditional PTA analyses. This connection is derived in Appendix \ref{appendix:psd_OU}. In short, one finds that Equations \eqref{eq:ornstein_for_at}--\eqref{eq:correlation} correspond to a PSD in the timing residuals which goes as $f^{-4}$ at high frequencies and $f^{-2}$ at low frequencies.

\subsection{State-space structure}\label{sec:state_space_strcuture}
In this section we relate explicitly the dynamical equations for PTAs presented in Section \ref{sec:spin_evolution} and Section \ref{sec:gw_background_modulation} to the general state space structure presented in Section \ref{sec:state_space}. We specify the latent and measurement state vectors, $\boldsymbol{x}(t)$ and $\boldsymbol{y}(t)$, the latent state dynamical equations involving $\boldsymbol{f}[t,\boldsymbol{x}(t)]$ and $\boldsymbol{w}(t)$ in Equation \eqref{eq:state_evolution}, and the measurement process involving $\boldsymbol{h}\left[t,\boldsymbol{x}(t)\right]$ and $\boldsymbol{v}(t)$, in Equation \eqref{eq:measurement_diagram}. \newline 

\noindent On the dynamics side of the problem, the latent state is a vector of length $2N$
\begin{equation}
\boldsymbol{x}(t) = \left[f_{\rm p}^{(1)}(t), \dots , f_{\rm p}^{(N)}(t),a^{(1)}(t), \dots , a^{(N)}(t)\right] \, .
\end{equation}
The deterministic function $\boldsymbol{f}[t,\boldsymbol{x}(t)]$ is set by Equation \eqref{eq:frequency_evolution} and takes the form
\begin{equation}
\boldsymbol{f}[t,\boldsymbol{x}(t)] =\boldsymbol{A} \boldsymbol{x}(t) + \boldsymbol{B}(t) \label{eq:fVector} \, .
\end{equation}
In Equation \eqref{eq:fVector}, $\boldsymbol{A}$ is a diagonal $2N \times 2N$ block matrix,
\begin{equation}
	\boldsymbol{A} = \begin{pmatrix}
		-\boldsymbol{\gamma}_{\rm p} & 0 \\
		0 & -\boldsymbol{\gamma}_{\rm a} 
	\end{pmatrix} 
	\ ,
\end{equation}
with
\begin{equation}
	\boldsymbol{\gamma}_{\rm p} = \text{diag} \left[\gamma^{(1)}, ... , \gamma^{(N)}\right] \, ,
\end{equation}
and 
\begin{equation}
	\boldsymbol{\gamma}_{\rm a} = \gamma_{\rm a} I_{N} \, ,
\end{equation}
where $I_{N}$ denotes the identity matrix of dimension $N$. In Equation \eqref{eq:fVector}, $\boldsymbol{B}(t)$ is a control vector of length $2N$
\begin{equation}
	\boldsymbol{B}(t) = \left[B^{(1)}(t), \dots , B^{(N)}(t),0, \dots , 0\right] \, ,
\end{equation}
with 
\begin{equation}
	B^{(n)}(t) =\gamma^{(n)} f_{\rm em}^{(n)} (t) + \dot{f}_{\rm em}^{(n)} (t)
\end{equation}
The stochastic process associated with the latent state dynamics is
\begin{equation}
	\boldsymbol{w}(t) = \left[\xi^{(1)}(t), \dots, \xi^{(N)}(t),\Xi_{\rm a}^{(1)}(t), \dots, \Xi_{\rm a}^{(N)}(t) \right] \, ,
\end{equation}
where $\xi^{(n)}(t)$ is defined by Equation \eqref{eq:xieqn_new},
and $\Xi_{\rm a}^{(n)}(t)$ is defined by Equation \eqref{eq:xieqn2}. \newline

\noindent On the measurement side of the problem, the data are packaged in a vector of length $N$,
\begin{equation}
	\boldsymbol{y}(t) = \left[f_{\rm m}^{(1)}(t), \dots , f_{\rm m}^{(N)}(t) \right] \, . \label{eq:yvector}
\end{equation}
The deterministic function  $\boldsymbol{h}\left[t,\boldsymbol{x}(t)\right]=\boldsymbol{h}\left[\boldsymbol{x}(t)\right]$, is set by Equation \eqref{eq:measurement} and takes the form
\begin{eqnarray}
\boldsymbol{h}\left[\boldsymbol{x}(t)\right] = \begin{pmatrix}
		\left[1 - a^{(1)}(t)\right] f_{\rm p}^{(1)}(t)  \vspace{1mm} \\
		\left[1 - a^{(2)}(t)\right] f_{\rm p}^{(2)}(t)  \\
		\vdots \\
		\left[1 - a^{(N)}(t)\right] f_{\rm p}^{(N)}(t) \\
	\end{pmatrix} \, , \label{eq:kalman_h_matrix}
\end{eqnarray}
The random variate associated with the measurement process, which includes measurement errors, is given by
\begin{equation}
	\boldsymbol{v}(t) = \left[\varepsilon^{(1)}(t), \dots, \varepsilon^{(N)}(t)\right] \, ,
\end{equation}
where  $\varepsilon^{(n)}(t)$ is defined by Equation \eqref{eq:vareps}.\newline 

A glossary defining the symbols used in this paper is included in Appendix \ref{sec:glossary}. The glossary highlights the similarities and differences with the symbols used in traditional PTA analyses for the reader's convenience.

\subsection{Parameters of the model}\label{sec:summary_of_static_parameters}
The state-space model described in Sections \ref{sec:spin_evolution} and \ref{sec:gw_background_modulation} comprises $4N$ parameters, that are specific to the pulsars in the array, viz.
\begin{equation}
	\boldsymbol{\theta}_{\rm psr} = \left \{ \gamma^{(n)},\sigma^{(n)}, f_{\rm em}^{(n)}(t_1),\dot{f}_{\rm em}^{(n)}(t_1) \right\}_{1\leq n \leq N} \ .  \label{eq:psrparams}
\end{equation}
It also comprises two parameters, that are specific to the stochastic GW background, viz. 
\begin{equation}
	\boldsymbol{\theta}_{\rm gw} = \left \{h_{\rm a}, \gamma_{\rm a} \right \} \ ,  \label{eq:params3}
\end{equation}
where we define $h_{\rm a} = \langle h^2\rangle^{1/2}$. The parameters $h_{\rm a}$ and $\gamma_{\rm a}$ can be respectively related to the amplitude of the PSD and the minimum GW frequency in a traditional PTA analysis, as explained in Section \ref{sec:gw_background_modulation} and Appendix \ref{appendix:psd_OU}. The complete set of $4N+2$ parameters is denoted by $\boldsymbol{\theta} =  \boldsymbol{\theta}_{\rm psr} \cup \boldsymbol{\theta}_{\rm gw}$. They are inferred from the TOA-derived data $f_{\rm m}^{(n)}(t)$ by combining a Kalman filter with a nested sampler, as described in Section \ref{sec:kfns}. The identifiable two-member set $\boldsymbol{\theta}_{\rm a}$ parameterising the Ornstein-Uhlenbeck approximation to the stochastic GW background is smaller than the non-identifiable set of $7M$ SMBHB source parameters used to generate synthetic data by superposition, as discussed in Section \ref{sec:gw_background_modulation}. None of the $4N+2$ parameters $\boldsymbol{\theta} $ vary with time in this paper.

\section{Bayesian inference with a Kalman filter}\label{sec:kfns}
In this section we briefly review how state-space algorithms for PTA data analysis estimate the parameters of the inference model, $\boldsymbol{\theta}$, and calculate the model evidence $\mathcal{Z}$. State-space algorithms use a two-step inference procedure whereby a Kalman filter \citep{Kalman1} tracks the intrinsic achromatic pulsar spin wandering simultaneously with the modulation induced by a gravitational wave. The Kalman filter then provides a likelihood for Bayesian inference algorithms such as nested sampling. \newline

A Kalman filter recovers a temporal sequence of stochastically evolving system state variables, $\boldsymbol{x}(t)$, which are not directly observed, given a temporal sequence of noisy measurements, $\boldsymbol{y}(t)$. Previous state-space analyses of single-source PTA experiments \citep[e.g.][]{KimpsonPTA1,KimpsonPTA2} and neutron star spin wandering \citep[e.g.][]{Myers2021MNRAS.502.3113M,Meyers2021,Melatos2023} employ a linear Kalman filter. In this paper the non-linear relation between ${\boldsymbol{X}}(t)$ and ${\boldsymbol{Y}}(t)$ in Equation \eqref{eq:measurement} necessitates the use of the extended (non-linear) Kalman filter \citep[EKF, see, e.g.][]{zarchan2000fundamentals}. Other non-linear filters are also suitable, such as the unscented Kalman filter \citep{882463van} or the particle filter \citep{Simon10}. \newline  

Implementation of the EKF for the PTA state-space model in Section \ref{sec:state_space_formulation}, including the full set of recursion relations, is presented in Appendix \ref{sec:kalman}. At each discrete timestep indexed by $1 \leq k  \leq t_{N_{t}}$, the Kalman filter returns an estimate of the state variables, $\hat{\boldsymbol{x}}_j = \hat{\boldsymbol{x}}(t_j)$, and the covariance of those estimates, ${\boldsymbol{P}}_j = \langle {\boldsymbol{\hat x}}_j {\boldsymbol{\hat x}}_j^{\rm T} \rangle$, where the superscript T denotes the matrix transpose. The filter tracks the error in its predictions of $\boldsymbol{x}_j$ by converting ${\boldsymbol{\hat x}}_j$ into predicted measurements ${\boldsymbol{\hat y}}_j$ via Equation \eqref{eq:measurement} and comparing with the actual, noisy measurements ${\boldsymbol{y}}_j$. It defines a residual $\boldsymbol{\epsilon}_j = \boldsymbol{y}_j  - \hat{\boldsymbol{y}}_j$, which is sometimes termed the innovation. The EKF also calculates the uncertainty in $\boldsymbol{\epsilon}_j$ via the innovation covariance $\boldsymbol{S}_i = \langle \boldsymbol{\epsilon}_j \boldsymbol{\epsilon}_j^{\rm T} \rangle$. At each timestep, the EKF returns a Gaussian log-likelihood 
\begin{eqnarray}
	\log \mathcal{L}_i =  -\frac{1}{2} \left (N \log 2 \pi + \log  \left | \boldsymbol{S}_i \right | + \boldsymbol{\epsilon}_i^{\intercal} \boldsymbol{S}_i^{-1}  \boldsymbol{\epsilon}_i \right ) \ .
\end{eqnarray}
The total log-likelihood for the entire sequence $1\leq i \leq J$ is
\begin{eqnarray}
	\log \mathcal{L} =  \sum_{i=1}^{J} \log \mathcal{L}_i \ . \label{eq:likelihood}
\end{eqnarray}
Given ${\boldsymbol{y}}_1, \dots, {\boldsymbol{y}}_J$, $\mathcal{L}$ is a function of the estimates ${\boldsymbol{\hat \theta}}$ of the parameters passed to the Kalman filter, i.e. $\mathcal{L}$ = $\mathcal{L}(\boldsymbol{y} | \boldsymbol{\hat \theta})$. \newline

The likelihood returned by the EKF for a particular $\boldsymbol{\hat \theta}$ is used to estimate the posterior distribution of $\boldsymbol{\theta}$ by Bayes's Rule,
\begin{equation}
	p(\boldsymbol{\theta} | \boldsymbol{y}) = \frac{\mathcal{L}(\boldsymbol{y} | \boldsymbol{\theta}) \pi(\boldsymbol{\theta})}{\mathcal{Z}} \ ,
\end{equation}
where $\pi(\boldsymbol{\theta})$ is the prior distribution on $\boldsymbol{\theta}$, and $\mathcal{Z}$ is the marginalised likelihood or evidence,
\begin{equation}
	\mathcal{Z} = \int d \boldsymbol{\theta} \mathcal{L}(\boldsymbol{y} | \boldsymbol{\theta})  \pi(\boldsymbol{\theta})  \ . \label{eq:model_evidence}
\end{equation}
In this paper, the calculation of $p(\boldsymbol{\theta} | \boldsymbol{y})$ and $\mathcal{Z}$ proceeds via a nested sampling algorithm \citep{Skilling,Ashton2022}. Multiple nested sampling algorithms and computational libraries exist \citep{Feroz2008,Feroz2009,Handley2015,dynesty2020,UltraNest2021}. In gravitational wave research it is common to use the \texttt{dynesty} sampler \citep{dynesty2020} via the \texttt{Bilby} \citep{bilby.507.2037A} front-end library. We follow this precedent and use \texttt{Bilby} in this paper. \newline

Appendix \ref{sec:kalman} outlines how to discretise the continuous dynamical and measurement equations in Section \ref{sec:state_space_formulation}, specifically Equations \eqref{eq:frequency_evolution}--\eqref{eq:spinevol}, \eqref{eq:ornstein_for_at}--\eqref{eq:correlation}, and \eqref{eq:measurement}--\eqref{eq:gfunc}.  Appendix \ref{sec:kalman} goes on to demonstrate how to solve the discretised equations using an EKF. Appendix \ref{sec:workflow} summarizes the state-space analysis workflow of the combined Kalman filter and nested sampler. We refer the reader to \cite{KimpsonPTA1} for additional details on the state-space analysis workflow.

\section{Detecting a synthetic background}\label{sec:representative_analysis_mdc}
In this section we test the state-space analysis algorithm on synthetic PTA data. The test has three goals: (i) to verify that the algorithm in Sections \ref{sec:state_space_formulation} and \ref{sec:kfns} can detect successfully a synthetic background; (ii) to determine the minimum background amplitude which the algorithm detects; and (iii) to determine the accuracy with which the parameters describing the background are estimated. We use synthetic data from the first IPTA MDC \footnote{\url{www.github.com/nanograv/mdc1}}. In Section \ref{sec:ipta_data} we specify the exact data used and explain how we convert the TOAs provided by the MDC to a $f_{\rm m}(t)$ time series to be ingested by the Kalman filter. In Section \ref{sec:pta_detection} we address the detection problem, that is, quantifying the evidence for or against a GW signal in the data. We go on to calculate the detectability of the GW background as a function of its amplitude. In Section \ref{sec:mdc_pe} we address the parameter estimation problem and calculate the joint posterior probability distribution of the parameters, $p(\boldsymbol{\theta} | \boldsymbol{Y})$. In Appendix \ref{appendix:exxtra_tests} additional tests are performed on synthetic data generated via an alternative procedure, with greater scope to vary the injection parameters in a controlled fashion, and the results are found be consistent.

\subsection{IPTA MDC}\label{sec:ipta_data}
\begin{table}
	\centering
		\begin{tabular}{lll}
			\toprule
			Parameter & Injected value & Units  \\
			\hline
			$A_{\rm gw}$ & $5 \times 10^{-14}$ & -  \\
			$\alpha$ & $-2/3$ & - \\
			$A_{\rm psr}$ & $5.77 \times 10^{-22}$ & $s^{1.3}$  \\
			$\alpha_{\rm psr}$ & $1.7$ & - \\
			$\sigma_{\rm TOA}$& $100$ & ns \\
			\bottomrule
		\end{tabular}
		\caption{Injected parameters used to generate synthetic data via the procedure outlined in Section \ref{sec:ipta_data}. The parameters $A_{\rm gw}$ and $\alpha$  specify the perturbations to the pulsar TOAs from the stochastic GW background, cf. Equation \eqref{eq:psd_gw}. The parameters $A_{\rm psr}$ and $\alpha_{\rm psr}$  specify the perturbations to the pulsar TOAs from the intrinsic pulsar spin noise, cf. Equation \eqref{eq:psd_psr}. $\sigma_{\rm TOA}$ is the uncertainty in the TOA due to the white instrument noise.}
		\label{tab:injected_parameters_mdc}
	\end{table}

\begin{figure}
	\includegraphics[width=\columnwidth, height =\columnwidth]{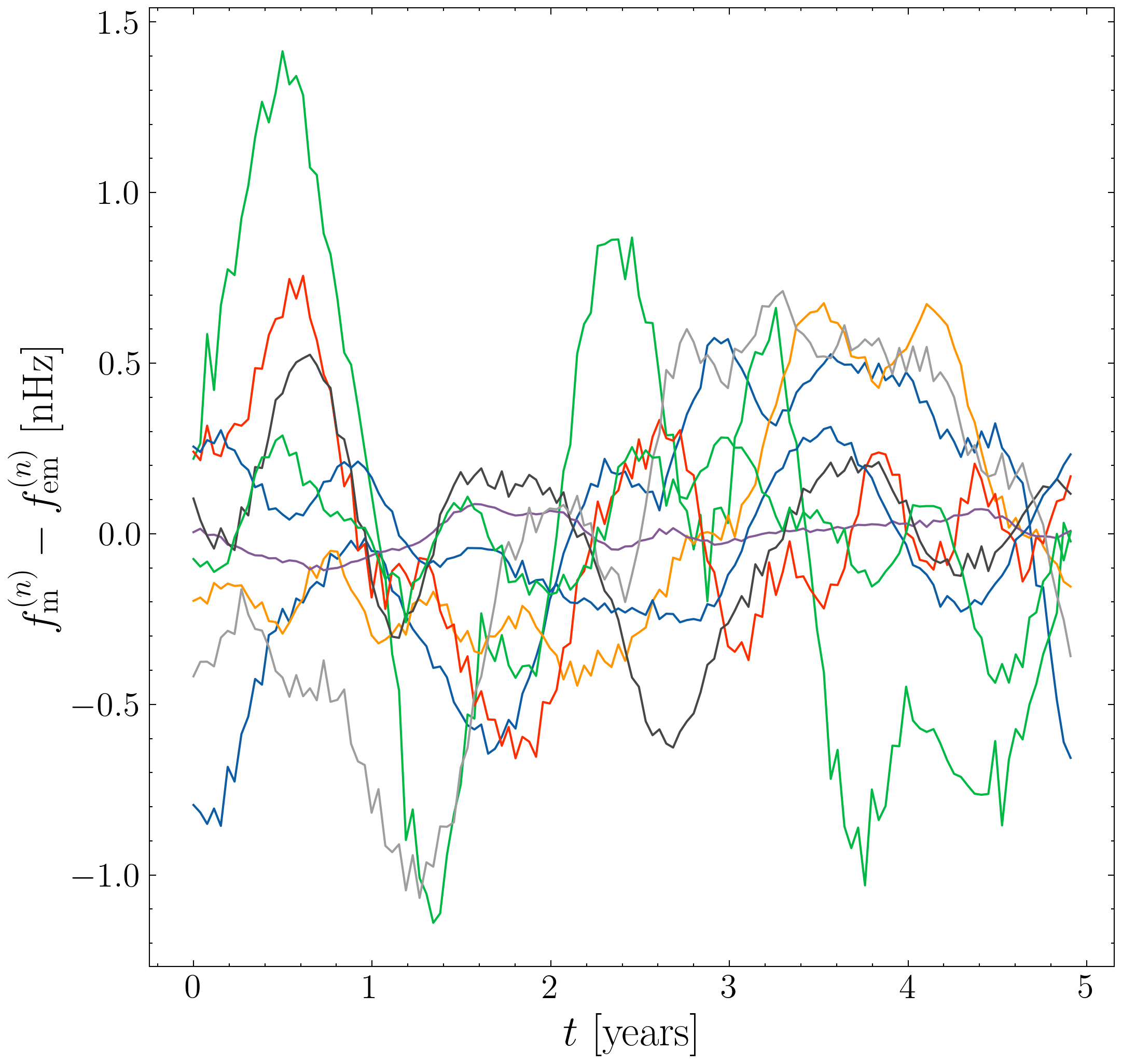} 	
	\caption{Synthetic pulse frequency time series generated by the procedure described in Section \ref{sec:ipta_data}, using the parameters summarised in Table \ref{tab:injected_parameters_mdc}. A subset (nine) of the $N=36$ time series are plotted, with curves coloured uniquely.  We plot the ephemeris-subtracted quantity $f_{\rm em}^{(n)}(t) - f_{\rm em}^{(n)}(t)$ to emphasise the stochastic variations in the time series, which are otherwise hard to discern visually.}
	\label{fig:mdc_fake_Data}
\end{figure}

State-space methods for PTA data analysis \citep[e.g.][]{KimpsonPTA1,KimpsonPTA2} accept as an input a discrete, temporal sequence of noisy measurements $\boldsymbol{y}(t)$ with $t \in \{t_1, ..., t_{N_{t}} \}$, where $N_{t}$ labels the number of sampling epochs. In this section we explain how to generate the synthetic data, namely a noisy pulse frequency time series ${\boldsymbol{y}}(t)$ given by Equation \eqref{eq:yvector}, for $N$ PTA pulsars as measured at Earth. \newline

In this paper we use synthetic data from the first IPTA MDC. Specifically, we use ``dataset 1'' from the open data challenge, which consists of $N=36$ pulsars sampled fortnightly for 5 years. Going forward, we use the notation MDC\_D1 to refer to this dataset. MDC\_D1 comprises of a set of $N$ \texttt{.tim} files, which contain the pulse TOAs, and $N$ \texttt{.par} files, which are used to construct an ephemeris via (e.g.) \texttt{tempo2} . There are two complications which must be overcome in order to run the state-space algorithm in this paper on the MDC data. First, the TOAs need to be corrected to account for the conversion to the inertial reference frame of the Solar system barycentre,  the pulsar's binary orbit, and so on; see Figure 7.1 and the associated discussion in \cite{nano_gw_astronomer}. Including these corrections directly into the state-space framework (instead of pre-processing, as in this paper) is straightforward in principle; they can be incorporated through the measurement equation that maps from $\boldsymbol{x}(t)$ to $\boldsymbol{y}(t)$. We postpone this extension to future work, to be implemented in tandem with the method being generalised to operate on TOAs instead of frequencies. Second, the TOAs need to be converted into a frequency time series, which is ingested by the Kalman filter. In order to overcome the above complications, we take the following steps. 
\begin{enumerate}[leftmargin=2em]
	\item Take $N$ pulsars from MDC\_D1, each of which has a \texttt{.tim} and \texttt{.par}  file.
	\item Use the timing solution defined by the \texttt{.par} file to correct the TOAs in the \texttt{.tim} file, such that the residuals are zero. 
	\item Perturb the TOAs by adding spin wandering red noise, instrumental white noise, and a stochastic GW background, producing non-zero residuals. 
	\item Convert from TOAs (i.e. phase) to frequency by finite differencing. 
\end{enumerate}
The result of steps (i)--(iv) is $\boldsymbol{y}(t)$, as defined in Equation \eqref{eq:yvector}, which is barycentered yet retains perturbations arising from measurement noise, intrinsic pulsar red noise (i.e. spin wandering), and a stochastic GW background. A subset (for nine pulsars) of the frequency $\boldsymbol{y}(t)$ is plotted in Figure \ref{fig:mdc_fake_Data}, with curves coloured uniquely. We plot the ephemeris-subtracted quantity $f_{\rm m}^{(n)}(t) - f_{\rm em}^{(n)}(t)$ to emphasise the stochastic variations in the time series, which are otherwise hard to discern visually on top of the secular spin-down trend $f_{\rm em}^{(n)}(t)$. \newline

In steps (ii) and (iii) above, we use the Python package \texttt{libstempo} \footnote{\url{https://github.com/vallis/libstempo/tree/master}} to correct the TOAs and then add back the noise perturbations. In \texttt{libstempo}, the GW background is generated according to a power-law PSD with index $\alpha$ and amplitude $A_{\rm gw}$, viz.
\begin{equation}
	P_{\rm GW}(f) = \frac{A_{\rm gw}^2}{12 \pi ^2} \left(\frac{f}{\text{yr}^{-1}}\right)^{- 3 + 2\alpha} \text {yr}^3 \, . \label{eq:psd_gw}
\end{equation}
Similarly, the red noise arising from spin wandering has PSD
\begin{equation}
	P_{\rm psr}(f) = A_{\rm psr}^2 \left(\frac{f}{\text{yr}^{-1}}\right)^{-\alpha_{\rm psr}} \text {yr}^3 \, . \label{eq:psd_psr}
\end{equation}
Both the PSDs in Equations \eqref{eq:psd_gw} and \eqref{eq:psd_psr} refer to the phase residuals. For the GW background, we take the same parameters as in MDC\_D1 and use $A_{\rm gw} = 5 \times 10^{-14}$ and $\alpha = -2/3$. The detectability of the GW background as a function of $A_{\rm gw}$ is investigated in Section \ref{sec:pta_detection}.  MDC\_D1 does not include spin wandering, so instead we set $A_{\rm psr} = 5.77 \times 10^{-22}$ and $\alpha_{\rm psr}=1.7$, copying ``dataset 3" of the open MDC. We follow MDC\_D1 and set the measurement uncertainty in the TOA, denoted by $\sigma_{\rm TOA}$, to be 100 ns for each pulsar. The parameters used to generate the synthetic data are summarised in Table \ref{tab:injected_parameters_mdc}.

\subsection{Detection}\label{sec:pta_detection}

	\begin{figure}
	\includegraphics[width=\columnwidth, height =\columnwidth]{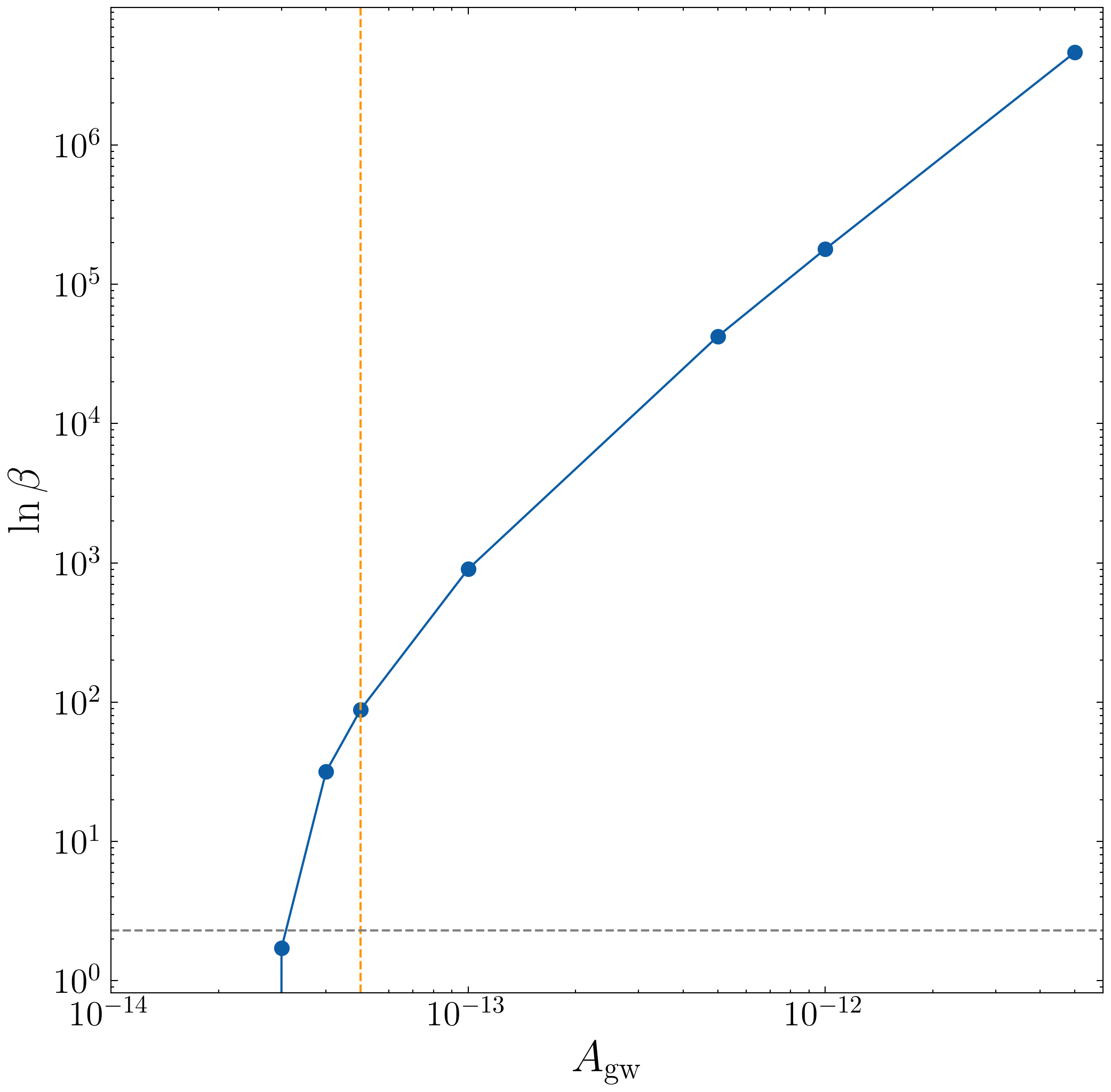} 	
	\caption{Log Bayes factor (odds ratio) $\ln \beta$ between the competing models $\mathcal{M}_{\rm gw}$ (GW present in data) and $\mathcal{M}_{\rm null}$ (GW not present in data) as a function of the GW amplitude, $A_{\rm gw}$, for the injected data in Table \ref{tab:injected_parameters_mdc}. The horizontal grey dashed line marks an arbitrary detection threshold, $\beta = 10$. The corresponding minimum detectable amplitude is $A_{\rm gw} \geq 3 \times 10^{-14}$ for $\beta \geq 10$. The vertical orange dashed line marks $A_{\rm gw}= 5 \times 10^{-14}$, the amplitude of the background in MDC\_D1, which is discussed in the main text. The axes are graduated on logarithmic scales. Consequently, for example, the lowest point on the vertical axis corresponds to $\ln\beta=10^{0}$ not $\beta=10^{0}$.}
	\label{fig:mdc_bayes}
\end{figure}
In this section we calculate the Bayesian evidence in favour of the presence of a stochastic GW background. We frame the detection problem in terms of Bayesian model selection. Specifically, we define $\mathcal{M}_{\rm gw}$ as the inference model that assumes a stochastic GW background exists in the data, i.e. the model described in Section \ref{sec:state_space_formulation}. We define $\mathcal{M}_{\rm null}$ as the null model that assumes no stochastic GW background exists in the data.  This is equivalent to setting $g^{(n)}(t) =1$ in Equation \eqref{eq:measurement}. Model $\mathcal{M}_{\rm gw}$ is parametrised by $\boldsymbol{\theta}$. Model $\mathcal{M}_{\rm null}$ is parametrised by $\boldsymbol{\theta}_{\rm psr}$. The support in the data for the presence of a stochastic GW background is quantified via the Bayes factor,
\begin{equation}
	\beta = \frac{\mathcal{Z}(\boldsymbol{y} | \mathcal{M}_{\rm gw})}{\mathcal{Z}(\boldsymbol{y} | \mathcal{M}_{\rm null})} \ . \label{eq:bayes}
\end{equation}
For the synthetic IPTA MDC data $\boldsymbol{y}$ generated by the procedure outlined in Section \ref{sec:ipta_data}, we infer with strong support the presence of a stochastic GW background, with $\ln \beta = 89$. \newline

Next, we consider $\beta$ as a function of the amplitude of the background,  $A_{\rm gw}$. The goal is to determine the minimum $A_{\rm gw}$ detectable by the NANOGrav-like PTA studied in this paper. We use the method outlined in Section \ref{sec:ipta_data} to produce multiple realisations of $\boldsymbol{y}(t)$ for different values of $A_{\rm gw}$. All other parameters are kept constant at the values summarised in Table \ref{tab:injected_parameters_mdc}. To control the test, the noise processes $\xi^{(n)}(t)$ and $\varepsilon^{(n)}(t)$ are identical realisations for every value of $A_{\rm gw}$; the only change from one $A_{\rm gw}$ value to the next is $A_{\rm gw}$ itself. The log Bayes factor $\ln \beta$ is plotted as a function of $A_{\rm gw}$ in Figure \ref{fig:mdc_bayes}. We vary the source amplitudes from $A_{\rm gw}= 10^{-14}$ (undetectable) to $A_{\rm gw} = 10^{-11}$ (easily detectable). The stochastic GW background is detectable with decisive evidence ($\beta \geq 10$) for $A_{\rm gw} \gtrsim 3 \times 10^{-14}$. The minimum detectable strain is particular to the system in Table \ref{tab:injected_parameters_mdc} and the specific realisation of $\boldsymbol{y}$. It is influenced in general by $T_{\rm obs}$, $T_{\rm cad}$ and ${\boldsymbol{\theta}}_{\rm psr}$. A full exploration of the sensitivity is postponed to future work, after the analysis scheme in this paper is upgraded to operate directly on pulse TOAs to enable a like-for-like comparison with standard PTA analyses. \newline

\subsection{Parameter estimation}\label{sec:mdc_pe}

	\begin{figure}
	\includegraphics[width=\columnwidth, height =\columnwidth]{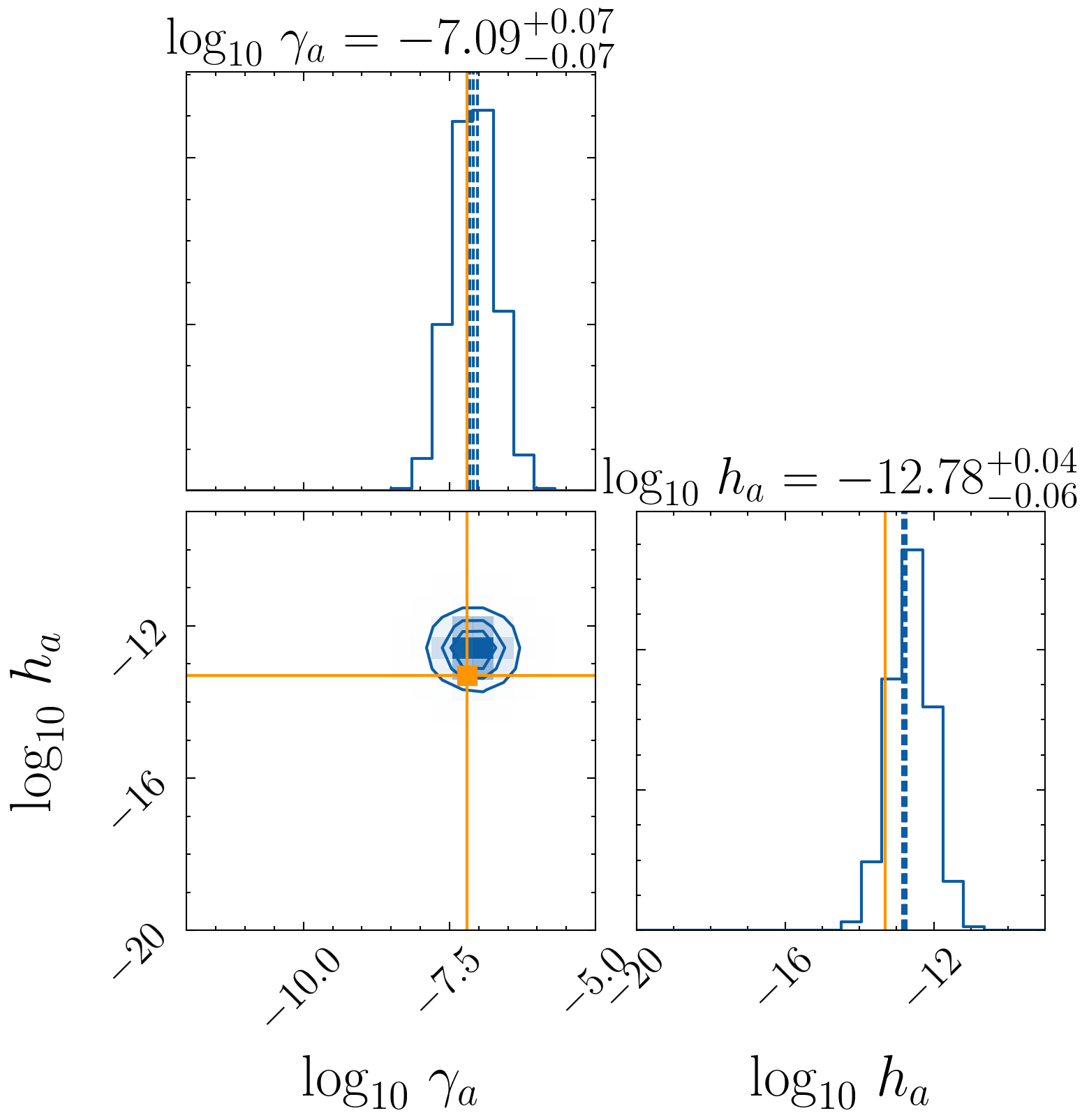} 	
	\caption{Posterior distribution $p \left( \boldsymbol{\theta}_{\rm gw} | \boldsymbol{y}\right)$ of the stochastic GW background parameters $\boldsymbol{\theta}_{\rm gw}$ for a single realisation of the synthetic data $\boldsymbol{y}$ generated for the injected parameters in Table \ref{tab:injected_parameters_mdc}. The contours in the two-dimensional histograms mark the (0.5, 1, 1.5, 2)-$\sigma$ levels. The one-dimensional histograms correspond to the joint posterior distribution marginalized over all but one parameter. The supertitles of the one-dimensional histograms record the median and the 0.16 and 0.84 quantiles. We plot the base-10 logarithm of the parameters, $\log_{10} \gamma_{\rm a}$ and $\log_{10} h_{\rm a}$. The state-space inference scheme converges to a well-behaved, unimodal posterior for both parameters in $\boldsymbol{\theta}_{\rm gw}$, as well as for the $3N$ parameters in $\boldsymbol{\theta}_{\rm psr}$ (not plotted). The horizontal axes span a subset of the prior domain for both parameters. The horizontal and vertical orange lines indicate the approximate injected values. The procedure for calculating the injected values is discussed in the main text.}
	\label{fig:corner_plot_mdc}
\end{figure}
In this section we apply the state-space analysis scheme to calculate the joint posterior probability distribution $p({\boldsymbol{\theta}} | {\boldsymbol{y}})$, where $\boldsymbol{y}$ is generated by the procedure described in Section \ref{sec:ipta_data}, using the injected parameters in Table \ref{tab:injected_parameters_mdc}. \newline

Figure \ref{fig:corner_plot_mdc} displays the posterior distribution for the two parameters in $\boldsymbol{\theta}_{\rm gw}$ in the form of a traditional corner plot. We plot base-10 logarithms on the horizontal and vertical axes. The histograms are the one-dimensional posteriors, marginalized over the other parameter. The two-dimensional contours mark the (0.5, 1, 1.5, 2)-sigma level surfaces. All histograms and contours are consistent with a unimodal joint posterior, with scant evidence of railing against the prior bounds. The modes of the one-dimensional posteriors are $-7.09_{+0.07}^{-0.07}$ for $ \log_{10} \gamma_{\rm a}$ and $-12.78_{+0.04}^{-0.06}$  for $ \log_{10} h_{\rm a}$, where the $\pm$ limits denote the 0.16 and 0.84 quantiles. The vertical and horizontal orange lines label the injected values. The inferred and injected values are in accord. \newline

The reader may wonder how the injected values are calculated, because the model used to generate the data, cf. Section \ref{sec:ipta_data}, is different from the model used for the purposes of inference, cf. Section \ref{sec:state_space_formulation}. From Equation \eqref{eq:gamma_a_min} we have $\gamma_{\rm a} \sim 2 \pi f_{\min}$, where $f_{\rm min}$ is the lower bound on the probability distribution of the frequencies over the SMBHB population and takes a default value in \texttt{libstempo} of $10^{-8}$ Hz.  From Equation \eqref{eq:sigma_a_expression} it is evident that $h_{\rm a}$ is an amplitude factor proportional to $A_{\rm gw}$. As a rough guide, and for illustrative purposes, we set $h_{\rm a}$ equal to $A_{\rm gw} = 5 \times 10^{-14}$ in Figure \ref{fig:corner_plot_mdc}. A fuller discussion relating the parameters of the inference model and the traditional parameters of PTA analyses is presented in terms of a comparison of PSDs in Appendix \ref{appendix:psd_OU}; see also Appendix \ref{appendix:justify_OU_process_for_background}. \newline

For the sake of brevity and visual clarity we do not display the calculated posterior distributions for the $3N$ parameters in $\boldsymbol{\theta}_{\rm psr}$, because inferring $\boldsymbol{\theta}_{\rm gw}$ is the main focus of this paper and most published PTA analyses. In general, $\boldsymbol{\theta}_{\rm psr}$ is recovered accurately. The estimates of $f_{\rm em}(t_1)$ and $\dot{f}_{\rm em}(t_1)$ are guided into narrow ranges by the narrow priors. The one-dimensional posteriors inferred for $\sigma^{(n)}$ are generally broad, consistent with the prior which covers 4 orders of magnitude. The $\sigma^{(n)}$ posteriors contain the injection within the 90\% credible interval in all cases. We remind the reader that $\gamma^{(n)}$ is not estimated in this paper. Astrophysically, it satisfies $\gamma^{(n)} \sim 10^{-5} T_{\rm obs}^{-1}$ \citep{Price2012,Myers2021MNRAS.502.3113M,Meyers2021,Vargas,2024MNRAS.tmp..891O}, so its influence is muted. 

\section{Computational cost}\label{sec:computation_costs}
The state-space algorithm in this paper, coupled with a generic nested sampler, runs roughly as fast as a traditional PTA analysis. Hence it is suitable for running in tandem as a cross-check. Moreover, the state-space algorithm scales favourably with increasing data volume, i.e. the number of pulsars in the array, or the number of time samples. In this section we benchmark the prototype software implementation of the state-space PTA analysis scheme in this paper, as a rough practical guide. We also discuss briefly the theoretical, asymptotic scaling of the computational cost as a function of $N$, the number of pulsars in the PTA, and $N_t$, the number of TOAs. We compare the asymptotic scaling with that of traditional PTA analysis methods, especially standard methods based on Gaussian processes \citep[][see also Appendix \ref{ap:kf_vs_gp}]{Haasteren}.\newline 

In this paper, as a first pass, we implement a prototype of the Kalman filter, without optimisation. As a benchmark, a single likelihood evaluation implemented in Python for the synthetic data presented in Section \ref{sec:validation} takes $T_{\mathcal{L}} \approx 0.05 \, {\rm s}$ of central processing unit (CPU) time on a 2.6 GHz  Intel Core i7 processor. Preliminary empirical tests yield $T_{\cal L} \propto N^{1.5} N_t$ roughly for $1 \leq N \leq 50$. This unoptimized scaling will be refined in future work. Translation into more performant languages such as C++ \citep{andrist2020c++} or Julia \citep{2012arXiv1209.5145B}, the use of Python pre-compilation libraries such as Numba \citep{2015llvm.confE...1L} or JAX \citep{jax2018github}, or additional optimisation \citep{gorelick2014high}, would reduce $T_{\mathcal{L}}$. \newline 

It is important to understand how $T_{\mathcal{L}}$ scales with both $N_t$ and $N$. Regarding the former, the theoretical time complexity of the Kalman filter (i.e. its asymptotic behaviour), is $\mathcal{O}(N_t)$, because the Kalman filter is an iterative algorithm. Regarding the latter, the rate-limiting step is set by matrix multiplication, e.g.\ in Equations \eqref{eq:EKF_predict2} and \eqref{eq:inn_covar} in the Kalman filter's predict and update steps respectively; see Appendix \ref{sec_kalman_general}. The dimension of the Kalman filter matrices, e.g. Equations \eqref{eq:kalman1}--\eqref{eq:update_2}, is set by $N$. Matrix multiplication without optimization scales as $\mathcal{O}(N^3)$ \citep{Daum2021}. Modern routines for matrix multiplication reduce the complexity to $\sim \mathcal{O}(N^{2.4})$ \citep{trefethen1997numerical}. Advanced techniques to further reduce the complexity include CKMS recursion \citep{kailath2000linear} or low-rank perturbation methods \citep{doi:10.1080/10618600.2012.760461}. We refer the reader to \cite{8861457} for further information about complexity reduction for the Kalman filter. The memory complexity of the Kalman filter  is $\mathcal{O}(N^{2})$, independent of $N_t$, because the algorithm is iterative. A naive implementation of Gaussian process regression \citep{rasmussen} has time complexity $\mathcal{O} \left(N_t^3\right)$ and memory complexity $\mathcal{O} \left(N_t^2\right)$ \newline

In traditional PTA analyses, approximation methods have been developed to improve the above scalings. They take advantage of the specific structure of the PTA covariance matrices and approximate the covariance function as a low rank matrix via a sum over $q$ Fourier modes \citep[e.g. Sections 3 \& 5, ][]{Haasteren}. These low-rank methods have time complexity $\mathcal{O} \left(q^2 N_t\right)$ and memory complexity $\mathcal{O} \left(q N_t\right)$ \newline 

\section{Conclusion}\label{sec:conclusion}
In this paper we demonstrate how to apply state-space methods for PTA data analysis to the stochastic GW background. Previous work on this topic focuses on individually resolvable SMBHBs \citep{KimpsonPTA1,KimpsonPTA2}. This paper applies state-space methods to the linear superposition of a large number of SMBHBs, which form a stochastic GW background.  The SMBHBs are not identifiable individually in the formal sense borrowed from engineering applications \citep{e5be7c83a0d24500826f6e1b414d1733}, but it is shown that their summed influence on the measured spin frequency of the $n$-th PTA pulsar can be modelled approximately through a random variable $a^{(n)}(t)$, which obeys a mean-reverting, Ornstein-Uhlenbeck process; see Equations \eqref{eq:ornstein_for_at}--\eqref{eq:correlation} and Appendix \ref{appendix:justify_OU_process_for_background}. Simultaneously, the rotational state of the $n$-th pulsar evolves according to another mean-reverting, Ornstein-Uhlenbeck process in the absence of a stochastic GW background but in the presence of achromatic timing noise (i.e. spin wandering); see Equations \eqref{eq:frequency_evolution}--\eqref{eq:spinevol} and Section \ref{sec:spin_evolution} \citep{Melatos2020ApJ...896...78M,Meyers2021,Myers2021MNRAS.502.3113M,Vargas,2024MNRAS.tmp..891O}. The measurement equations \eqref{eq:measurement}--\eqref{eq:vareps} and dynamical equations, Equations \eqref{eq:frequency_evolution}--\eqref{eq:spinevol} and \eqref{eq:ornstein_for_at}--\eqref{eq:correlation}, are in the correct mathematical form to be analysed by a Kalman filter, which inverts TOA data to infer the most probable state sequence $a^{(n)}(t)$. The Kalman filter is combined with a nested sampler to estimate the posterior distributions of the parameters of the state-space model, $\boldsymbol{\theta}_{\rm psr}$ and $\boldsymbol{\theta}_{\rm gw}$, with a particular focus on the elements of  $\boldsymbol{\theta}_{\rm gw}$, namely $h_{\rm a}$ and $\gamma_{\rm a}$. The associated Bayesian evidence $\mathcal{Z}$ of the model is also calculated. \newline  

The state-space model is tested on synthetic data obtained from the first IPTA MDC. We start by considering a stochastic background with amplitude $A_{\rm gw} =5 \times 10^{-14}$, observed by $N=37$ pulsars with $T_{\rm obs} = 5 \, {\rm years}$ and $T_{\rm cad} = 2$ weeks. The detection of the stochastic GW background with a state-space model is formulated in terms of Bayesian model selection. At $A_{\rm gw} =5 \times 10^{-14}$, the background is detected with strong support, with log Bayes factor $\ln \beta = 89$. It is found that the inference scheme can successfully detect injected signals with $\beta \geq 10$ for $A_{\rm gw} \geq 3 \times 10^{-14}$. The state-space model successfully converges to a well-behaved, unimodal posterior which does not rail against the prior bounds and matches the injected parameters used to generate the synthetic data.  \newline

State-space methods for PTA data analysis complement traditional techniques.  Traditional techniques average over the ensemble of possible timing noise realisations, to form and estimate the phase residual PSD. In contrast, state-space methods track the actual, astrophysical, time-ordered realisation of the timing noise in every PTA pulsar by harnessing the adaptive gain of a Kalman filter. The noise model utilised in state-space methods is related, but different, to the noise model in traditional techniques. Traditional techniques assume a stationary Gaussian process described by an ensemble-averaged, power-law PSD, whose amplitude and exponent are adjustable. In this paper, the state-space model assumes a mean-reverting Ornstein-Uhlenbeck process. The Ornstein-Uhlenbeck process also maps onto a stationary Gaussian process, whose PSD is a power law $\propto f^{-4}$ at high frequencies and $\propto f^{-2}$ at low frequencies, and whose amplitude and exponent can be adjusted by modifying the form of the damping term and Langevin driver in Equation \eqref{eq:frequency_evolution}; we refer the reader to Appendix \ref{appendix:psd_OU} for a discussion on the PSD of the Ornstein-Uhlenbeck process and a comparison with the GW PSD used in traditional analyses. We also refer the interested reader to Appendix B in \cite{KimpsonPTA1} for a pedagogical comparison of state-space and traditional PTA analysis methods. An in-depth study clarifying the similarities and difference between various approaches promises to be a fruitful avenue of future work. \newline

 It will be valuable in future to generalize the method in this paper in at least two ways.

\begin{enumerate}[leftmargin=2em]
	\item Real PTA analyses operate on data which contain both a stochastic GW background and multiple, individually resolvable SMBHB sources. State-space analysis schemes for PTA data now exist for both the stochastic GW background and individual SMBHBs, as set out in this paper and by \citet{KimpsonPTA1,KimpsonPTA2}. It is feasible to combine the two state-space approaches and separate the two signal contributions, as long as the individual SMBHBs are few enough to avoid formal identifiability obstacles \citep{e5be7c83a0d24500826f6e1b414d1733}. \newline  
	
	\item Published state-space analyses operate on a sequence of pulse frequencies $f_{\rm m}^{(n)}(t)$, instead of a sequence of pulse TOAs, as happens in traditional PTA analyses \citep[e.g.][]{Zhupulsarterms,Chen2022,2023ApJ...951L...8A,2023arXiv230616214A,2023ApJ...951L...6R,2023RAA....23g5024X,Arzoumanian2023,2023arXiv230616226A}. The generalisation of the state-space framework to accept pulse TOAs, necessary for application to real astrophysical data, is a subtle task and will be presented in a forthcoming paper.\newline

\end{enumerate}

%
%
%
%

\section*{Acknowledgements}
This research was supported by the Australian Research Council Centre of Excellence for Gravitational Wave Discovery (OzGrav), grant number CE170100004. The numerical calculations were performed on the OzSTAR supercomputer facility at Swinburne University of Technology. The OzSTAR program receives funding in part from the Astronomy National Collaborative Research Infrastructure Strategy (NCRIS) allocation provided by the Australian Government.

\section*{Data Availability}
No new data were generated or analysed in support of this research.

\bibliographystyle{mnras}
\bibliography{example} 

\appendix

\section{Stochastic GW background as an Ornstein-Uhlenbeck process}\label{appendix:justify_OU_process_for_background}

%
%
%
%
%
%
%
%

A central assumption in Section \ref{sec:state_space_formulation} is that the stochastic GW background can be approximated by an Ornstein-Uhlenbeck process through Equation \eqref{eq:ornstein_for_at}. In this appendix we justify the assumption. We show that a linear combination of $M$ individual, quasi-monochromatic GW sources can be represented at the $n$-th pulsar by a noisy time series $a^{(n)}(t)$ in the limit of large $M$, and that the statistics of $a^{(n)}(t)$ can be reproduced by a phenomenological Ornstein-Uhlenbeck model. The appendix is organised as follows. In Appendix \ref{app:review_of_individual_equations} we review the measurement equation for an individual, monochromatic planar GW used in \cite{KimpsonPTA1,KimpsonPTA2}, and demonstrate how this extends to $M$ GWs to construct $a(t)$. In Appendix \ref{app:stats_of_at} we derive the ensemble averaged statistics (mean and variance) of $a^{(n)}(t)$. In Appendix \ref{app:OU_rep} we demonstrate how the statistics of $a^{(n)}(t)$ can be represented by an Ornstein-Uhlenbeck process. In Appendix \ref{sec:posteriors_multiple} we outline how the parameters of the Ornstein-Uhlenbeck inference model $\boldsymbol{\theta}_{\rm gw}$ are related to the parameters of the data generation model, $\boldsymbol{\theta}_{\rm a}$, so as to identify approximate ``injection'' values for $\boldsymbol{\theta}_{\rm gw}$. In Appendix \ref{sec:dispersion} we discuss the dispersion in $p(\boldsymbol{\theta}_{\rm} | \boldsymbol{Y})$ for multiple realisations of $\boldsymbol{Y}$, cf. Figure \ref{fig:corner_plot_2}.

\subsection{Review of GW measurement equation for individual sources}\label{app:review_of_individual_equations}
In the presence of an individual, quasi-monochromatic GW, the measurement function $g^{(n)}(t)$ in Equation \eqref{eq:measurement} is given by \citep[e.g.][]{Maggiore,KimpsonPTA1,KimpsonPTA2}
\begin{align}
	g^{(n)}(t) =& 1 - \frac{ H_{ij}[q^{(n)}]^i [q^{(n)}]^j }{2 [1 + \boldsymbol{n}\cdot \boldsymbol{q}^{(n)}] } \nonumber \\
	& \times \Big[\cos\left(-\Omega t +\Phi_0\right) \nonumber \\
	&- \cos \left \{-\Omega t +\Phi_0 + \Omega \left[1 + \boldsymbol{n}\cdot \boldsymbol{q}^{(n)} \right]  d^{(n)} \right \} \Big ] \ ,
	\label{eq:g_func_trig}
\end{align}
where $[q^{(n)}]^i$ labels the $i$-th Cartesian component of the $n$-th pulsar's position vector $\boldsymbol{q}^{(n)}$, $\Omega$ is the constant angular frequency of the GW, $\boldsymbol{n}$ is a unit vector specifying the direction of propagation of the GW, $H_{ij}$ is the spatial part of the GW amplitude tensor, and $\Phi_0$ is the phase offset of the GW with respect to some reference time. \footnote{Equation \eqref{eq:g_func_trig} assumes that the GW is linearly polarised. In general, the GW emitted by an individual SMBHB is elliptically polarised \citep{Maggiore}, and Equation \eqref{eq:g_func_trig}  contains two additional terms, in which the cosines are replaced by sines, e.g. \cite{Arzoumanian2023}. In this paper, we copy some traditional Hellings-Downs analyses and omit the sine terms, because the contribution of the cross polarisation to second-order moments like $\langle a^{(n)}(t) a^{(n')}(t') \rangle$ averages to zero for an isotropic stochastic GW background \citep{2009PhRvD..79h4030A,2015AmJPh..83..635J,KimpsonPTA1}}
 \newline

In the presence of $M$ individual GW sources labelled by $1\leq m \leq M$, the measurement function becomes a linear superposition, viz.
\begin{align}
	g^{(n)}(t) =& 1 - a^{(n)}(t),
	\label{eq:g_func_sum}
\end{align}
with
\begin{equation}
	a^{(n)}(t) = \sum_{m=1}^{M} B^{(n,m)} C^{(n,m)}(t) \ ,  \label{eq:asummation}
\end{equation}
\begin{equation}
	B^{(n,m)}= \frac{ H_{ij}^{(m)}[q^{(n)}]^i [q^{(n)}]^j }{1 + \boldsymbol{n}^{(m)}\cdot \boldsymbol{q}^{(n)}} \ ,
\end{equation}
\begin{align}
	C^{(n,m)} =&\frac{1}{2}\cos\left[-\Omega^{(m)} t +\Phi_0\right] \nonumber \\
	&- \frac{1}{2}\cos \bigg\{-\Omega^{(m)} t +\Phi_0^{(m)} \nonumber \\ 
	&+ \Omega^{(m)} \left[1 + \boldsymbol{n}^{(m)}\cdot \boldsymbol{q}^{(n)} \right]  d^{(n)} \bigg \} \ . \label{eq:C_func_sum}
\end{align}
Equations \eqref{eq:g_func_sum}--\eqref{eq:C_func_sum} are equivalent to Equations \eqref{eq:gfunc}--\eqref{eq:delta_h}. For $M=1$, $a^{(n)}(t)$ is fully deterministic and the GW parameters $H_{ij}^{(1)}$, ${\boldsymbol{n}}^{(1)}$, $\Omega^{(1)}$, and $\Phi_0^{(1)}$ are formally identifiable \citep{e5be7c83a0d24500826f6e1b414d1733}; that is, they can be inferred uniquely from the TOA data. This is the situation considered in previous state-space analyses \citep[e.g.][]{KimpsonPTA1,KimpsonPTA2}. For $M \gg 1$, $a^{(n)}(t)$ is deterministic in principle, but in practice it is ``wiggly'' and behaves approximately like a random variable forming the stochastic GW background. The latter situation is considered in this paper.

\subsection{Ensemble-averaged statistics of the random variable $a^{(n)}(t)$ for $M\gg 1$}\label{app:stats_of_at}
In this section we derive the ensemble-averaged statistics of $a^{(n)}(t)$, namely the mean
$\langle a^{(n)}(t) \rangle$ and the variance $\langle a^{(n)}(t) a^{(n')}(t') \rangle$. The ensemble corresponds to many random realizations of the set of SMBHB source parameters $\boldsymbol{\theta}_{\rm a}$. The PDF of $\boldsymbol{\theta}_{\rm a}$ is discussed in Section \ref{sec:generate_a} and Appendix \ref{sec:tong}. \newline

Consider first $\langle a^{(n)}(t) \rangle$. In the absence of an astrophysical requirement to the contrary, we assume that $0\leq \Phi_0^{(m)} \leq 2\pi$ is distributed uniformly in its domain. \footnote{The same assumption holds for the polarization angle $\psi^{(m)}$ implicit in $H_{ij}^{(m)}$, if the SMBHBs are distributed isotropically \citep{2015AmJPh..83..635J,Maggiore}, and leads to the same conclusion, viz. $\langle a^{(n)}(t) \rangle = 0$.} The phase $\Phi_0^{(m)}$ enters $C^{(n,m)}$ but not $B^{(n,m)}$, so the ensemble average $\langle \dots \rangle_{\Phi_0^{(m)}}$ factorizes according to
\begin{align}
	\langle B^{(n,m)} C^{(n,m)} (t)\rangle_{\Phi_0^{(m)}}  &= B^{(n,m)} \langle C^{(n,m)} (t)\rangle_{\Phi_0^{(m)}}  \\ &= 0 \, . \label{eq:bc_average}
\end{align}
Equation \eqref{eq:bc_average} implies
\begin{eqnarray}
\langle a^{(n)}(t) \rangle = 0 \, . \label{eq:a_average_tot}
\end{eqnarray}
Equation \eqref{eq:a_average_tot} makes sense physically; $a^{(n)}(t)$ is a linear superposition of $M$ harmonic, zero-mean signals with random phases from $M$ SMBHBs.\newline

Next we consider the ensemble variance $\langle a^{(n)}(t) a^{(n')}(t') \rangle$, which can be expressed as
\begin{align}
	&\langle a^{(n)}(t) a^{(n')}(t') \rangle = \\  
	&=\left\langle \sum_{m=1}^{M} B^{(n,m)} C^{(n,m)}(t) \sum_{m'=1}^{M} B^{(n',m')} C^{(n',m')}(t') \right\rangle  \\
	&= \sum_{m,m'=1}^{M} \left\langle \left\langle B^{(n,m)} B^{(n',m')} \right \rangle_{\psi} \left\langle C^{(n,m)}(t) C^{(n',m')}(t') \right \rangle_{\Phi_0} \right\rangle_{h,\Omega,\boldsymbol{n}} \ . \label{eq:ensemble_varaince}
\end{align}
The nine Cartesian components of the random parameter $H_{ij}$ (suppressing henceforth the superscript $(m)$ for readability) can be expressed in terms of two independent quantities: a scalar amplitude $h$ and a polarization angle $\psi$. Both $h$ and $\psi$ appear in $B^{(n,m)}$ but not $C^{(n,m)}$, so it is tempting to average $B^{(n,m)} B^{(n',m')}$ over $h$ and $\Omega$ first without involving the factor $C^{(n,m)} C^{(n',m')}$. In general, however, $h$ and $\Omega$ are correlated; $h$ increases with $\Omega$, as a SMBHB inspiral proceeds, all else being equal. Therefore one should average $B^{(n,m)} B^{(n',m')}$ over $\psi$ by itself first, then average all four factors in Equation \eqref{eq:ensemble_varaince} jointly over $h$ and $\Omega$. Likewise, the phase $\Phi_0$ appears in $C^{(n,m)}$ but not $B^{(n,m)}$, so it is safe to average $C^{(n,m)}C^{(n',m')}$ over $\Phi_0$ by itself first. The direction of propagation ${\boldsymbol{n}}$ appears in both $B^{(n,m)}$ and $C^{(n,m)}$, so one should average all four factors in Equation \eqref{eq:ensemble_varaince} over ${\boldsymbol{n}}$. \newline 

In this paper, for the sake of simplicity, we assume that $h$ and $\Omega$ are uncorrelated in the inference model, so that their joint PDF factorizes, with $p(h,\Omega) = p(h) p(\Omega)$. In Section \ref{sec:average1} we calculate $\left\langle C^{(n,m)}(t) C^{(n',m')}(t') \right \rangle_{ \Phi_0,\Omega}$ and demonstrate that the result is approximately independent of $\boldsymbol{n}$. In Section \ref{sec:average2} we calculate $\left\langle \left\langle B^{(n,m)} B^{(n',m')} \right \rangle_{h, \psi} \right\rangle_{\boldsymbol{n}}$. In Section \ref{sec:average3} we bring together the two results and calculate $\langle a^{(n)}(t) a^{(n')}(t') \rangle$ by averaging over $h$ and $\Omega$. 

\subsubsection{Ensemble average $\left\langle C^{(n,m)}(t) C^{(n',m')}(t') \right \rangle_{\Phi_0,\Omega}$}\label{sec:average1}

We assume that the GW phase offset associated with every SMBHB is distributed uniformly in the interval $0 \leq \Phi_0^{(m)} \leq 2\pi$. The same assumption is made implicitly in standard Hellings-Downs analyses \citep{Maggiore}. The average over $\Phi_0^{(m)}$ and $\Phi_0^{(m')}$ then reduces to
\begin{align}
	&\left\langle C^{(n,m)}(t) C^{(n',m')}(t') \right \rangle_{\Phi_0,\Omega}  \nonumber \\
	=&   \frac{\delta_{m,m'}}{4} \left \langle \cos \left[-\Omega^{(m)} (t-t')\right] \right \rangle_{\Omega^{(m)}} \nonumber \\ 
	 &-\frac{\delta_{m,m'}}{4} \left \langle \cos \left \{ -\Omega^{(m)} \left[ t-t' +d^{(n')} +d^{(n')} \boldsymbol{n}^{(m)} \cdot \boldsymbol{q}^{(n')} \right] \right \} \right \rangle_{\Omega^{(m)}} \, .
	\label{eq:ensemble_varaincec}
\end{align}

The PDF of $\Omega$ after factorization, namely $p(\Omega)$, and the lower and upper bounds of its domain ($\Omega_{\rm min}$ and $\Omega_{\rm max}$ respectively) are defined by the power law in Equation \eqref{eq:power_law_omega}. We define the integral
\begin{eqnarray}
	I_1(t-t') = \int_{\Omega_{\rm min}^{(m)}}^{\Omega_{\rm max}^{(m)}} d \Omega^{(m)} \left[\Omega^{(m)} \right]^{-\kappa} \cos \left[-\Omega^{(m)}(t-t')\right] \, .\label{eq:I1integral}
\end{eqnarray}
With the definition in Equation \eqref{eq:I1integral}, Equation \eqref{eq:ensemble_varaincec} simplifies to
\begin{align}
	&\left\langle C^{(n,m)}(t) C^{(n',m')}(t') \right \rangle_{\Phi_0, \Omega}  \nonumber \\
	&=  \frac{K_1 \delta_{m,m'} }{4} \left \{I_1(t-t') -I_1 \left[t-t' +d^{(n')} +d^{(n')} \boldsymbol{n}^{(m)} \cdot \boldsymbol{q}^{(n')}\right] \right \} \ . \label{eq:apprenix_CC}
\end{align}
Consider the argument of $I_1$ in the second term in Equation \eqref{eq:apprenix_CC}. For a typical PTA, one has $10^3 \, {\rm yr} \sim d^{(n')} \gg | t - t'| \sim 10 \, {\rm yr}$ and hence, $d^{(n')} +d^{(n')} \boldsymbol{n}^{(m)} \cdot \boldsymbol{q}^{(n')} \gg t-t'$ for all but a small set of orientations with $\boldsymbol{n}^{(m)} \cdot \boldsymbol{q}^{(n')} \approx -1$. Consequently to a good approximation Equation \eqref{eq:apprenix_CC} becomes
\begin{equation}
	\left\langle C^{(n,m)}(t) C^{(n',m')}(t') \right \rangle_{\Phi_0,\Omega}  \approx \frac{K_1}{4} I_1(t-t') \, .\label{eq:final_C_eqn}
\end{equation}
Equation \eqref{eq:final_C_eqn} is independent of $\boldsymbol{n}$. Consequently, Equation \eqref{eq:ensemble_varaince} can be written as
\begin{align}
	&\langle a^{(n)}(t) a^{(n')}(t') \rangle \nonumber \\  
	&= \left\langle \left\langle B^{(n,m)} B^{(n',m')} \right \rangle_{\psi, h} \right\rangle_{\boldsymbol{n}}  \left\langle C^{(n,m)}(t) C^{(n',m')}(t') \right \rangle_{\Phi_0,\Omega}  \ .
	\label{eq:ensemble_varaince2}
\end{align}

\subsubsection{Ensemble average $\left\langle \left\langle B^{(n,m)} B^{(n',m')} \right \rangle_{H_{ij}} \right\rangle_{\boldsymbol{n}}$ }\label{sec:average2}

The familiar Hellings-Downs cross-correlation between the pulse frequency fluctuations of the $n$-th and $n'$-th pulsars is proportional to the ensemble average of $B^{(n,m)} B^{(n',m')}$ over $H_{ij}$ and ${\boldsymbol{n}}$ \citep{Hellings,Maggiore,2015AmJPh..83..635J}. Specifically, assuming that the background is homogeneous and isotropic and contains an equal admixture of plus and cross polarizations, one obtains the familiar quadrupolar sky pattern
\begin{align}
	&\left\langle \left\langle B^{(n,m)} B^{(n',m')} \right \rangle_{H_{ij}} \right\rangle_{\boldsymbol{n}} = \frac{1}{6} \left \langle h^2 \right \rangle \rho \left[\theta^{(n,n')}\right]
	\label{eq:ensemble_varaince123}
\end{align}
where $\rho \left[\theta^{(n,n')}\right]$ is defined by Equation \eqref{eq:correlation} and $\theta^{(n,n')}$ is the angle between the lines of sight to the $n$-th and $n'$-th pulsars in the array. 

\subsubsection{Ensemble average $\langle a^{(n)}(t) a^{(n')}(t') \rangle$ }\label{sec:average3}
Equations \eqref{eq:ensemble_varaince}, \eqref{eq:final_C_eqn}, \eqref{eq:ensemble_varaince2} and \eqref{eq:ensemble_varaince123} combine to give
\begin{align}
	\langle a^{(n)}(t) a^{(n')}(t') \rangle = \frac{M}{24} \left \langle h^2 \right \rangle \rho \left[\theta^{(n,n')}\right]  \frac{I_1(t-t')}{I_1(0)} \ . \label{eq:a_variance}
\end{align}
where we use the normalisation in Equation \eqref{eq:omega_norm} to write $I_1(0) = K_1^{-1}$, and we perform the double sum over $m$ and $m'$ in Equation \eqref{eq:ensemble_varaince} with the aid of the Kronecker deltas in Equation \eqref{eq:apprenix_CC}. We define $\langle h^2 \rangle $ as the total mean square wave strain from $M$ sources as measured at Earth, and assume an isotropic background of GW sources.

\subsection{Ornstein-Uhlenbeck approximation}\label{app:OU_rep}

\begin{figure}
	\includegraphics[width=\columnwidth, height=\columnwidth]{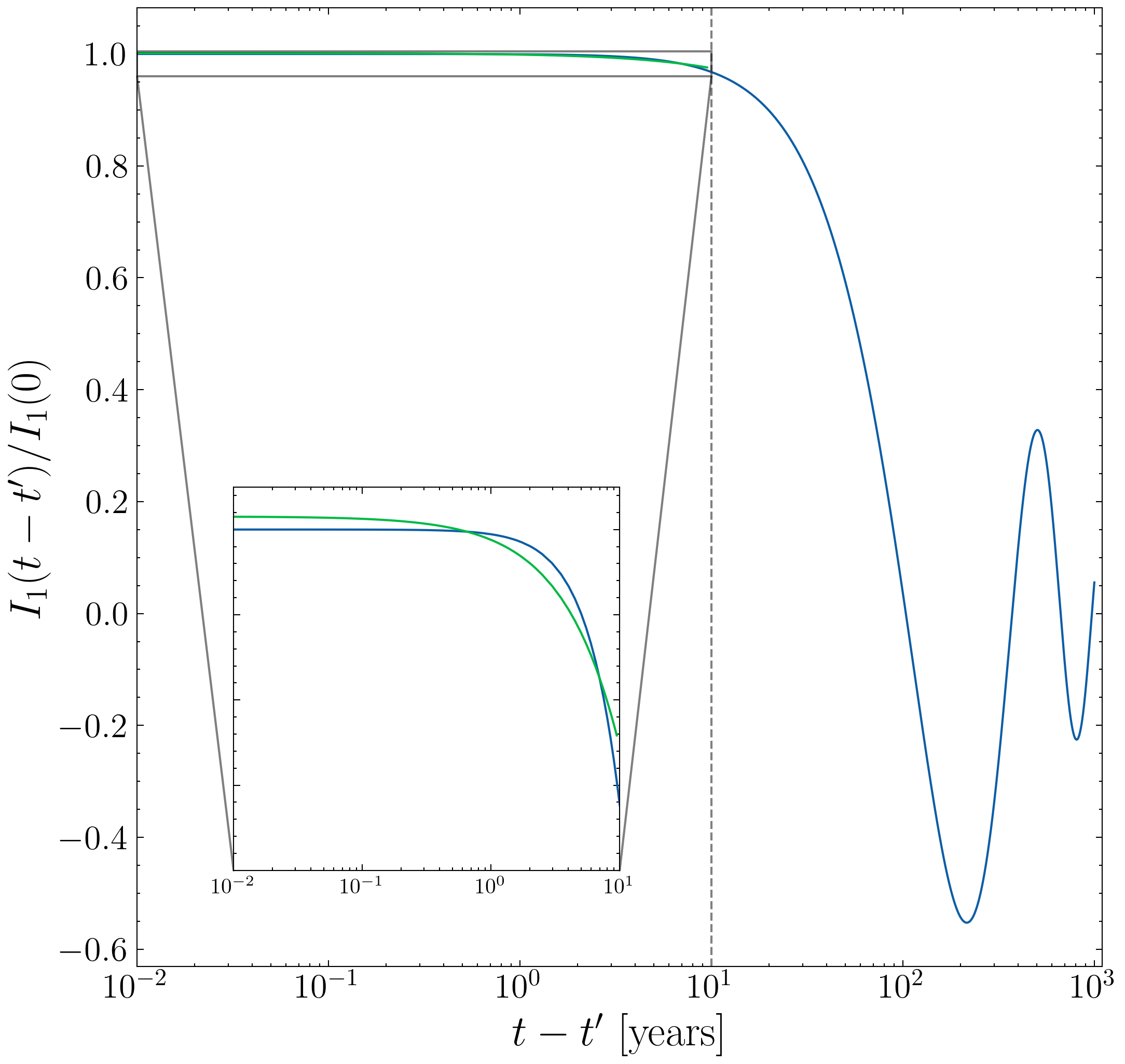} 	
	\caption{Normalized cross-correlation $\langle a^{(n)}(t) a^{(n')}(t') \rangle \propto I_1(t-t')/I_1(0)$ as a function of the time lag $t-t'$ (units: yr) for $10^{-2} \leq (t-t')/(1\, {\rm yr}) \leq 10^3$ (main panel) and $10^{-2} \leq (t-t')/(1 \, {\rm yr}) \leq 10$ (inset panel; magnified view). The exact numerical integral given by Equation \eqref{eq:I1integral} and the exponential approximation given by Equation \eqref{eq:I1integral_approx1} are plotted as solid blue and green curves respectively. The region to the left of the vertical grey dashed line is relevant to the current generation of PTA experiments. Physical parameters: $\Omega_{\rm min} = 10^{-2}$ years$^{-1}$,  $\Omega_{\rm max} = 10^{3}$ years$^{-1}$, $\kappa = 3$.}
	\label{fig:rollover_of_I1}
\end{figure}
We now show how the ensemble-averaged statistics of $a^{(n)}(t)$ derived in Appendix \ref{app:stats_of_at}, namely Equation \eqref{eq:a_average_tot} and Equation \eqref{eq:a_variance}, are consistent with approximating $a^{(n)}(t)$ as an Ornstein-Uhlenbeck process. \newline 

First consider the normalized integral $I_1(t-t') / I_1(0)$ given by Equation \eqref{eq:I1integral}. In the main panel of Figure \ref{fig:rollover_of_I1} the numerical solution to Equation \eqref{eq:I1integral} is plotted (solid blue curve) for $10^{-2} \leq (t - t')/(1 \, {\rm yr}) \leq 10^3$, a generous domain which more than covers the current generation of PTA experiments. To the left of the vertical grey dashed line, one has $\Omega^{(m)} |t - t'|  \lesssim \pi$, and $I_1(t-t')/I_1(0)$ decreases monotonically. To the right of the vertical grey dashed line, one has $\Omega^{(m)} | t - t' |$, and $I_1(t-t')/I_1(0)$ oscillates due to the cosine in the integrand of Equation \eqref{eq:I1integral}. Current PTA experiments operate to the left of the vertical grey dashed line, i.e.\ $|t-t'| \lesssim 10 \, {\rm yr}$, where $I_1(t-t') / I_1(0)$ can be approximated as an exponential function
\begin{eqnarray}
	\frac{I_1(t-t')}{I_1(0)} = e^{- 2.7 \times 10^{-2} \, \Omega_{\rm min} (t-t')} \, . \label{eq:I1integral_approx1}
\end{eqnarray}
The approximation is plotted as a solid green curve in Figure \ref{fig:rollover_of_I1}. The inset panel in Figure \ref{fig:rollover_of_I1} displays a magnified view of Equation \eqref{eq:I1integral} and Equation \eqref{eq:I1integral_approx1} for $10^{-2} \leq (t-t')/(1 \, {\rm yr}) \leq 10$. The value of the prefactor in the exponential, $2.7 \times 10^{-2}$ is obtained by using a non-linear least squares algorithm to fit Equation \eqref{eq:I1integral_approx1} to the solution given by Equation \eqref{eq:I1integral} (solid blue curve) in the regime $10^{-2} \leq (t-t')/(1 \, {\rm yr}) \leq 10$. The approximation is accurate to better than $0.5 \%$ across the domain plotted in the inset panel. By combining Equations  \eqref{eq:a_variance} and \eqref{eq:I1integral_approx1} we obtain approximately
\begin{equation}
	\langle a^{(n)}(t) a^{(n')}(t') \rangle = \frac{1}{24} \left \langle h^2 \right \rangle \rho \left[\theta^{(n,n')}\right]  e^{-2.7 \times 10^{-2} \, \Omega_{\rm min}  (t-t')} \, , \label{eq:a_variance22}
\end{equation}
for $|t-t'| \lesssim 10\, {\rm ys}$. \newline

How do the ensemble-averaged statistics of $a^{(n)}(t)$ in Appendix \ref{app:stats_of_at}, as well as the cross-correlation in Equation \eqref{eq:a_variance22}, relate to an Ornstein-Uhlenbeck process? Consider a description of a random variable $a^{(n)}_{\rm OU}(t)$ that obeys an Ornstein-Uhlenbeck process, described by the Langevin equation
\begin{eqnarray}
	\frac{d a^{(n)}_{\rm OU}}{dt} = -\gamma^{(n)}_{\rm a} a^{(n)}_{\rm OU} + \Xi_{\rm a}^{(n)}(t) \, , \label{eq:ornstein_for_at1}
\end{eqnarray}
which is identical to Equation \eqref{eq:ornstein_for_at}. The definitions of $\gamma_{\rm a}^{(n)}$ and $\Xi_{\rm a}^{(n)}(t)$ are given in Section \ref{sec:gw_background_modulation}. We use the subscript ``OU'' temporarily to emphasise that $a_{\rm OU}^{(n)}(t)$ is not the same as $a^{(n)}(t)$, which is generated by summing the deterministic, harmonic signals from $M$ SMBHBs according to Equation \eqref{eq:asummation}. We now show that the ensemble-averaged statistics of $a^{(n)}(t)$ and $a^{(n)}_{\rm OU}(t)$ are approximately equal, starting from the standard textbook result \citep{gardiner2009stochastic,calin2015informal}, 
\begin{align}
	\langle a^{(n)}_{\rm OU}(t) a^{(n')}_{\rm OU}(t') \rangle =&  \frac{[\sigma^{(n,n')}_{\rm a}]^2}{2 \gamma^{(n)}_{\rm a}} \nonumber \\ 
	& \times \left[e^{-\gamma^{(n)}_{\rm a}|t-t'|} - e^{-\gamma^{(n)}_{\rm a}(t+t')} \right] \ . \label{eq:variance_of_ou_process}
\end{align}
In general the Ornstein-Uhlenbeck process is not stationary, as evidenced by the $t+t'$ term in Equation \eqref{eq:variance_of_ou_process}; the statistics are not exclusively a function of the lag, $|t-t'|$. However, if the process $a^{(n)}(t)$ starts in the remote past, it is approximately stationary for $t, t' \gg [\gamma_{\rm a}^{(n)}]^{-1}$, consistent with standard PTA analyses \citep[e.g.][]{Maggiore,NanoGrav2018,Arzoumanian2023}. The stationary component of Equation \eqref{eq:variance_of_ou_process} for $t,t' \gg [ \gamma_{\rm a}^{(n)} ]^{-1}$ is given by
\begin{eqnarray}
	\langle a^{(n)}_{\rm OU}(t) a^{(n')}_{\rm OU}(t') \rangle  \approx  \frac{[\sigma^{(n,n')}_{\rm a}]^2}{2 \gamma^{(n)}_{\rm a}} e^{-\gamma_{\rm a}|t-t'|} \ . \label{eq:variance_of_ou_process2}
\end{eqnarray}
By comparing Equation \eqref{eq:a_variance22} with Equation \eqref{eq:variance_of_ou_process2} it is evident that the ensemble-averaged statistics of $a^{(n)}(t)$ from $M \gg 1$ SMBHBs can be reproduced approximately by an Ornstein-Uhlenbeck process with
\begin{align}
\gamma^{(n)}_{\rm a} &=  2.7 \times 10^{-2} \, \Omega_{\rm min} \, , \label{eq:lamnda_gamma} \\
\left[ \sigma^{(n,n')} \right]^2 &= \frac{\left \langle h^2\right \rangle }{12} \gamma_{\rm a}^{(n)} \rho \left[ \theta^{(n,n')} \right] \, .
\end{align}
In Equation \eqref{eq:lamnda_gamma} $\gamma^{(n)}_{\rm a}$ is the same for all $n$. It is written as $\gamma_{\rm a}$ without the superscript in the main text, e.g.\ in Section \ref{sec:gw_background_modulation}.

\section{Extended Kalman filter} \label{sec:kalman}
In this appendix, we describe the extended Kalman filter (EKF) algorithm used in this paper. The Kalman filter ingests a temporal sequence of noisy measurements, $\boldsymbol{Y}(t)$, and recovers a temporal sequence of stochastically evolving state variables, $\boldsymbol{X}(t)$, which are not directly observed. Previous state-space analyses of individual continuous GW sources have used a linear Kalman filter \citep[e.g.][]{KimpsonPTA1,KimpsonPTA2}. In this paper the non-linear relation between ${\boldsymbol{X}}(t)$ and ${\boldsymbol{Y}}(t)$ necessitates the use of the EKF \citep{zarchan2000fundamentals}. General recursion relations for the discrete-time EKF are written down for an arbitrary dynamical system in Appendix \ref{sec_kalman_general}. The mapping onto the specific continuous-time state-space model in Section \ref{sec:state_space_formulation} is written down in Appendix \ref{sec_kalman_specific}. The reader who wishes to reproduce the results in this paper can do so entirely by using the equations in Appendices \ref{sec_kalman_general} and \ref{sec_kalman_specific}.

\subsection{General recursion equations}\label{sec_kalman_general}
The EKF operates on temporally discrete, noisy measurements $\boldsymbol{Y}_k$, which are related to a set of unobservable discrete system states $\boldsymbol{X}_k$, via a general differentiable function, viz.
\begin{equation}
	\boldsymbol{Y}_k = h \left(\boldsymbol{X}_k\right) + \boldsymbol{v}_k \ ,\label{eq:kalman1}
\end{equation}
where $h(\boldsymbol{X}_k)$ is a differentiable measurement function acting on the state, $\boldsymbol{v}_k \sim {\cal N}(0,\boldsymbol{R}_k)$
is a zero-mean Gaussian measurement noise variable with covariance $\boldsymbol{R}_k$, and the subscript $k$ labels the time-step. Similarly, the underlying states evolve as
\begin{equation}
	\boldsymbol{X}_k = f \left(\boldsymbol{X}_{k-1}\right) + \boldsymbol{w}_{k-1} \ ,\label{eq:kalman1}
\end{equation}
where $f(\boldsymbol{X}_k)$ is a differentiable state transition function, and $\boldsymbol{w}_k \sim {\cal N}(0,\boldsymbol{Q}_k)$ is a zero-mean Gaussian process noise variable, with covariance $\boldsymbol{Q}_k$. \newline

As with the linear Kalman filter, the EKF is a recursive estimator involving a ``predict'' stage and an ``update'' stage. The predict stage predicts $\hat{\boldsymbol{X}}_{k|k-1}$, the estimate of the state at step $k$, given the state estimate from step $k-1$. Specifically, the predict step proceeds as
\begin{align}
	\hat{\boldsymbol{X}}_{k|k-1} &=  f \left( \hat{\boldsymbol{X}}_{k-1|k-1}\right) \, , \label{eq:EKF_predict1}\\
	\hat{\boldsymbol{P}}_{k|k-1} &=  \boldsymbol{F}_k \hat{\boldsymbol{P}}_{k-1|k-1} \boldsymbol{F}_k^\intercal + \boldsymbol{Q}_k  \label{eq:EKF_predict2} \ ,
\end{align}
where $\hat{\boldsymbol{P}}_{k|k-1}$ is the covariance of the prediction and $\boldsymbol{F}_k$ is the Jacobian
\begin{equation}
	\boldsymbol{F}_k = \left. \frac{\partial f}{\partial \boldsymbol{X} } \right \rvert_{	\hat{\boldsymbol{X}}_{k|k-1}} \ .
\end{equation}

The predict stage is independent of the measurements. The information in $\boldsymbol{Y}_k$ is included by updating the prediction during the update stage as follows:
\begin{align}
	\boldsymbol{\epsilon}_{k} &= \boldsymbol{Y}_k - h \left(\hat{\boldsymbol{X}}_{k|k-1} \right)\ , \label{eq:update_1}\\
	\boldsymbol{S}_k &= \boldsymbol{H}_k \hat{\boldsymbol{P}}_{k|k-1} \boldsymbol{H}_k^\intercal + \boldsymbol{R}_k \ , \label{eq:inn_covar} \\
	\boldsymbol{K}_k &= \hat{\boldsymbol{P}}_{k|k-1} \boldsymbol{H}_k^\intercal \boldsymbol{S}_k^{-1} \ ,\label{eq:kalman gain} \\
	\hat{\boldsymbol{X}}_{k|k} &=\hat{\boldsymbol{X}}_{k|k-1} +\boldsymbol{K}_k  \boldsymbol{\epsilon}_{k}  \ , \label{eq:kalmangainupdate} \\
	\hat{\boldsymbol{P}}_{k|k} &= \left( \boldsymbol{I} - \boldsymbol{K}_k \boldsymbol{H}_k \right) 	\hat{\boldsymbol{P}}_{k|k-1} \, ,
\end{align}
where $\boldsymbol{H}_k$ is the Jacobian
\begin{equation}
	\boldsymbol{H}_k = \left. \frac{\partial h}{\partial \boldsymbol{X} } \right \rvert_{	\hat{\boldsymbol{X}}_{k|k-1}} \ . \label{eq:update_2}
\end{equation}
For a full review of the EKF, including its derivation, we refer the reader to \cite{zarchan2000fundamentals}. \newline

To apply the EKF in practice means specifying the vectors $\boldsymbol{X}_k$
and $\boldsymbol{Y}_k$, the two functions $h(\boldsymbol{X}_k)$ and $f(\boldsymbol{X}_k)$, their associated Jacobians $\boldsymbol{H}_k$ and $\boldsymbol{F}_k$, and the covariance matrices $\boldsymbol{Q}_k$ and $\boldsymbol{R}_k$. In Section \ref{sec_kalman_specific} we describe how these components are determined for the state-space model in Section \ref{sec:state_space_formulation}.

\subsection{Application to a PTA analysis}\label{sec_kalman_specific}
We apply the Kalman recursion relations in Appendix \ref{sec_kalman_general} to the state-space model of a PTA composed of $N$ pulsars described in Section \ref{sec:state_space_formulation} as follows. \newline 

Let us define for convenience the ephemeris-subtracted variables
\begin{eqnarray}
	f_{\rm p}^{*(n)} = f_{\rm p}^{(n)} - f_{\rm em}^{(n)}(t) \, , \label{eq:ephsub1}\\
		f_{\rm m}^{*(n)} = f_{\rm m}^{(n)} - f_{\rm em}^{\diamond(n)}(t) \label{eq:ephsub2}  \, ,
\end{eqnarray}
where  $f_{\rm em}^{(n)}(t)$ is given by Equation \eqref{eq:spinevol} and $f_{\rm em}^{\diamond(n)}(t)$ is based on the ephemeris returned by {\sc tempo2}, which is measured to high accuracy in practice. For synthetic data we can set $ f_{\rm em}^{\diamond (n)} = f_{\rm em}^{(n)}$ without loss of generality, but this is impossible generally for astronomical observations, because the ephemeris is only known approximately. We emphasise that the change of variables in Equations \eqref{eq:ephsub1} and \eqref{eq:ephsub1} is a convenient device to simplify the Kalman filter equations and to bring the numerical values into a reasonable dynamic range,  avoiding numerical errors due to finite precision arithmetic and sidestepping the requirement for excessively long floating point formats (e.g. long double, quadruple). It does not remove any degrees of freedom nor does it involve an approximation. A similar reparameterisation is used in other states-space analyses of PTA experiments \citep[e.g.][]{KimpsonPTA1,KimpsonPTA2}. \newline

We now identify $\boldsymbol{X}(t)$ with a vector of length $2N$ composed of the intrinsic ephemeris-subtracted pulsar frequency state $f^{*(n)}_{\rm p}(t)$, and the stochastic GW background $a^{(n)}(t)$, viz.
\begin{equation}
	\boldsymbol{X}(t) = \left[f_{\rm p}^{*(1)}(t), \dots , f_{\rm p}^{*(N)}(t),a^{(1)}(t), \dots , a^{(N)}(t)\right] \ .
\end{equation}
We package the ephemeris-subtracted, measured pulsar frequencies as
\begin{equation}
	\boldsymbol{Y}(t) = \left[f_{\rm m}^{*(1)}(t), \dots , f_{\rm m}^{*(N)}(t) \right] \ .
\end{equation}
In Section \ref{sec:state_space_formulation} we assume that the states evolve linearly according to the continuous stochastic (Langevin) differential equation 
\begin{equation}
	\frac{d \boldsymbol{X}}{dt} = \boldsymbol{A} \boldsymbol{X} +  \boldsymbol{\Sigma}(t) \ , \label{eq:kalmn2}
\end{equation}
according to Equations \eqref{eq:frequency_evolution} and \eqref{eq:ornstein_for_at}. In Equation \eqref{eq:kalmn2}, $\boldsymbol{A}$ is a diagonal $2N \times 2N$ block matrix,
\begin{equation}
	\boldsymbol{A} = \begin{pmatrix}
		-\boldsymbol{\gamma}_{\rm p} & 0 \\
		0 & -\boldsymbol{\gamma}_{\rm a} 
	\end{pmatrix} 
\ ,
\end{equation}
with
\begin{equation}
\boldsymbol{\gamma}_{\rm p} = \text{diag} \left[\gamma^{(1)}, ... , \gamma^{(N)}\right] \, ,
\end{equation}
and 
\begin{equation}
	\boldsymbol{\gamma}_{\rm a} = \gamma_{\rm a} I_{N} \, ,
\end{equation}
where $I_{N}$ denotes the identity matrix of dimension $N$. Furthermore, $\boldsymbol{\Sigma}(t)$ is a vector of length $2N$,
 \begin{equation}
 	\boldsymbol{\Sigma}(t) = \left[\xi^{(1)}(t), \dots, \xi^{(N)}(t),\Xi_{\rm a}^{(1)}(t), \dots, \Xi_{\rm a}^{(N)}(t) \right]
 \end{equation}
 where $\xi^{(n)}(t)$ is defined by Equation \eqref{eq:xieqn_new}
 and $\Xi_{\rm a}^{(n)}(t)$ is defined by Equation \eqref{eq:xieqn2}. \newline

The general solution to Equation \eqref{eq:kalmn2} is given by \citep{gardiner2009stochastic},
\begin{equation}
	\boldsymbol{X}(t) = e^{\boldsymbol{A} t} \boldsymbol{X}(0) + \int_0^t dt' e^{\boldsymbol{A}(t-t')} \boldsymbol{\Sigma}(t') \ . \label{eq:gardenier}
\end{equation} 
From Equation \eqref{eq:gardenier} we construct the discrete, recursive solution for $\boldsymbol{X}(t_k) = \boldsymbol{X}_k$ following the template in Appendix \ref{sec_kalman_general}. The result is
\begin{equation}
	\boldsymbol{X}_k = \boldsymbol{F}_k \boldsymbol{X}_{k-1} + \boldsymbol{w}_k \ , \label{eq:kalman2}
\end{equation}
with 
\begin{align}
	\boldsymbol{F}_k &= e^{\boldsymbol{A} \Delta t } \, ,  \\
		\boldsymbol{w}_k &= \int_{t_k}^{t_{k+1}} d \boldsymbol{B}(t') e^{\boldsymbol{A}\left( t_{k+1} - t' \right)} \boldsymbol{\Sigma}(t') \ .  \label{eq:appendix_noise}
\end{align}
where $d \boldsymbol{B}(t')$ denotes a standard Brownian increment and we write $\Delta t = t_{k+1} - t_{k}$. From Equation \eqref{eq:appendix_noise} the process noise covariance matrix is
\begin{align}
	\boldsymbol{Q}_k \boldsymbol{\delta}_{kj}&= \langle \boldsymbol{w}_k \boldsymbol{w}_j^\intercal \rangle \\
	&= \begin{pmatrix}
		\boldsymbol{Q}_{\rm p} & 0  \\
		0 & \boldsymbol{Q}_{\rm a} 
	\end{pmatrix} \, , \label{eq:qdelta}
\end{align}
with 
\begin{equation}
	\boldsymbol{Q}_{\rm p} = \text{diag} \left[Q^{(1)}, \dots, Q^{(N)} \right] \, , \label{eq:q_vector}
\end{equation}
and
\begin{equation}
	\boldsymbol{Q}_{\rm a} = \frac{\langle h^2\rangle}{12} \boldsymbol{\Gamma}  \left(1 - e^{-2 \gamma_{\rm a} \Delta t}\right) 
	\, . \label{eq:q_matrix_term}
\end{equation}
 In Equation \eqref{eq:qdelta}, the Einstein summation convention is suppressed temporarily, i.e.\ there is no summation implied over the repeated index $k$. In Equation \eqref{eq:q_vector},	$Q^{(n)}$ is given by 
\begin{equation}
	Q^{(n)} = \frac{[\sigma^{(n)}]^2}{2 \gamma^{(n)}} \left[ 1 - e^{-2 \gamma^{(n)} \Delta t}\right] \ .
\end{equation}
In Equation \eqref{eq:q_matrix_term}, $\boldsymbol{\Gamma}$ is given by
\begin{eqnarray}
	\boldsymbol{\Gamma} = \begin{Bmatrix}
		1 & \Gamma \left[\theta^{(1,2)}\right] & \dots & \Gamma\left[\theta^{(1,N)}\right] \\
		\Gamma\left[\theta^{(2,1)}\right]  &1 & \dots & \Gamma\left[\theta^{(2,N)}\right] \\
		\dots  &\dots & \dots & \dots \\
		\Gamma\left[\theta^{(N,1)}\right]  &\Gamma\left[\theta^{(N,2)}\right] & \dots & 1
	\end{Bmatrix}&
	\, , \label{eq:q_matrix_term2}
\end{eqnarray}
where the scalar Hellings-Downs correlation function, $\Gamma(\theta)$, is defined by Equation \eqref{eq:correlation}. \newline

Due to the linear evolution of $\boldsymbol{X}(t)$, the non-linear predict stage of the EKF, viz. Equations \eqref{eq:EKF_predict1} and \eqref{eq:EKF_predict2}, can be replaced by the predict step from the linear Kalman filter, viz.
\begin{align}
	\hat{\boldsymbol{X}}_{k|k-1} &=  \boldsymbol{F}_k \hat{\boldsymbol{X}}_{k-1|k-1}  \ , \\
	\hat{\boldsymbol{P}}_{k|k-1} &=  \boldsymbol{F}_k \hat{\boldsymbol{P}}_{k-1|k-1} \boldsymbol{F}_k^\intercal + \boldsymbol{Q}_k  \, .
\end{align}
The relation between $\boldsymbol{X}(t)$ and $\boldsymbol{Y}(t)$ remains non-linear, viz.
\begin{equation}
	f_{\rm m}^{*(n)}(t) =  \left[1 - a^{(n)}(t)\right] f_{\rm p}^{*(n)}  -a^{(n)}(t)f_{\rm em}^{(n)}(t)+  \varepsilon^{(n)}(t)\ , \label{eq:hfunc_final}
\end{equation}
from Equation \eqref {eq:measurement}. As such, the non-linear EKF algorithm for the update stage, Equations 
\eqref{eq:update_1}--\eqref{eq:update_2}, is retained. The remaining unspecified component matrices of the EKF are the measurement function $h(\boldsymbol{X}_k)$, its Jacobian $\boldsymbol{H}_k$, and the measurement covariance matrix $\boldsymbol{R}_k$.  These are derived straightforwardly from Equation \eqref{eq:hfunc_final}, with 
\begin{eqnarray}
h(\boldsymbol{X}_k) = \begin{pmatrix}
	 \left[1 - a^{(1)}(t)\right] f_{\rm p}^{*(1)}(t)  -a^{(1)}(t)f_{\rm em}^{(1)}(t) \vspace{1mm} \\
	 \left[1 - a^{(2)}(t)\right] f_{\rm p}^{*(2)}(t)   -a^{(2)}(t)f_{\rm em}^{(2)}(t) \\
	  \vdots \\
	 \left[1 - a^{(N)}(t)\right] f_{\rm p}^{*(N)}(t)   -a^{(N)}(t)f_{\rm em}^{(N)}(t) \\
\end{pmatrix} \, , \label{eq:kalman_h_matrix}
\end{eqnarray}
\begin{eqnarray}
	\boldsymbol{H}_k= \begin{pmatrix}
\boldsymbol{V}_{k} &\boldsymbol{W}_{k}
	\end{pmatrix} \, ,
\end{eqnarray}
\begin{eqnarray}
	\boldsymbol{R}_k = E \left[  \varepsilon^{(n)}(t)  \varepsilon^{(n)}(t)^{\rm T} \right] = \sigma^2_{\rm m}\boldsymbol{I}_k \, ,
\end{eqnarray}
\begin{eqnarray}
	\boldsymbol{V}_{k} &= \text{diag} \left[1 - a^{(1)}, \dots, 1- a^{(N)} \right] \, ,
\end{eqnarray}
\begin{eqnarray}
		\boldsymbol{W}_{k} &= \text{diag} \left[-f_{\rm p}^{(1)}, \dots, -f_{\rm p}^{(N)} \right] \, , \label{eq:kalman_h_matrix_end}
\end{eqnarray}
Equations \eqref{eq:appendix_noise}, \eqref{eq:qdelta}--\eqref{eq:q_matrix_term2}, and \eqref{eq:kalman_h_matrix}--\eqref{eq:kalman_h_matrix_end} contain all the required components in order to apply the EKF to the state space model in Section \ref{sec:state_space_formulation}.

\section{PSD for the Ornstein-Uhlenbeck process}\label{appendix:psd_OU}
In standard PTA analyses, the stochastic GW background is described by a stationary Gaussian process, with an ensemble-averaged, power-law PSD, parameterised by an amplitude, $A_{\rm gw}$, and exponent, $\gamma_{\rm gw}$. Specifically, the TOA residuals PSD used in standard analyses is \citep[e.g.][]{Haasteren,nano_gw_astronomer}
\begin{equation}
	P(f) = \frac{A_{\rm gw}^2}{12 \pi ^2} \left(\frac{f}{\text{yr}^{1}}\right)^{- \gamma_{\rm gw}} \text {yr}^3 \, . \label{eq:psr_gw_appendix}
\end{equation}
In this paper the influence of the stochastic GW background on the $n$-th pulsar, $a^{(n)}(t)$, is modelled as an Ornstein-Uhlenbeck process, governed by Equation \eqref{eq:ornstein_for_at} and defined formally in Appendix \ref{appendix:justify_OU_process_for_background}. The Ornstein-Uhlenbeck process also maps onto a stationary Gaussian process. The PSD of the Gaussian process follows from Equations \eqref{eq:ornstein_for_at} and \eqref{eq:xieqn2} and is given by
\begin{equation}
	P^{(n)}_{\rm a}(f) = \frac{ \left[ \sigma_{\rm a}^{(n,n)} \right]^2}{\gamma_{\rm a}^2 + \left(2 \pi f\right)^2} \, ,
\end{equation}
where $\sigma_{\rm a}^{(n,n)} $ is given by Equation \eqref{eq:correlation}. Here $P^{(n)}_{\rm a}(f)$ is the Fourier transform of the temporal autocorrelation function of $a^{(n)}(t)$. \newline

We can relate $P^{(n)}_{\rm a}(f)$ to the PSD of the TOA residuals, $P_{\rm \delta t}(f)$, as follows. We use the notation $P_{\rm \delta t}(f)$ to refer to the PSD of the TOA residuals generated by the Ornstein-Uhlenbeck process and $P(f)$ to refer to the TOA residuals PSD used in standard analyses, cf. Equation \eqref{eq:psr_gw_appendix}. For conciseness we drop the $(n)$-superscript temporarily. Let the measured phase be $\phi_{\rm m}$. With $d\phi_{\rm m} /dt = f_{\rm m}$, Equation \eqref{eq:measurement} implies
\begin{equation}
	\frac{d \phi_{\rm m}}{dt} =  f_{\rm p} - f_{\rm p} a \, . \label{eq:dphi_dt}
\end{equation}
In this appendix, we focus on the effect of the stochastic GW background and assume for simplicity that $f_{\rm p}$ is approximately constant over the autocorrelation time-scale of $a(t)$. Letting the random part of $\phi_{\rm m}$ be $\delta \phi_{\rm m}$, Equation \eqref{eq:dphi_dt} implies
\begin{equation}
	\frac{d \left(\delta \phi_{\rm m}\right)}{dt} =- f_{\rm p} a \, . \label{eq:dphidt_stoch}
\end{equation}
Taking the Fourier transform of Equation \eqref{eq:dphidt_stoch} (denoted with a circumflex) and evaluating the squared modulus gives
\begin{equation}
	| \delta \hat{\phi}_{\rm m} |^2 = \omega^{-2} f_{\rm p}^2 |\hat{a}|^2 \, , \label{eq:phi_ft}
\end{equation}
with $\omega = 2 \pi f$. If a TOA has a residual $\delta t > 0$ then the pulse arrives late, and the measured phase is slightly less than expected, with $\delta\phi_{\rm m} \approx f_{\rm p} \delta t < 0$ and hence
\begin{equation}
	| \delta \hat{t} |^2 = \omega^{-2} |\hat{a}|^2 \, . \label{eq:deltat_ft}
\end{equation}
Now, the PSD $P_{x}(\omega)$ of a general function $x(t)$ is given by
\begin{equation}
	P_{x}(\omega) = \lim\limits_{T \to \infty} \frac{1}{T} |\hat{x}_{T}(\omega)|^2 \, , \label{eq:psr_defn1}
\end{equation}
with $x_{T}(t) = x(t) \text{ for } t_0 \leq t \leq t_0+T \text{ and } x_{T}(t)=0 \text{ otherwise}$. Alternatively, the Wiener-Khintchin theorem gives
\begin{equation}
	\langle \hat{x}(\omega)\hat{x}^{*}(\omega') \rangle = 2 \pi \delta \left(\omega - \omega '\right) P_{x} \left(\omega\right) \, . \label{eq:psr_defn2}
\end{equation}
Equations \eqref{eq:deltat_ft}--\eqref{eq:psr_defn2} imply that the TOA residuals PSD generated from the Ornstein-Uhlenbeck process for $a(t)$ is
\begin{equation}
	P_{\delta t}(\omega) = \omega^{-2} P_{a}(\omega) \, ,
\end{equation}
or equivalently
\begin{equation}
	P_{\delta t}(f) =  \frac{\langle h^2\rangle\gamma_{\rm a}}{48 \pi^2 f^2 \left[\gamma_{\rm a}^2 + \left(2 \pi f\right)^2 \right]} \, . \label{eq:PSD_delta_t}
\end{equation}
In the limit $f \gg \gamma_{\rm a}$, Equation \eqref{eq:PSD_delta_t} reduces to
\begin{equation}
	P_{\rm \delta t}(f)  = \frac{\langle h^2\rangle  \gamma_{\rm a} f^{-4} }{192 \pi^4} \, .
\end{equation}
In the limit $f \ll \gamma_{\rm a}$, Equation \eqref{eq:PSD_delta_t} reduces to
\begin{equation}
	P_{\rm \delta t}(f)  = \frac{\langle h^2\rangle  \gamma_{\rm a} f^{-2}}{48 \pi^2}  \, .
\end{equation}

 \section{Glossary of symbols}\label{sec:glossary}
 In this appendix, for the reader's convenience, we present a glossary of certain key symbols used in this paper and, where appropriate, relate them to symbols or quantities used in standard PTA analyses. 
 
 \begin{itemize}

 	 	\item $a^{(n)}(t)$. Fractional change in $f_{\rm m}^{(n)}(t)$ caused by the stochastic GW background. It is described by a stochastic Ornstein-Uhlenbeck process, parameterised by $\boldsymbol{\theta}_{\rm gw} = \left \{h_{\rm a}, \gamma_{\rm a} \right \}$.

 	 	\item[]
 	
 	\item $f_{\rm em}^{(n)}(t)$. The deterministic electromagnetic spin-down of the $n$-th pulsar's intrinsic spin frequency.

 	 	\item[]
 	 	
 	 \item $f_{\rm m}^{(n)}(t)$. The pulse frequency of the $n$-th pulsar, as measured on Earth. One has $f_{\rm m}^{(n)}(t) \neq f_{\rm p}^{(n)}(t)$ in general because of measurement errors $\varepsilon^{(n)}(t)$ and modulation by the stochastic GW background; see Equations \eqref{eq:measurement}--\eqref{eq:vareps}. The Kalman filter solves the inverse problem of estimating $f_{\rm p}^{(n)}(t)$ given $f_{\rm m}^{(n)}(t)$.

 	 \item[]

 	 \item $f_{\rm p}^{(n)}(t)$. The intrinsic spin frequency of the $n$-th pulsar in the array, as evaluated in the pulsar's rest frame. It obeys a stochastic Ornstein-Uhlenbeck process, with mean-reverting, random fluctuations; see Equations \eqref{eq:frequency_evolution}--\eqref{eq:xieqn_new}. The evolution of  $f_{\rm p}^{(n)}(t)$ forms a hidden Markov sequence, cf. Equation \eqref{eq:state_evolution}.
 	
 	 	 	\item[]
 	\item $h_{\rm a}$. Amplitude parameter of the Ornstein-Uhlenbeck description of the stochastic GW background. It is analogous, but not equal, to the amplitude factor $A_{\rm gw}$ in the PSD of traditional PTA analyses, cf. Equation \eqref{eq:psr_gw_appendix}.
 	
 	  	\item[]
 	\item $\gamma_{\rm a}$. Mean reversion timescale of the stochastic GW background in the Ornstein-Uhlenbeck model. It is set approximately by the minimum SMBHB orbital frequency, i.e. $\gamma_{\rm a} \sim 2 \pi f_{\rm min}$.
 	 	\item[]
 	\item $\chi^{(n)}(t)$. A white noise stochastic process, parameterised by $\sigma^{(n)}$.

 \end{itemize}

\section{Workflow summary}\label{sec:workflow}
For the reader's convenience we summarise the workflow for a representative PTA analysis using the Kalman filter in conjunction with a Bayesian inference algorithm for parameter estimation and model selection. In this paper nested sampling is used for the Bayesian inference algorithm, but any likelihood-based sampling scheme can be used. 

\begin{enumerate}[leftmargin=2em]
	\item From a PTA composed of $N$ pulsars, obtain $N$ measurement time series, $f_{\rm m}^{(n)}(t)$, collectively labelled $\boldsymbol{Y}$.
	\item Specify a state-space model $\mathcal{M}$ (see Section \ref{sec:bayes}) with parameters $\boldsymbol{\theta}$.
	\item Specify a prior distribution $\pi(\boldsymbol{\theta})$.
	\item For each iteration in the Bayesian inference algorithm:
	\begin{enumerate}[leftmargin=2em]
		\item Sample from $\pi(\boldsymbol{\theta})$ to obtain $\boldsymbol{\theta}_{\rm sample}$
		\item Pass $\boldsymbol{\theta}_{\rm sample}$ to the Kalman filter.
		\item Iterate over $\boldsymbol{Y}$ using the Kalman filter and obtain a single likelihood value $\mathcal{L} \left(\boldsymbol{Y} | \boldsymbol{\theta}_{\rm sample}\right)$ value.
		\item Update $p\left(\boldsymbol{\theta}|\boldsymbol{Y}\right)$ and $\mathcal{Z}$
	\end{enumerate}
	\item Repeat steps (a)--(d) until a user-selected convergence criterion are satisfied.
\end{enumerate}
In order to compute the odds ratio $\beta$ in Equation \eqref{eq:bayes} the above workflow is repeated for a two models $\mathcal{M}_1$ and $\mathcal{M}_2$, and the two $\mathcal{Z}$ values are divided. We remind the reader that the above workflow differs from a realistic PTA analysis in one important respect, namely that the data are input as frequency time series $f_{\rm m}^{(n)}(t)$ instead of pulse TOAs. The generalization to TOAs is subtle and will be tackled in a forthcoming paper.

\section{Additional tests on synthetic data}\label{appendix:exxtra_tests}
In this appendix we augment the IPTA MDC validation test in Section \ref{sec:representative_analysis_mdc}, with additional tests on synthetic data generated by an alternative, tunable procedure. The aim of testing on alternative data is to check that the method operates successfully on a variety of datasets across a range of systematically selected parameters. The additional testing also minimizes the possibility that the set-up or parameters of any particular test (including the IPTA MDC) are fortuitously favourable to the method. In section \ref{sec:validation} we describe how the alternative synthetic data are generated. Specifically, we generate frequency time series $f_{\rm m}^{(n)}(t)$ by solving Equations \eqref{eq:measurement}--\eqref{eq:delta_h}. In section \ref{sec:representative_analysis} we repeat the analysis of the main text on the alternative synthetic data. Summarising the results in advance, we find that the method performs equally well on the alternative synthetic data as on the IPTA MDC.

\subsection{Validation procedure with synthetic data}\label{sec:validation}
\begin{table}
	\centering
		\begin{tabular}{lcll}
			\toprule
			Set&Parameter & Injected value & Units  \\
			\hline
			\vspace{1mm}& $f_{\rm em}^{(n)} (t_1)$       & $f_{\rm ATNF}^{(n)}$ & Hz  \\
			\multirow{2}{2mm}{$\boldsymbol{\theta}_{\rm psr}$} & $\dot{f}_{\rm em}^{(n)} (t_1)$       & $\dot{f}_{\rm ATNF}^{(n)}$ & s$^{-2}$  \\
			& $\sigma^{(n)}$              & $\sigma_{\rm SC}^{(n)}$ & s$^{-3/2}$ \\
			& $\gamma^{(n)}$              & $10^{-13}$ & s$^{-1}$  \\
			\hline 
			\multirow{7}{2mm}{$\boldsymbol{\theta}_{\rm a}$}& $\Omega^{(m)}$       & $p \left( \Omega\right)$ & years$^{-1}$ \\
			& $\alpha^{(m)}$          & Uniform($0, 2 \pi $)  & rad \\
			& $\delta^{(m)}$              & Cosine($-\pi/2, \pi/2$) & rad \\
			& $\psi^{(m)}$              & Uniform($0, \pi $) & rad  \\
			\vspace{1mm}& $\Phi_0^{(m)}$          & Uniform($0, 2 \pi $)  & rad  \\
			\vspace{1mm}& $h_0^{(m)}$            & DiracDelta$\left(h_0\right)$  \\
			& $\iota^{(m)}$             & Cosine($0, \pi$) & rad  \\ 
			\hline
			& $\sigma_{\rm m}^{(n)}$ & $10^{-11}$ & Hz  \\ 
			\bottomrule
		\end{tabular}
		\caption{Injected parameters used to generate synthetic data via the procedure outlined in Section \ref{sec:validation}, when validating the analysis scheme in Section \ref{sec:representative_analysis}. The top and middle sections of the table contain $\boldsymbol{\theta}_{\rm psr}$ ($4N$ parameters for $N$ PTA pulsars) and $\boldsymbol{\theta}_{\rm a}$ (stochastic GW background parameters) respectively. The bottom section of the table contains $\sigma_{\rm m}^{(n)}$. In the top section, the subscript ``ATNF'' denotes values obtained from the ATNF pulsar catalogue. The subscript ``SC'' on $\sigma^{(n)}$ indicates that the injected value is calculated from Equation \eqref{eq:sigmap_f} and the empirical timing noise model for MSPs in \protect \cite{Shannon2010}. In the middle section, the injected values are sampled from the specified probability distributions. The probability distribution for $\Omega$, viz. $p \left(\Omega\right)$, is given by Equation \eqref{eq:power_law_omega}. The PDF for $h_0^{(m)}$ is a Dirac delta function; all $M$ GW sources have equal strain for the purposes of testing in Section \ref{sec:representative_analysis} (the astrophysical distribution of $h_0$ is different). Different values of $h_0$ are investigated and specified in the text. We set $M=10^4$ throughout this appendix.}
		\label{tab:injected_parameters}
	\end{table}
Generating synthetic data $f_{\rm m}^{(n)}(t)$ synthetically requires three independent ingredients: $f_{\rm p}^{(n)}(t)$ (see Section \ref{sec:generate_fp}), $a^{(n)}(t)$ (see Section \ref{sec:generate_a}), and $\varepsilon^{(n)}(t)$ (see Section \ref{sec:generate_eps}). For notational convenience going forward, we drop the explicit time dependence of $\boldsymbol{Y}(t)$ and write $\boldsymbol{Y} = \boldsymbol{Y}(t)$. Throughout this appendix we take $T_{\rm obs} = t_{N_{t}} - t_{1} = 10$ years and $T_{\rm cad} = t_{k+1} - t_{k} = 1$ week for $1 \leq k \leq N_{t} -1$.
	
	\subsubsection{Generating $f_{\rm p}^{(n)}(t)$} \label{sec:generate_fp}
	We integrate Equations \eqref{eq:frequency_evolution}--\eqref{eq:spinevol} numerically using a Runge-Kutta It$\hat{\text{o}}$ integrator implemented in the \texttt{sdeint} python package \footnote{\url{https://github.com/mattja/sdeint}}. This produces a random realisation of $f_{\rm p}^{(n)}(t)$ for $1\leq n \leq N$. The numerical solutions depend on $\boldsymbol{\theta}_{\rm psr}$, which specifies the configuration of a synthetic PTA. 
	In this appendix we adopt for consistency the same $\boldsymbol{\theta}_{\rm psr}$ values as in previous related work \citep{KimpsonPTA1,KimpsonPTA2}. Specifically, we work with the $N=47$ MSPs in the 12.5-year NANOGrav dataset \citep{2020ApJ...905L..34A}. Fiducial values for $f_{\rm em}^{(n)}(t_1)$ and $\dot{f}^{(n)}_{\rm em}(t_1)$ are read from the Australia Telescope National Facility (ATNF) pulsar catalogue \citep{Manchester2005} using the \texttt{psrqpy} package \citep{psrqpy}. \newline

	No direct measurements exist for $\gamma^{(n)}$ or $\sigma^{(n)}$. Regarding the former, the mean reversion timescale typically satisfies $[\gamma^{(n)}]^{-1} \gg T_{\rm obs}$ \citep{Price2012,Myers2021MNRAS.502.3113M,Meyers2021,Vargas,2024MNRAS.tmp..891O}. In this appendix, for the sake of simplicity, we fix $\gamma^{(n)} = 10^{-13}$ s$^{-1}$ for all $n$. Regarding the latter, $\sigma^{(n)}$ is related to the root mean square TOA noise $\sigma^{(n)}_{\rm TOA}$ accumulated over an interval of length $T_{\rm cad}$ by \citep{KimpsonPTA1}
	\begin{eqnarray}
		\sigma^{(n)} \approx \sigma_{\rm TOA}^{(n)} f_{\rm p}^{(n)}(t_1) {T_{\rm cad}}^{-3/2} \ . \label{eq:sigmap_f}
	\end{eqnarray}
	For consistency with previous work, we calculate $\sigma^{(n)}_{\rm TOA}$ by applying the empirical timing noise model for MSPs from \cite{Shannon2010ApJ...725.1607S} in the manner described in Section 4 in \cite{KimpsonPTA1}. For the 12.5-year NANOGrav dataset used in this appendix, representative values of $\sigma^{(n)}$ calculated using Equation \eqref{eq:sigmap_f} are $\text{median} [\sigma^{(n)}] = 5.51 \times 10^{-24} $ s$^{-3/2}$, $\min [ \sigma^{(n)} ] = 1.67 \times 10^{-26}$s$^{-3/2}$ for PSR J0645+5158 and $\max [ \sigma^{(n)} ] = 2.56 \times 10^{-19}$ s$^{-3/2}$ for PSR J1939+2134. \newline 
	
	We emphasise that Equations \eqref{eq:frequency_evolution}--\eqref{eq:spinevol} specify a stochastic process. Accordingly, different random realisations of $f_{\rm p}^{(n)}(t)$ are generated for the same choice of $\boldsymbol{\theta}_{\rm psr}$ by starting with a different random seed. Different realisations of $f_{\rm p}^{(n)}(t)$ are needed to test the analysis algorithm in Section \ref{sec:representative_analysis}, because of course the actual, astronomical realisations of the fluctuating rotation histories of the NANOGrav pulsars (and indeed the stochastic GW background itself) are unknown. By analysing an ensemble of $f_{\rm p}^{(n)}(t)$ realizations, we quantify the irreducible dispersion in inference outcomes (``cosmic variance'') and hence the accuracy of the parameter estimation scheme in Section  \ref{sec:representative_analysis}. The injected elements of $\boldsymbol{\theta}_{\rm psr}$ are summarised in the top section of Table \ref{tab:injected_parameters}.

	\subsubsection{Generating $a^{(n)}(t)$} \label{sec:generate_a}

	The multiplicative modulation of $f_{\rm m}^{(n)}(t)$ by the linear superposition of $M$ SMBHB sources is given by Equations \eqref{eq:measurement}--\eqref{eq:xieqn2}. For the $n$-th pulsar, $a^{(n)}(t)$ in Equations \eqref{eq:ornstein_for_at}--\eqref{eq:xieqn2} is parameterised by $7M$ parameters,
	\begin{eqnarray}
		\boldsymbol{\theta}_{\rm a} = \left \{h_0^{(m)}, \iota^{(m)}, \psi^{(m)}, \delta^{(m)}, \alpha^{(m)}, \Omega^{(m)}, \Phi_0^{(m)} \right \}_{1\leq m \leq M} \ ,  \label{eq:params31}
	\end{eqnarray}
	where $h_0^{(m)}$ is the characteristic wave strain, $\iota^{(m)}$ is the orbital inclination, $\psi^{(m)}$ is the polarisation angle, $\delta^{(m)}$ is the declination and $\alpha^{(m)}$ is the right ascension. These parameters enter 
	Equations \eqref{eq:afunc} -- \eqref{eq:delta_h} via $H_{ij}^{(m)} = H_{ij}^{(m)} \left[h_0^{(m)}, \iota^{(m)}, \psi^{(m)}, \delta^{(m)}, \alpha^{(m)} \right]$ and $\boldsymbol{n}^{(m)}=\boldsymbol{n}^{(m)} \left[\delta^{(m)},\alpha^{(m)}\right]$. To generate a stochastic background of GW sources we randomly sample the $7M$ parameters in $\boldsymbol{\theta}_{\rm a}$ from the probability density functions (PDFs) specified in the middle section of Table \ref{tab:injected_parameters}. For the five angles $ \{\iota, \psi, \delta, \alpha, \Phi_0 \}$, uniform distributions are used \citep[e.g.][]{Bhagwat2021}. The five angles are independent. In contrast, $h_0$ and $\Omega$ are not independent astrophysically; an SMBHB nearer the end of its inspiral (higher $\Omega$) is likely to be louder at Earth (higher $h_0$). However, the joint PDF $p(h_0,\Omega)$ is complicated and uncertain to calculate in a realistic astrophysical manner without large-scale cosmological simulations \citep[e.g.][]{2018NatCo...9..573M,2019ApJ...887...35M,2022MNRAS.511.5241S,2023arXiv230518293M}, which lie outside the scope of this paper. For the purposes of validating the algorithm in Sections \ref{sec:state_space_formulation} and \ref{sec:kfns}, and for the sake of simplicity, it is sufficient to treat $h_0$ and $\Omega$ as independent random variables and factorize $p(h_0,\Omega) = p(h_0) p(\Omega)$. We fix $h_0$ as a constant for all sources, so that \ $p(h_0)$ is a delta function in Table \ref{tab:gw_priors}, and treat it as an adjustable parameter to tune the strength of the synthetic stochastic GW background. For $\Omega$ we adopt a power-law PDF, on the bounded domain $\Omega_{\rm min} \leq \Omega \leq \Omega_{\rm max}$, viz.
	\begin{eqnarray}
		p\left( \Omega\right) =  K_1 \Omega^{-\kappa} \label{eq:power_law_omega} \, ,
	\end{eqnarray}
	with 
	\begin{eqnarray}
		K_1 = \frac{1 - \kappa}{\Omega_{\rm max}^{1 - \kappa} - \Omega_{\rm min}^{1 - \kappa}} \, . \label{eq:omega_norm}
	\end{eqnarray}
	In Equations \eqref{eq:power_law_omega} and \eqref{eq:omega_norm} we set $\kappa = 3$, $\Omega_{\rm min} = T_{\rm obs}^{-1}$ and $\Omega_{\rm max} =  T_{\rm cad}^{-1}$ for the sake of definiteness. The value of $\kappa = 3$ and the power-law form of Equation \eqref{eq:power_law_omega} are chosen arbitrarily but are astrophysically reasonable given current assembly models of SMBHBs \citep[e.g.][]{10.1111/j.1365-2966.2008.13682.x,2009ApJ...698..198G,2013ARA&A..51..511K,2017MNRAS.470.1738C,2018NatCo...9..573M} and the scale invariance of the inspiral physics \citep{PhysRev.136.B1224,2001astro.ph..8028P}. The values of $\Omega_{\rm min}$ and $\Omega_{\rm max}$ are upper and lower bounds respectively, set by the effective bandwidth of the PTA. The true, physical values satisfy $\Omega_{\rm min} \ll T_{\rm obs}^{-1}$ (start of inspiral) and $\Omega_{\rm max} \gg T_{\rm cad}^{-1}$ (end of inspiral), but SMBHBs with $\Omega \ll T_{\rm obs}^{-1}$ and $\Omega \gg T_{\rm cad}^{-1}$ do not contribute measurably to the stochastic GW background in the PTA band. Appendix \ref{sec:tong} demonstrates that the main results of this paper are insensitive to the exact value of $\kappa$. \newline

	The complete procedure for generating $a^{(n)}(t)$ is as follows:
	\begin{enumerate}[leftmargin=2em]
		\item Specify a synthetic PTA, determining $q^{(n)}$ and $d^{(n)}$.
		\item For $M$ (here $M=10^4$) discrete quasi-monochromatic GW sources, randomly sample the $7M$ parameters in $\boldsymbol{\theta}_{\rm a}$ from the probability distributions summarised in Table \ref{tab:injected_parameters}. 
		\item Evaluate Equations \eqref{eq:afunc}--\eqref{eq:delta_h} to obtain $a^{(n)}(t)$.
	\end{enumerate}
	We emphasise that $a^{(n)}(t)$ depends on the pulsar-specific quantities $\boldsymbol{q}^{(n)}$ and $d^{(n)}$, specified in Section \ref{sec:generate_fp} as well as ${\boldsymbol{\theta}}_{\rm a}$. We also emphasize that we do not approximate $a^{(n)}(t)$ as a stochastic, Ornstein-Uhlenbeck process represented by Equations \eqref{eq:ornstein_for_at}--\eqref{eq:xieqn2} (see Appendix \ref{appendix:justify_OU_process_for_background}) when generating the synthetic data. The Ornstein-Uhlenbeck approximation is reserved for the inference model in Section \ref{sec:state_space_formulation}, to circumvent the otherwise insurmountable identifiability challenge posed by the $7M$ parameters in ${\boldsymbol{\theta}}_{\rm a}$.

	\subsubsection{Generating $\varepsilon^{(n)}(t)$} \label{sec:generate_eps}
	The  measurement noise $\varepsilon^{(n)}(t)$ is assumed for simplicity to be additive, Gaussian, and white. It excludes chromatic disturbances in PTA TOAs, such as those induced by propagation through the turbulent interstellar medium \citep[e.g.][]{Goncharov2021}. Its variance $\sigma_{\rm m}^{(n)}$ can be related to $\sigma_{\rm TOA}^{(n)}$ (in units of $s$) by 
	\begin{equation}
		\sigma_{\rm m}^{(n)} \approx f_{\rm p}^{(n)}(t_1) \sigma_{\rm TOA}^{(n)} \ {T_{\rm cad}}^{-1} \ . \label{eq:sigma_m_eqn}
	\end{equation}
	For an MSP with $f_{\rm p}^{(n)} \sim 0.1$ kHz, $T_{\rm cad} = 1 \, {\rm week}$, and $\sigma_{\rm TOA}^{(n)} \sim 1 \mu$s,  Equation \eqref{eq:sigma_m_eqn} implies $\sigma_{\rm m}^{(n)} \sim 10^{-10}$ Hz. The most accurately timed pulsars have $\sigma_{\rm TOA}^{(n)} \sim 10 $ ns and $\sigma_{\rm m}^{(n)} \sim 10^{-12}$ Hz. We fix $\sigma_{\rm m}^{(n)} = 10^{-11}$ Hz for all $n$, copying \cite{KimpsonPTA1}, and take it as known \textit{a priori} rather than a parameter to be inferred. Despite $\sigma_{\rm m}^{(n)}$ taking the same value for every pulsar, $f_{\rm m} ^{(n)}(t)$ is constructed from a different random realisation of $\varepsilon^{(n)}(t)$ generated by a different random seed for each pulsar. We refer the reader to Section 4 in \cite{KimpsonPTA1} for additional discussion on the specification of astrophysically representative values of $\sigma_{\rm m}^{(n)}$.

\subsubsection{Justification of the injection PDF}\label{sec:tong}
\begin{figure}
	\includegraphics[width=\columnwidth, height =\columnwidth]{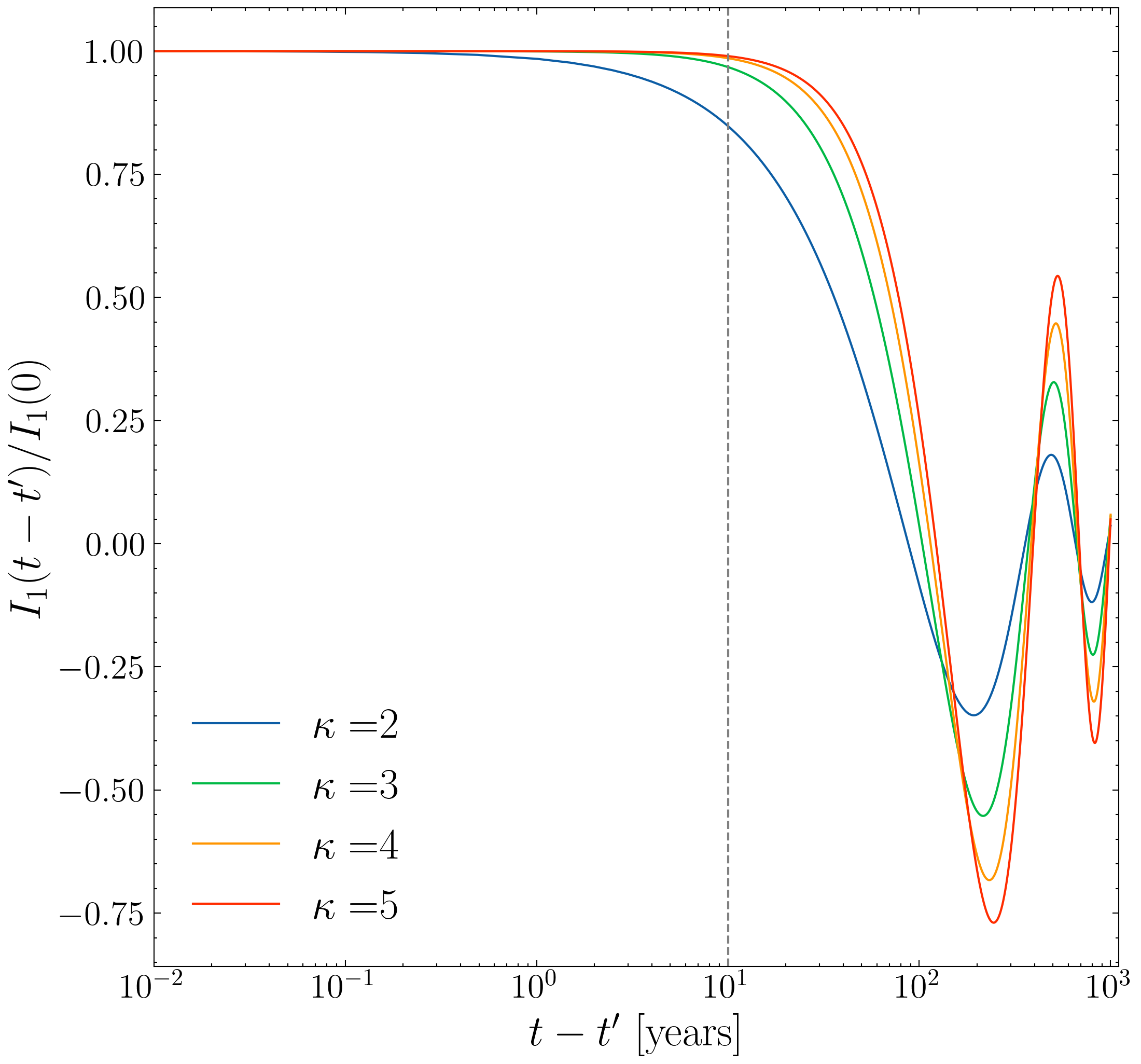} 	
	\caption{Normalised cross-correlation integral $I_1(t-t')/I_1(0)$, from Equation \eqref{eq:I1integral}, as a function of $\tau = |t-t'|$ (units: yr) for four values of $2\leq \kappa \leq 5$, colour coded per the legend. The grey dashed vertical line indicates the typical 10-yr duration of a PTA observation. All coloured curves are approximated accurately by Equation \eqref{eq:I1integral_approx1}. Physical parameters: $\Omega_{\rm min} = 10^{-2}$ years$^{-1}$,  $\Omega_{\rm max} = 10^{3}$ years$^{-1}$.}
	\label{fig:different_exponents}
\end{figure}
When generating synthetic data in Section \ref{sec:generate_a}, it is necessary to sample the $7M$ parameters of $\boldsymbol{\theta}_{\rm a}$ from the probability distributions specified in Table \ref{tab:injected_parameters}. For the five angles $ \{\iota, \psi, \delta, \alpha, \Phi_0 \}$, we draw from uncorrelated and uniform PDFs, consistent with the standard assumption that the stochastic GW background is isotropic \citep{Hellings,Maggiore}. In contrast, the parameters $h_0$ and $\Omega$ are correlated; both $h_0$ and $\Omega$ increase, as a SMBHB inspiral proceeds. Moreover, $\Omega$ and hence $h_0$ are not distributed uniformly; a SMBHB spends more time in the earlier than in the later stages of inspiral, so lower values of $\Omega$ are more common \citep{2001astro.ph..8028P,10.1111/j.1365-2966.2008.13682.x,2018NatCo...9..573M}. In general, therefore, one should sample $h_0$ and $\Omega$ from the joint PDF $p(h_0,\Omega)$, which is not uniform and correlates $h_0$ positively with $\Omega$. A realistic and accurate calculation of $p(h_0,\Omega)$ requires large-scale cosmological $N$-body simulations \citep[see e.g.][and references therein]{2018NatCo...9..573M} and falls well outside the scope of this paper. Instead, we factorize the joint PDF in the form $p(h_0,\Omega) = p(h_0)p(\Omega)$ and set $p(h_0) = \delta(h_0 - h_{0,{\rm inj}})$ for simplicity, as a practical way of tuning the mean square amplitude of the background $\langle h^2 \rangle$ in Equation \eqref{eq:sigma_a_expression} and hence $h_{\rm a}$ in $\boldsymbol{\theta}_{\rm gw}$ in Equation \eqref{eq:params3}. For the remaining parameter $\Omega$, the PDF is a power law, Equation \eqref{eq:power_law_omega}, with exponent $-\kappa$. In this paper we take $\kappa=3$, consistent with standard assembly models of SMBHBs \citep[e.g.][]{10.1111/j.1365-2966.2008.13682.x,2009ApJ...698..198G,2013ARA&A..51..511K,2017MNRAS.470.1738C,2018NatCo...9..573M}. \newline 

We now show that modifying $\kappa$ over an astrophysically plausible range does not hamper the ability of the state-space inference scheme in Sections \ref{sec:state_space_formulation} and \ref{sec:kfns} to detect a stochastic GW background, although of course it does modify the temporal cross-correlation properties of the background through $\langle a^{(n)}(t) a^{(n')}(t') \rangle$. The value of $\kappa$ controls the shape of Equation \eqref{eq:I1integral}. In Figure \ref{fig:different_exponents} we evaluate Equation \eqref{eq:I1integral} for four astrophysically plausible values of $2 \leq \kappa \leq 5$. Current PTA experiments operate in the region to the left of the grey dashed vertical line, where $I_1(t-t')$ decreases monotonically with $|t-t'|$, and Equation \eqref{eq:I1integral} is approximated accurately by Equation \eqref{eq:I1integral_approx1}. Therefore the state-space inference scheme can be applied safely without modification for $2 \leq \kappa \leq 5$.

\subsection{Detecting a synthetic background}\label{sec:representative_analysis}
In this section we test the state-space analysis algorithm on synthetic data $\boldsymbol{Y}$ produced by the method described in Section \ref{sec:validation}. The test has three goals: (i) to verify that the algorithm in Sections \ref{sec:state_space_formulation} and \ref{sec:kfns} can detect successfully a synthetic background of the form constructed in Section \ref{sec:validation}; (ii) to determine the minimum background amplitude $h_a$ which the algorithm detects, assuming a NANOGrav-like PTA configured as described in Section \ref{sec:validation}; and (iii) to determine the accuracy with which the parameters describing the background are estimated. In Section \ref{sec:priors} we define the priors on $\boldsymbol{\theta}$. In Section \ref{sec:posterioirs} we calculate the joint posterior probability distribution of the parameters, $p(\boldsymbol{\theta} | \boldsymbol{Y})$. In Section \ref{sec:bayes} we calculate the detectability of the GW background as a function of $h_0$. To quantify the impact of cosmic variance, the algorithm is validated on multiple random realisations of the stochastic processes that generate $\boldsymbol{Y}$: the pulsar process noise $\xi^{(n)}(t)$, the telescope measurement noise $\varepsilon^{(n)}(t)$ and the realisation of the $M$ synthetic sources parameterised by $\boldsymbol{\theta}_{\rm a}$. A real PTA analysis witnesses a unique realisation of $\boldsymbol{Y}$ --- the actual, astronomical one --- but one never knows where this realisation lies within the admissible statistical ensemble.

\subsubsection{Prior distribution}\label{sec:priors}
\begin{table}
	\centering
		\begin{tabular}{lcl}
			\toprule
			Set&Parameter & Prior  \\
			\hline
			\vspace{1mm}& $f_{\rm em}^{(n)} (t_1)$       & Uniform$\left[f_{\rm ATNF}^{(n)} - 10^3 \eta^{(n)}_{f}, f_{\rm ATNF}^{(n)} + 10^3 \eta^{(n)}_{f} \right]$ \\
			\multirow{2}{2mm}{$\boldsymbol{\theta}_{\rm psr}$} & $\dot{f}_{\rm em}^{(n)} (t_1)$    & Uniform$\left[ \dot{f}_{\rm ATNF}^{(n)} - 10^3 \eta^{(n)}_{\dot{f}}, \dot{f}_{\rm ATNF}^{(n)} + 10^3 \eta^{(n)}_{\dot{f}} \right]$ \\
			& $\sigma^{(n)}$  & LogUniform$ \left [10^{-2} \sigma_{\rm SC}^{(n)}, 10^2 \sigma_{\rm SC}^{(n)} \right ]$ \\
			& $\gamma^{(n)}$  &  --- \\
			\hline 
			\multirow{2}{2mm}{$\boldsymbol{\theta}_{\rm gw}$} & $\gamma_a$     &  LogUniform$ \left [10^{-13} \text{s}^{-1},10^{-7} \text{s}^{-1} \right ]$\\
			& $h_{\rm a}$              & LogUniform$ \left [10^{-1} h_0,10^{3} M^{1/2} h_0 \right ]$ \\
			\bottomrule
		\end{tabular}
		\caption{Priors on the parameters of the model assumed in Section \ref{sec:state_space_formulation} for the purposes of Bayesian inference. The top and bottom sections of the table contain $ \pi\left(\boldsymbol{\theta}_{\rm psr} \right)$ and $ \pi \left(\boldsymbol{\theta}_{\rm gw} \right)$ respectively. In the top section, the quantities $\eta^{(n)}_{f}$ and $\eta^{(n)}_{\dot{f}}$ are the uncertainties in $f^{(n)}_{\rm em} (t_1)$ and $\dot{f}^{(n)}_{\rm em} (t_1)$ respectively, as quoted in the ATNF pulsar catalogue. We do not infer $\gamma^{(n)} \sim 10^{-5} T_{\rm obs}$ for simplicity, so no prior is set. In the bottom section, the limits on the prior of $h_{\rm a}$ are determined by $M^{1/2} h_0$; several values of $M^{1/2} h_0$ are investigated as specified in the text.}
		\label{tab:gw_priors}
	\end{table}
	The Bayesian priors on each parameter are summarised in Table \ref{tab:gw_priors}. We assume no {\em a priori} knowledge about $\boldsymbol{\theta}_{\rm gw}$, except that plausible values for both its elements span several decades according to cosmological simulations \citep[e.g.][]{10.1111/j.1365-2966.2008.13682.x,2009ApJ...698..198G,2013ARA&A..51..511K,2017MNRAS.470.1738C,2018NatCo...9..573M}. We therefore choose broad uniform priors spanning five and six decades in $\gamma_a$ and $h_a$ respectively. The relationship between $\boldsymbol{\theta}_{\rm gw}$ and the $7M$ parameters in $\boldsymbol{\theta}_{\rm a}$, which specify the $M$ synthetic SMBHB sources, is explained in detail in Appendix \ref{appendix:justify_OU_process_for_background}. Electromagnetic observations offer \textit{a priori}  information on $\boldsymbol{\theta}_{\rm psr}$. We adopt constrained uniform priors on $f_{\rm em}^{(n)}(t_1)$ and $\dot{f}_{\rm em}^{(n)}(t_1)$, which extend $\pm 10^3 \eta_f^{(n)}$ and $\pm 10^3 \eta_{\dot{f}}^{(n)}$ respectively about the central, injected values drawn from the ATNF pulsar catalogue, where $\pm \eta_f^{(n)}$ and $\pm \eta_{\dot{f}}^{(n)}$ denote the error bars quoted in the ATNF pulsar catalogue. By using wider-than-necessary priors we expose the analysis scheme to a more stringent test. We also set $\pi[\sigma^{(n)} / (1 \, {\rm s^{-3/2}})] \sim$ LogUniform$ \left [10^{-2} \sigma_{\rm SC}^{(n)}, 10^2 \sigma_{\rm SC}^{(n)} \right ]$,
	where $\sigma_{\rm SC}^{(n)}$ is the noise amplitude for pulsar $n$ inferred from
	Equation \eqref{eq:sigmap_f}. We do not set a prior on $\gamma^{(n)}$, because one typically has $\gamma^{(n)} T_{\rm obs} \sim 10^{-5}$ astrophysically, and $\gamma^{(n)}$ is effectively ``unobservable'' for $T_{\rm obs} \sim 10 \, {\rm years}$. For validation purposes it is sufficient to carry $\gamma^{(n)}$ through the analysis at its injected value, reducing the total dimension of the parameter space to $2 + 3N$. \newline

	\subsubsection{Posterior distribution and its accuracy}\label{sec:posterioirs}
	
	\begin{figure}
		\includegraphics[width=\columnwidth, height =\columnwidth]{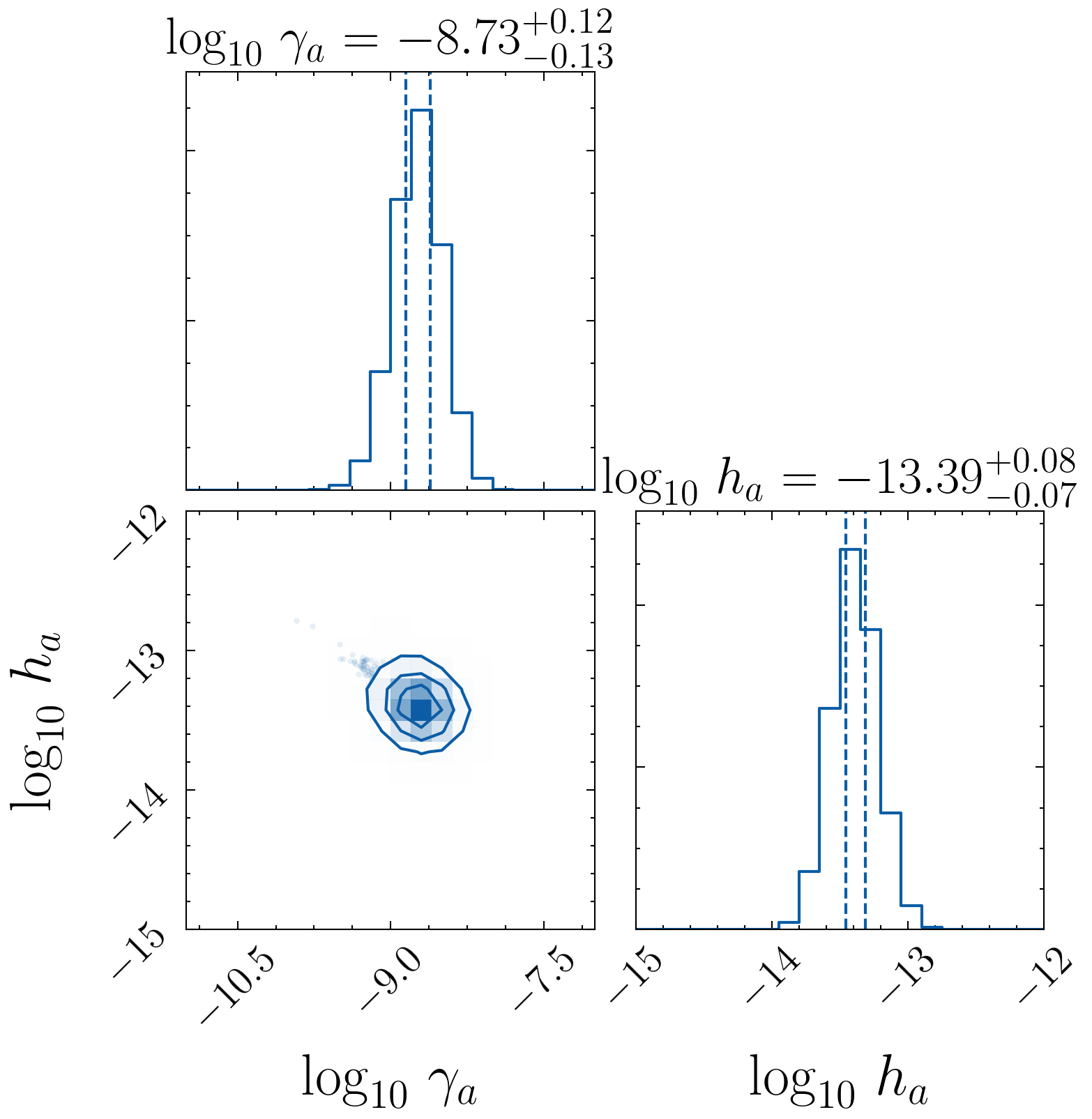} 	
		\caption{Posterior distribution $p \left( \boldsymbol{\theta}_{\rm gw} | \boldsymbol{Y}\right)$of the stochastic GW background parameters $\boldsymbol{\theta}_{\rm gw}$ for a single realisation of the synthetic data $\boldsymbol{Y}$ generated for the injected parameters in Table \ref{tab:injected_parameters}, with $h_0 = 10^{-16}$ and $M=10^{4}$. The contours in the two-dimensional histograms mark the (0.5, 1, 1.5, 2)-$\sigma$ levels. The one-dimensional histograms correspond to the joint posterior distribution marginalized over all but one parameter. The supertitles of the one-dimensional histograms record the median and the 0.16 and 0.84 quantiles. We plot the base-10 logarithm of the parameters, $\log_{10} \gamma_{\rm a}$ and $\log_{10} h_{\rm a}$. The state-space inference scheme converges to a well-behaved, unimodal posterior for both parameters in $\boldsymbol{\theta}_{\rm gw}$, as well as for the $3N$ parameters in $\boldsymbol{\theta}_{\rm psr}$ (not plotted). The horizontal axes span a subset of the prior domain for both parameters. The peak of the posterior is consistent with theoretical expectation for the $M$ injected sources parametrized by $\boldsymbol{\theta}_{\rm a}$, as discussed in Section \ref{sec:posterioirs} and Appendix \ref{appendix:justify_OU_process_for_background}.}
		\label{fig:corner_plot_1}
	\end{figure}

	\begin{figure}
		\includegraphics[width=\columnwidth, height =\columnwidth]{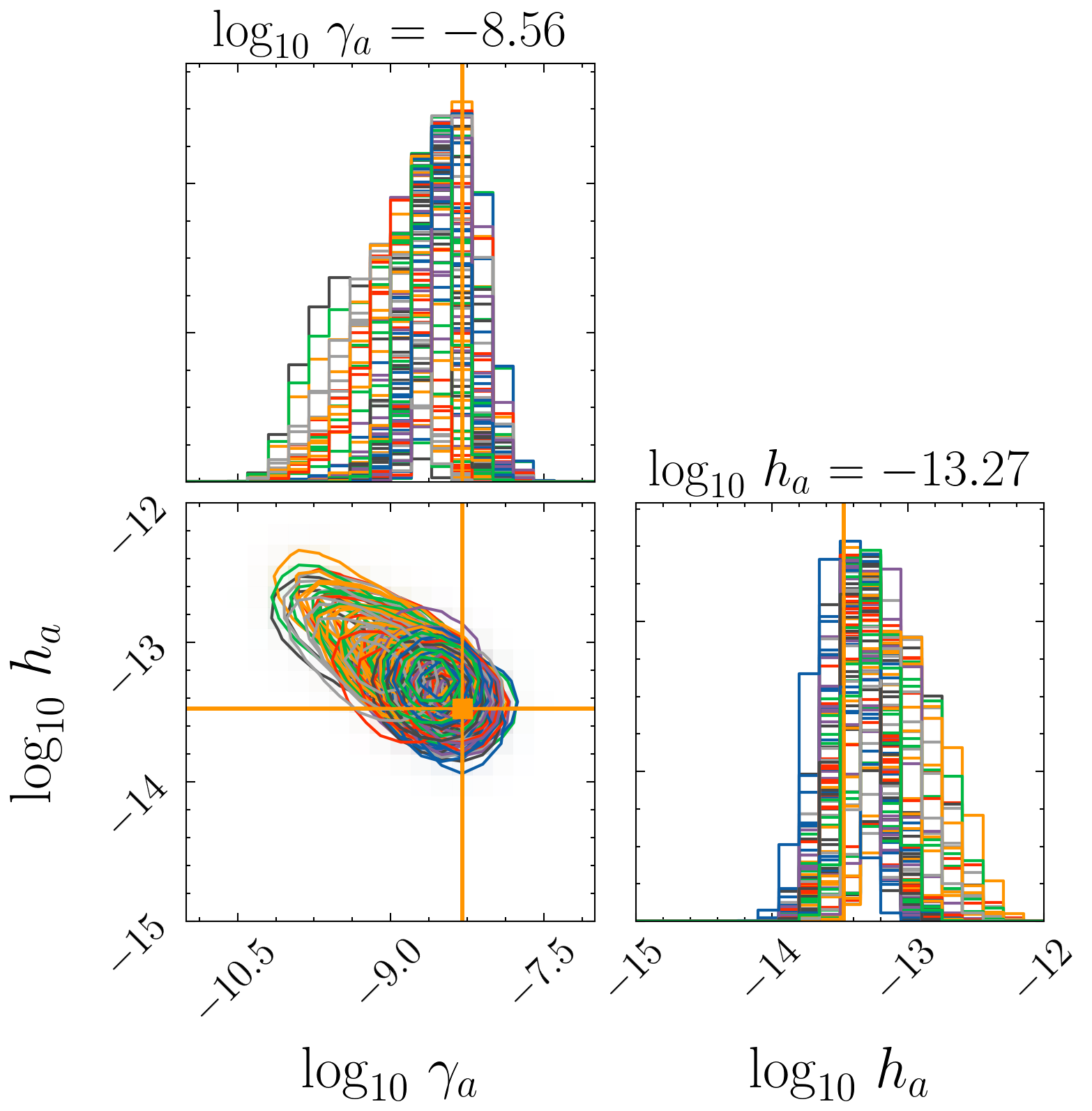} 	
		\caption{Same as Figure \ref{fig:corner_plot_1}, for $10^2$ realisations of $\boldsymbol{Y}$, with curves coloured uniquely per realisation. The supertitles of the one-dimensional histograms record the average of the one-dimensional posterior modes, $\mu(\theta)$, via Equation \eqref{eq:mean_rel_error}. The horizontal and vertical orange lines indicate the injected values. The procedure for obtaining the injected values of the model parameters $\boldsymbol{\theta}_{\rm gw}$ from $10^6$ realisations of the data parameters $\boldsymbol{\theta}_{\rm gw}$, cf. Equation \eqref{eq:params3}, is described in Appendix \ref{sec:posteriors_multiple}. The state-space analysis scheme converges to a well-behaved, unimodal posterior for each parameter in $\boldsymbol{\theta}_{\rm gw}$, for each realisation in the ensemble. There is significant dispersion in the estimated posteriors between realisations, quantified in Section \ref{sec:posterioirs}.}
		\label{fig:corner_plot_2}
	\end{figure}
	In this section we apply the state-space analysis scheme to calculate the joint posterior probability distribution $p({\boldsymbol{\theta}} | {\boldsymbol{Y}})$, where $\boldsymbol{Y}$ is generated by the procedure described in Section \ref{sec:validation}, using the injected parameters in Table \ref{tab:injected_parameters}, with $h_0 = 10^{-16}$ and $M=10^4$. The example is representative; other parameters and $h_0$ values lead to similar outcomes. \newline

	We consider first a single realisation of $\boldsymbol{Y}$. Figure \ref{fig:corner_plot_1} displays the posterior distribution for the two parameters in $\boldsymbol{\theta}_{\rm gw}$ in the form of a traditional corner plot. We plot base-10 logarithms on the horizontal and vertical axes. The histograms are the one-dimensional posteriors, marginalized over the other parameter. The two-dimensional contours mark the (0.5, 1, 1.5, 2)-sigma level surfaces. All histograms and contours are consistent with a unimodal joint posterior, with scant evidence of railing against the prior bounds. The modes of the one-dimensional posteriors are $-8.73_{+0.12}^{-0.13}$ for $ \log_{10} \gamma_{\rm a}$ and $-13.39_{+0.08}^{-0.07}$  for $ \log_{10} h_{\rm a}$, where the $\pm$ limits denote the 0.16 and 0.84 quantiles. \newline

	The reader may wonder why the injected $\gamma_{\rm a}$ and $h_{\rm a}$ values are not marked in Figure \ref{fig:corner_plot_1} for comparison with the inferred posterior. The reason is that $\gamma_{\rm a}$ and $h_{\rm a}$ are not injected directly per se; they represent an approximate parametrisation of the random realization of the SMBHB ensemble constructed for $\boldsymbol{\theta}_{\rm gw}$ according to the recipe in Appendix \ref{sec:posteriors_multiple}. There is no  unique, deterministic equation relating $\boldsymbol{\theta}_{\rm gw}$ (the parameters of the inference model) and $\boldsymbol{\theta}_{\rm a}$ (the parameters of the data generation model) for a single realisation of $\boldsymbol{Y}$. We refer the reader to Appendix \ref{appendix:justify_OU_process_for_background} for a detailed discussion about modelling the stochastic GW background as an Ornstein-Uhlenbeck process, parameterised by $\boldsymbol{\theta}_{\rm gw}$. Nonetheless, it is clear that the mode of the posterior $p(\boldsymbol{\theta}_{\rm gw}|\boldsymbol{Y})$ is roughly the correct order of magnitude, as predicted in Appendix \ref{appendix:justify_OU_process_for_background}, with $h_{\rm a} \sim M^{1/2} h_0$ and $\gamma_a \sim \Omega_{\rm min}$, given the PDFs on $\boldsymbol{\theta}_{\rm a}$ in Table \ref{tab:injected_parameters}. If one repeats the state-space analysis scheme to calculate $p({\boldsymbol{\theta}} | {\boldsymbol{Y}})$ for multiple realisations of $\boldsymbol{Y}$ and overplots the results on Figure \ref{fig:corner_plot_1}, the mode of the one-dimensional posteriors should cluster around the ensemble averaged value of $\boldsymbol{\theta}_{\rm gw}$, set by the ensemble average over the $\boldsymbol{\theta}_{\rm a}$ PDFs in Table \ref{tab:injected_parameters}. We perform this test in the next paragraph. \newline

	We now consider multiple realisations of $\boldsymbol{Y}$. By performing the ensemble average described above, one obtains approximate injection values for $\boldsymbol{\theta}_{\rm gw}$, denoted $\boldsymbol{\theta}_{\rm gw, inj}$, for comparison with the multiple posteriors inferred from the synthetic data. The procedure for calculating $\boldsymbol{\theta}_{\rm gw, inj}$ from multiple realisations of $a^{(n)}(t)$ is presented and justified in Appendix \ref{sec:posteriors_multiple}. In short, it relies on approximating the ensemble statistics of $a^{(n)}(t)$ as an Ornstein-Uhlenbeck process parametrised by $\boldsymbol{\theta}_{\rm gw}$. The procedure allows one to check the accuracy of --- and, importantly, the dispersion between --- the posteriors inferred from the synthetic data. \newline 
	
	Figure \ref{fig:corner_plot_2} displays the posterior distributions for the two parameters in $\boldsymbol{\theta}_{\rm gw}$, analogous to Figure \ref{fig:corner_plot_1}, for $10^2$ noise realisations. Each coloured curve corresponds to a different realisation. The injected values $\boldsymbol{\theta}_{\rm gw, inj}$ obtained via the procedure described in Appendix \ref{sec:posteriors_multiple}, viz. $\gamma_{\rm a, inj} = 5 \times 10^{-9}$ s$^{-1}$ and $h_{\rm a, inj} = 3 \times 10^{-14}$ (dimensionless), are marked by solid orange lines. Figure \ref{fig:corner_plot_2} displays three key features. First, for each individual realisation, the corresponding individual histograms and contours are consistent with a unimodal joint posterior. That is, the state-space inference scheme converges as expected across multiple realisations of the data. Second, on average the estimated posteriors agree closely with the injected values. We define $\mu(\theta)$ as the average of $10^2$ one-dimensional posterior modes for $\theta \in \boldsymbol{\theta}_{\rm gw}$, viz.
	\begin{eqnarray}
		\mu(\theta) = 10^{-2} \sum_{j=1}^{10^2} \underset{\theta}{\text{argmax }} p\left(\theta| \boldsymbol{Y}_j\right) \ , \label{eq:mean_rel_error}
	\end{eqnarray}
	where $p\left(\theta| \boldsymbol{Y}_j\right)$ is the one-dimensional posterior for parameter $\theta$ returned by nested sampling for the $j$-th realisation of the data, $\boldsymbol{Y}_j$. For the data in Figure \ref{fig:corner_plot_2}, we find $\mu(\gamma_{\rm a}) = 2.8 \times 10^{-9}$ s$^{-1}$ (cf. $\gamma_{\rm a, inj} = 5 \times 10^{-9}$ s$^{-1}$) and $\mu(h_{\rm a}) = 5.4 \times 10^{-14}$ (cf. $h_{\rm a, inj} = 3 \times 10^{-14}$). Third, there is significant dispersion between the posterior modes for different $\boldsymbol{Y}_j$. We define $\sigma(\theta)$ as the standard deviation in the one-dimensional modes of the $10^2$ posteriors for $\theta \in \boldsymbol{\theta}_{\rm gw}$, viz.
	\begin{eqnarray}
		\sigma(\theta) =  10^{-1} \sqrt{\sum_{j=1}^{10^2} \left[\underset{\theta}{\text{argmax }} p\left(\theta| \boldsymbol{Y}_j\right)-\mu(\theta)\right]^2} \, ,
	\end{eqnarray}
	where $\mu(\theta)$ is given by Equation \eqref{eq:mean_rel_error}. For the data in Figure \ref{fig:corner_plot_2}, we find $\sigma(\gamma_{\rm a}) = 1.2 \times 10^{-9}$ s$^{-1}$ and $\sigma(h_{\rm a}) = 2.0 \times 10^{-14}$. We define a coefficient of variation,
	\begin{equation}
		C(\theta) = \frac{\sigma(\theta)}{\mu(\theta)}  \, , \label{eq:cov}
	\end{equation}
	which evaluates to $C(\gamma_{\rm a}) = 0.45$ and $C(h_{\rm a}) = 0.38$. The dispersion quantified by $C(\theta)$ stems from the irreducible “cosmic” variance in the data itself \citep[e.g.][]{2023PhRvD.107d3018A}. That is to say, the actual astronomical realisation of the stochastic background that is observed by PTAs is just one, unknown member of a set of admissible realisations within the statistical ensemble of SMBHB sources. Additional discussion about the dispersion is presented in Appendix \ref{sec:dispersion}. \newline 
	
	For the sake of brevity and visual clarity we do not display the calculated posterior distributions for the $3N$ parameters in $\boldsymbol{\theta}_{\rm psr}$, because inferring $\boldsymbol{\theta}_{\rm gw}$ is the main focus of this paper and most published PTA analyses. In general, $\boldsymbol{\theta}_{\rm psr}$ is recovered accurately. The estimates of $f_{\rm em}(t_1)$ and $\dot{f}_{\rm em}(t_1)$ are guided into narrow ranges by the narrow priors. The one-dimensional posteriors inferred for $\sigma^{(n)}$ are generally broader consistent with the broader prior, but contain the injection within the 90\% credible interval in all cases. We remind the reader that $\gamma^{(n)}$ is not estimated in this paper. Astrophysically, it satisfies $\gamma^{(n)} \sim 10^{-5} T_{\rm obs}^{-1}$ \citep{Price2012,Myers2021MNRAS.502.3113M,Meyers2021,Vargas,2024MNRAS.tmp..891O}, so its influence is muted. 
	\subsubsection{Detectability vs. $h_0$ and $h_{\rm a}$}\label{sec:bayes}
	\begin{figure}
		\includegraphics[width=\columnwidth, height =\columnwidth]{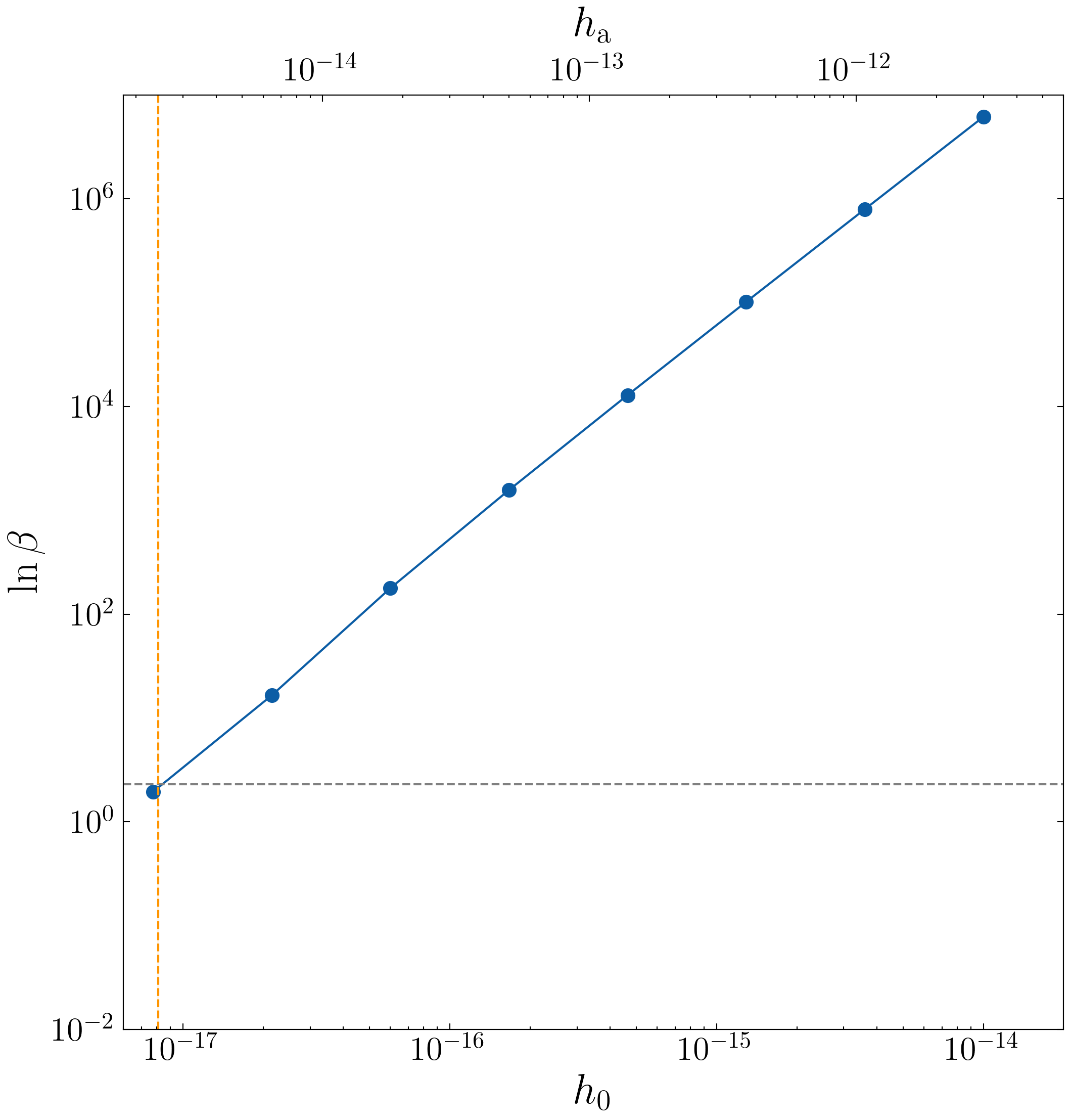} 	
		\caption{Log Bayes factor (odds ratio) $\ln \beta$ between the competing models $\mathcal{M}_{\rm gw}$ (GW present in data) and $\mathcal{M}_{\rm null}$ (GW not present in data) as a function of the amplitude, $h_0$ (bottom axis) or $h_{\rm a}$ (top axis) for the injected data in Table \ref{tab:injected_parameters}. The horizontal grey dashed line marks an arbitrary detection threshold, $\beta = 10$. The vertical orange dashed line labels the corresponding minimum detectable strain, viz. $h_0 \geq 8.1 \times 10^{-18}$ or $h_{\rm a} \geq 2.4 \times 10^{-15}$ for $\beta \geq 10$. The axes are graduated on logarithmic scales. Consequently, for example, the lowest point on the vertical axis corresponds to $\ln\beta=10^{-2}$ not $\beta=10^{-2}$.}
		\label{fig:bayes1}
	\end{figure}

	\begin{figure}
		\includegraphics[width=\columnwidth, height =\columnwidth]{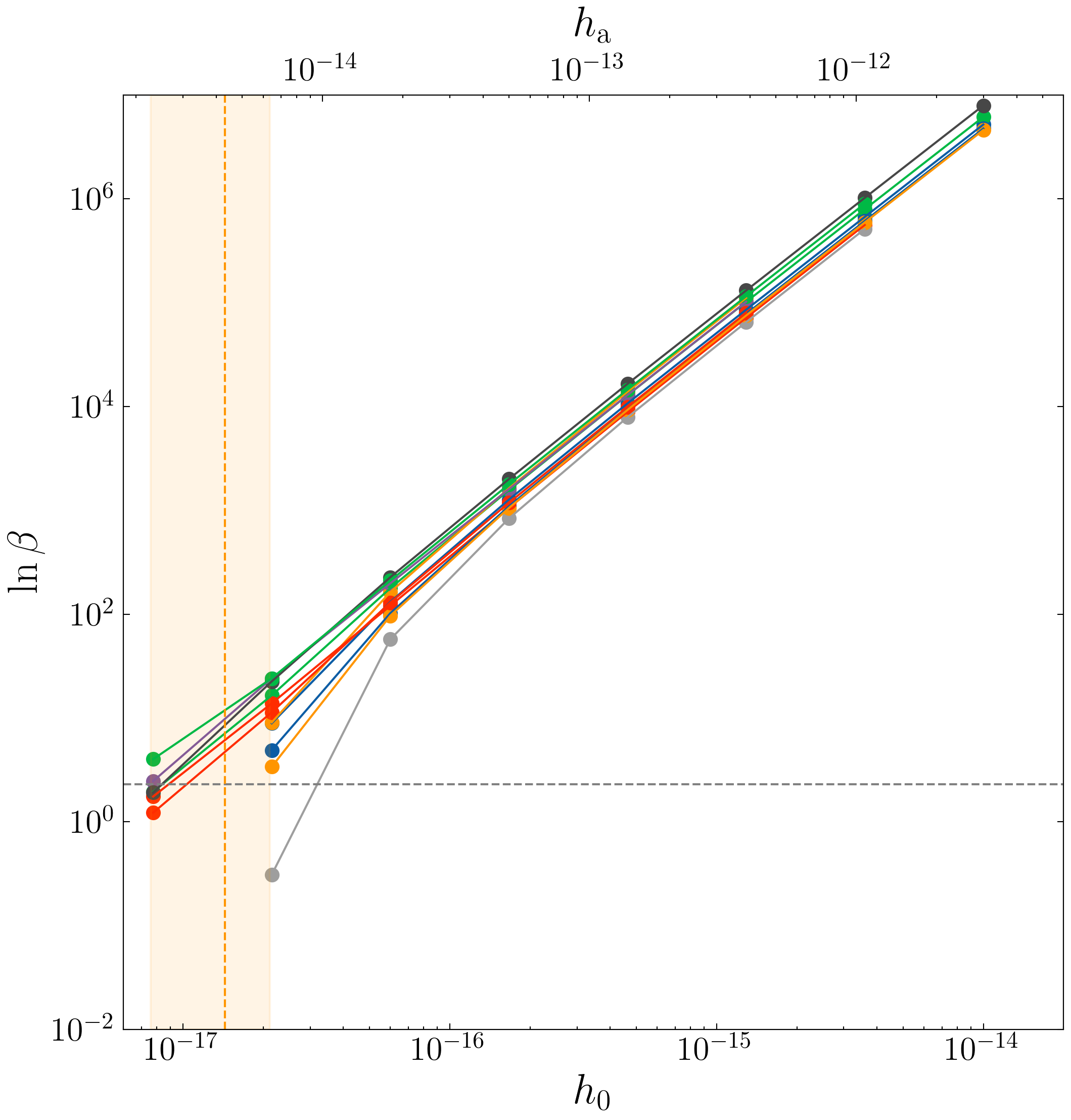} 	
		\caption{Same as Figure \ref{fig:bayes1}, but for 10 realisations of the data, with curves coloured uniquely per realisation. The vertical orange dashed line labels the minimum detectable strain, averaged over the 10 curves, viz. $h_0 \geq 1.4 \times 10^{-17}$ ($h_{\rm a} \geq 4.3 \times 10^{-15}$) for $\beta\geq 10$. The shaded vertical orange region spans plus or minus one standard deviation in the minimum detectable strain, viz.\ $h_0 \pm 6.8 \times 10^{-18}$ ($h_{\rm a} \pm 2.0 \times 10^{-15}$).}
		\label{fig:bayes2}
	\end{figure}
	Next we calculate the Bayesian evidence in favour of the presence of a stochastic GW background as a function of the amplitude of the background, parameterised by $h_0$ (or equivalently $h_{\rm a}$). The goal is to determine the minimum $h_0$ or $h_{\rm a}$ detectable by the NANOGrav-like PTA studied in this paper. We remind the reader that $h_0$ and $h_a$ are equivalent but distinct. Specifically, $h_0$ is a parameter of the data generation procedure in Section \ref{sec:validation} and is the same for all SMBHB sources in the tests conducted in this paper (i.e.\ it is used to tune the amplitude of the background and cannot be measured astronomically). In contrast $h_{\rm a}$ is a parameter in the inference model, which describes the root-mean-square amplitude of the background in terms of the Ornstein-Uhlenbeck approximation in Section \ref{sec:state_space_formulation} and can be measured astronomically. \newline  
	
	We frame the detection problem in terms of Bayesian model selection. Specifically, we define $\mathcal{M}_{\rm gw}$ as the inference model that assumes a stochastic GW background exists in the data, i.e. the model described in Section \ref{sec:state_space_formulation}. We define $\mathcal{M}_{\rm null}$ as the null model that assumes no stochastic GW background exists in the data.  This is equivalent to setting $g^{(n)}(t) =1$ in Equation \eqref{eq:measurement}. Model $\mathcal{M}_{\rm gw}$ is parametrised by $\boldsymbol{\theta}$. Model $\mathcal{M}_{\rm null}$ is parametrised by $\boldsymbol{\theta}_{\rm psr}$. The support in the data for the presence of a stochastic GW background is quantified via the Bayes factor,
	\begin{equation}
		\beta = \frac{\mathcal{Z}(\boldsymbol{Y} | \mathcal{M}_{\rm gw})}{\mathcal{Z}(\boldsymbol{Y} | \mathcal{M}_{\rm null})} \ . \label{eq:bayes}
	\end{equation}
	We consider first a single realisation of $\boldsymbol{Y}$. The log Bayes factor $\ln \beta$ is plotted as a function of $h_0$ (bottom axis) and $h_{\rm a}$ (top axis) in Figure \ref{fig:bayes1}. We vary the source amplitudes from $h_0 = 10^{-18}$ for all sources (undetectable) to $h_0 = 10^{-14}$ (easily detectable). To control the test, the noise processes $\xi^{(n)}(t)$ (Equation \eqref{eq:frequency_evolution}) and $\varepsilon^{(n)}(t)$ (Equation \eqref{eq:measurement}) in the synthetic data are identical realisations for each value of $h_0$; the only change from one $h_0$ value to the next is $h_0$ itself. Moreover, the $M$ sources that compose the stochastic GW background are identical, apart from $h_0$. The value for $h_{\rm a}$ is obtained via the procedure described in Appendix \ref{sec:posteriors_multiple}, cf. Section \ref{sec:posterioirs}. Figure \ref{fig:bayes1} reveals an approximately quadratic relationship $\ln \beta \propto h_0^2$, consistent with state-space analyses of individual SMBHBs \citep{KimpsonPTA1,KimpsonPTA2}. The stochastic GW background is detectable with decisive evidence ($\beta \geq 10$) for $h_0 \gtrsim 8.1 \times 10^{-18}$ ($h_{\rm a} \gtrsim 2.4 \times 10^{-15}$). The minimum detectable strain is particular to the system in Table \ref{tab:injected_parameters} and the realisation of $\boldsymbol{Y}$. It is influenced in general by $T_{\rm obs}$, $T_{\rm cad}$ and ${\boldsymbol{\theta}}_{\rm psr}$. A full exploration of the sensitivity of state-space analyses to a stochastic GW background is postponed to future work, after the analysis scheme in this paper is upgraded from ingesting pulse frequencies to pulse TOAs to enable a like-for-like comparison with standard PTA analyses. \newline 
	
	Let us now repeat the analysis in Figure \ref{fig:bayes1} for 10 realisations of $\boldsymbol{Y}$. It is of interest to quantify the dispersion between the minimum detectable $h_0$ or $h_{\rm a}$, in light of the dispersion observed in the estimated one-dimensional posteriors, by analogy with Figure \ref{fig:corner_plot_2}. Figure \ref{fig:bayes2} displays the log Bayes factor $\ln \beta$ as a function of $h_0$ and $h_{\rm a}$, like in Figure \ref{fig:bayes1}, for the 10 realisations of $\boldsymbol{Y}$. Each coloured curve corresponds to a different realisation. All realisations yield $\ln \beta \propto h_0^2$ approximately. The average minimum detectable $h_0$ (above which one has $\beta \geq 10$) across the 10 realisations is $1.4 \times 10^{-17}$, with standard deviation $6.8 \times 10^{-18}$. The corresponding average minimum detectable $h_{\rm a}$ is $4.3 \times 10^{-15}$, with standard deviation $2.0 \times 10^{-15}$. The vertical dashed orange line marks the average, and the shaded vertical orange band denotes the $\pm 1 \sigma$ standard deviation. The coefficient of variation (cf. Equation \eqref{eq:cov}) in the minimum detectable strain is 47\%.

\subsubsection{Effective injection parameters} \label{sec:posteriors_multiple}

In Section  \ref{sec:posterioirs}, we test the accuracy of the state-space algorithm in Sections \ref{sec:state_space_formulation} and \ref{sec:kfns} by performing Monte Carlo simulations on synthetic data. The synthetic data are generated by summing the signals from $M=10^4$ SMBHBs as in Equation \eqref{eq:asummation}. Each random realization of $M$ SMBHBs is defined by the seven parameters in $\boldsymbol{\theta}_{\rm a}$, drawn from the PDFs in Table \ref{tab:injected_parameters}. Needless to say, the ensemble-averaged parameters $\boldsymbol{\theta}_{\rm gw} = \{ h_{\rm a}, \gamma_{\rm a} \}$ are ill-defined for a single random realization. This poses a challenge: how can one compare the posterior generated by the state-space inference scheme for a single random realization of the stochastic GW background with the ``effective'' injected values $\boldsymbol{\theta}_{\rm gw,inj}$ of $\boldsymbol{\theta}_{\rm gw}$? One procedure is as follows.
\begin{enumerate}[leftmargin=2em]
	\item Generate $J$ realisations of  $a^{(n)(j)}(t)$  and  $a^{(n')(j)}(t)$ via Equation \eqref{eq:asummation}, for an arbitrary pair of pulsars $n$ and $n'$, where the superscript $1 \leq j \leq J$ labels the $j$-th realisation. 
	\item For $1 \leq j \leq J$, calculate 
	\begin{eqnarray}
		L^{(j)}(\tau) = a^{(n)(j)}(t_{\rm c}) a^{(n')(j)}(t') \label{eq:Ltau_j} \, ,
	\end{eqnarray}
	where $t_{\rm c}$ is some fixed time and we define $\tau = |t' - t_{\rm c}|$
	\item Calculate the ensemble average 
	\begin{eqnarray}
		L(\tau) = \frac{1}{J} \sum_{j=1}^{J} L^{(j)}(\tau) \label{eq:Ltau} \, .
	\end{eqnarray}
	\item Fit Equation \eqref{eq:variance_of_ou_process2} to $L(\tau)$ by nonlinear least squares.
	\item The best-fitting $h_{\rm a}$ and $\gamma_{\rm a}$ are the injection parameters $h_{\rm a, inj}$ and $\gamma_{\rm a, inj}$.
\end{enumerate}
For the tests in Section \ref{sec:posterioirs} we take $J=10^6$ and $t_c = t_1$. We set $J$ to be greater than the number of realisations ($10^2$) of the synthetic data used in Section \ref{sec:posterioirs} in order to take advantage of large number statistics when evaluating $\boldsymbol{\theta}_{\rm gw, inj}$.

\subsubsection{Dispersion}\label{sec:dispersion}
\begin{figure}
	\includegraphics[width=\columnwidth, height=2\columnwidth]{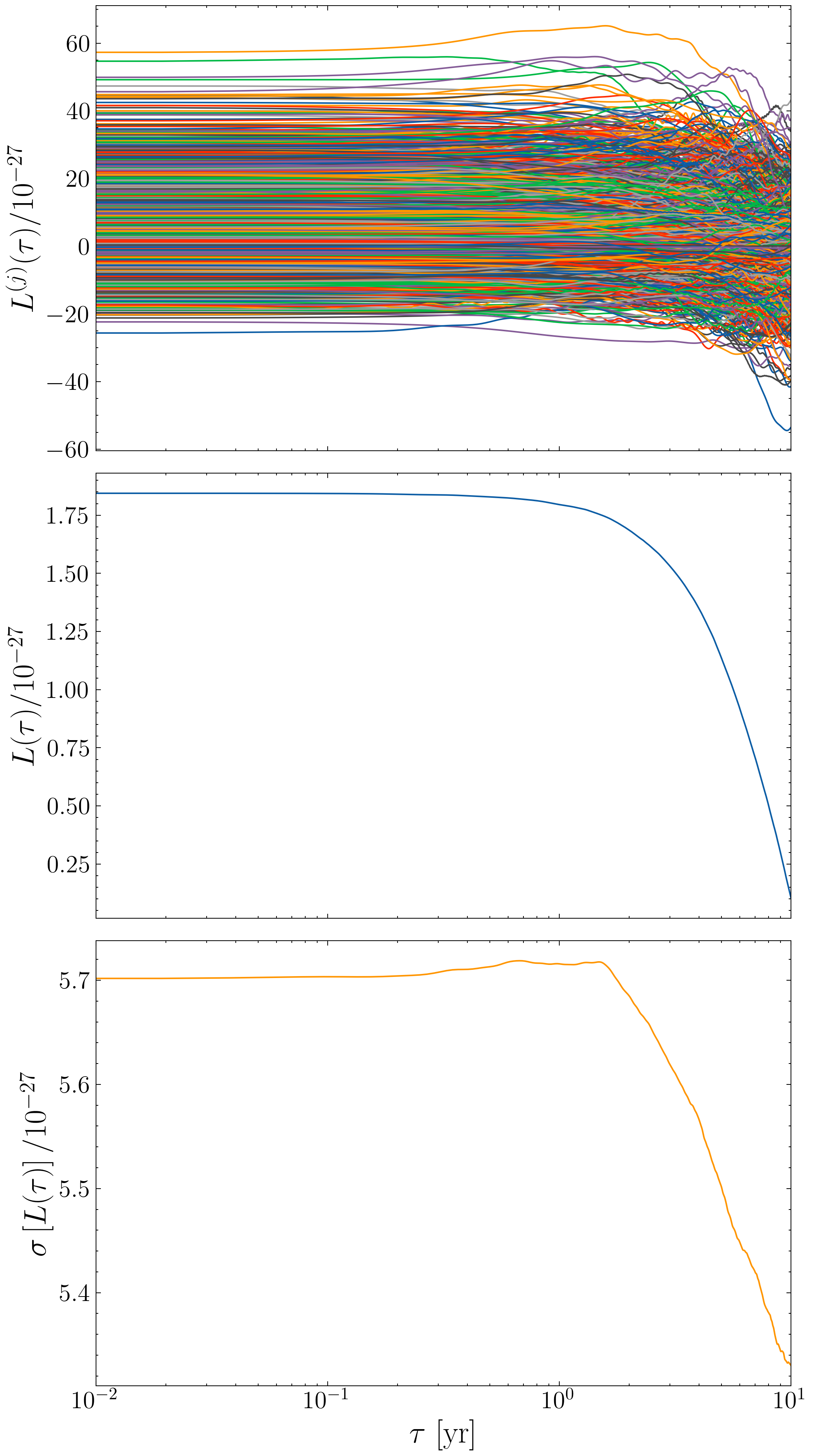} 	
	\caption{Dispersion of temporal cross-correlation $L^{(j)}(\tau) = \langle a^{(n)(j)}(t) a^{(n')(j)}(t') \rangle$ as a function of $\tau = |t-t'|$ (units: yr) for $1\leq j \leq J=2\times 10^4$ random realizations of $M=10^4$ SMBHB sources. Top panel: $L^{(j)}(\tau)$ for every realization from Equation \eqref{eq:Ltau_j}, coloured differently. Middle panel: sample mean $L(\tau)$ from Equation \eqref{eq:Ltau}. Bottom panel: sample standard deviation $\sigma[L(\tau)]$ from Equation \eqref{eq:L_var}. Note the ordering $\text{max}_j |L^{(j)}(\tau)| \gg \sigma[L(\tau)] \gg L(\tau)$, i.e. the dispersion is significant.}
	\label{fig:dispersion}
\end{figure}
In Section \ref{sec:posterioirs} we observe dispersion in the estimated posteriors $p\left( \theta_{\rm gw} | \boldsymbol{Y} \right)$ between different realisations of $\boldsymbol{Y}$. Here we analyse the phenomenon. \newline

Figure \ref{fig:dispersion} comprises three panels. The top panel shows $L^{(j)}(\tau)$, for $J = 2 \times 10^4$ realisations, with curves coloured uniquely per realisation. The middle panel shows the ensemble average $L(\tau)$, cf. Equation \eqref{eq:Ltau}, calculated for the $J$ realisations. The bottom panel shows the sample standard deviation in $L(\tau)$ as a function of $\tau$, calculated for the $J$ realisations. Explicitly, the bottom panel displays the quantity
\begin{eqnarray}
	\sigma \left[L(\tau)\right] =  \sqrt{\frac{\sum_{j=1}^{J} \left[L(\tau) - L^{(j)}(\tau) \right]^2}{J}} \label{eq:L_var}\, .
\end{eqnarray}
The logarithmic horizontal axis is the same in all three panels, spanning $10^{-2} \leq \tau / (1 \, {\rm yr}) \leq 10$ with $\tau = |t-t'|$.\newline


The main conclusion to be drawn from Figure \ref{fig:dispersion} is that dispersion is significant. One has the ordering $\text{max}_j |L^{(j)}(\tau) | \gg \sigma[L(\tau)] \gg L(\tau)$, so that the spread of $L^{(j)}(\tau)$ exceeds its mean at fixed $\tau$. Whilst $L(\tau)$ in the middle panel can be approximated accurately by an exponential parametrised by $\boldsymbol{\theta}_{\rm gw, inj}$, as in Equation \eqref{eq:variance_of_ou_process2}, an exponential parametrised by $\boldsymbol{\theta}_{\rm gw, inj}$ may not be a good approximation to $L^{(j)}(\tau)$ for all $j$. Accordingly, when the state-space analysis scheme ingests an individual realisation of data, $\boldsymbol{Y}_j$, and returns $p\left(\boldsymbol{\theta}_{\rm gw} | \boldsymbol{Y}_j \right)$ there is no reason to expect $p\left(\boldsymbol{\theta}_{\rm gw} | \boldsymbol{Y}_j \right)$ to lie close to $\boldsymbol{\theta}_{\rm gw, inj}$. This is what we observe in Section \ref{sec:posterioirs}; the estimated posteriors agree with $\boldsymbol{\theta}_{\rm gw, inj}$ in the aggregate, but $p\left(\boldsymbol{\theta}_{\rm gw} | \boldsymbol{Y}_j \right)$ may lie far from $\boldsymbol{\theta}_{\rm gw, inj}$ for some $j$.

\section{Comparison of Kalman filtering and Gaussian process regression} \label{ap:kf_vs_gp}
The Kalman filter used in this paper is a special case of the Bayesian filter \citep{sarkka} for Gaussian probability distributions. Generally, the Bayesian filter recursively computes the posterior distribution of the Markov state $\boldsymbol{X}_k$ at each time step $k$, given the history of the measurements up to that timestep, via Bayes's rule, viz.
\begin{equation}
	p \left ( \boldsymbol{X}_k | \boldsymbol{Y}_{1:k} \right ) \propto p \left (  \boldsymbol{Y}_{1:k} | \boldsymbol{X}_k \right ) p \left ( \boldsymbol{X}_k | \boldsymbol{Y}_{1:k-1} \right ) \, .
\end{equation}
For the case where the prior, likelihood and posterior are Gaussian, the Kalman filter equations presented in Appendix \ref{sec:kalman} apply \citep[see e.g. ][]{sarkka}. One may ask if there is a connection between the Kalman filter and Gaussian process (GP) regression \citep{rasmussen}, a technique which is already commonly used in PTA data analysis \citep{Haasteren}. The answer is yes; there is a clear, established connection between the Kalman filter and GPs. In this appendix we discuss this connection in more detail. In Appendix \ref{sec:GPs} we briefly review GPs. In Appendix \ref{sec:gp_vs_kf} we discuss the similarities and differences between GPs and the Kalman filter. For more detail on the connection between GPs and the Kalman filter, we refer the reader to \cite{Hartikainen2010,reeceandRoberts} and \cite{Sarkka2012}

\subsection{Gaussian processes}\label{sec:GPs}
A GP \citep{rasmussen} is a Bayesian framework for solving the problem of estimating the value of a function $f$ at a series of arbitrary times $\boldsymbol{t}_{*}$, given a set of observations $\boldsymbol{y}$ at times $\boldsymbol{t}$. The $i$-th observation is assumed to be the function value obscured by Gaussian noise, viz.
\begin{equation}
	y_i = f(t_i) + \epsilon_i \, , \label{eq:gp_y}
\end{equation}
with $\epsilon_i \sim \mathcal{N}(0,\sigma^2_{m})$. GPs assume that the joint prior distribution of the function values is Gaussian, viz.
\begin{equation}
	f \left(\boldsymbol{t} \right) \sim \mathcal{N}\left[\boldsymbol{0},\boldsymbol{K}(\boldsymbol{t}, \boldsymbol{t}', \boldsymbol{\theta}_{\rm hyper})\right] \label{eq:gp_prior}
\end{equation} 
for covariance matrix $\boldsymbol{K}(\boldsymbol{t}, \boldsymbol{t}',\boldsymbol{\theta}_{\rm hyper})$, with hyperparameters $\boldsymbol{\theta}_{\rm hyper}$. Going forwards, for notational convenience, we drop the explicit $ \boldsymbol{\theta}_{\rm hyper}$ dependence  and take it to be implicit, i.e. $\boldsymbol{K}(\boldsymbol{t}, \boldsymbol{t}', \boldsymbol{\theta}_{\rm hyper})=\boldsymbol{K}(\boldsymbol{t}, \boldsymbol{t}')$. From Equation \eqref{eq:gp_y} and \eqref{eq:gp_prior} it follows that the posterior and the prior are conjugate, satisfying
\begin{equation}
	p \left[f(\boldsymbol{t}_* | \boldsymbol{y})\right] \sim \mathcal{N} \left[\boldsymbol{\mu}(\boldsymbol{t}_*), \boldsymbol{\Sigma}(\boldsymbol{t}_*) \right] \, ,
\end{equation}
with mean
\begin{equation}
	\boldsymbol{\mu}(\boldsymbol{t}_*) =  \boldsymbol{K}(\boldsymbol{t}_*, \boldsymbol{t}) \left[ \boldsymbol{K}(\boldsymbol{t}, \boldsymbol{t}) + \sigma_{\rm m}^2 \boldsymbol{I} \right]^{-1}\boldsymbol{y} \, , \label{eq:gp_mean}
\end{equation}
and covariance
\begin{equation}
	\boldsymbol{\Sigma}(\boldsymbol{t}_*) = \boldsymbol{K}(\boldsymbol{t}_*, \boldsymbol{t}_*) -  \boldsymbol{K}(\boldsymbol{t}_*, \boldsymbol{t}) \left[ \boldsymbol{K}(\boldsymbol{t}, \boldsymbol{t}) + \sigma_{\rm m}^2 \boldsymbol{I} \right]^{-1} \boldsymbol{K}(\boldsymbol{t}, \boldsymbol{t}_*) \, . \label{eq:gp_covar}
\end{equation}
The choice of the kernel function $\boldsymbol{K}(\boldsymbol{t}, \boldsymbol{t}')$ encodes the relevant physics; see e.g. Section 3, \cite{Haasteren} for the choice of $\boldsymbol{K}(\boldsymbol{t}, \boldsymbol{t}')$ in the context of PTA data analysis. For a particular choice of $\boldsymbol{K}(\boldsymbol{t}, \boldsymbol{t}')$, the hyperparameters $\boldsymbol{\theta}_{\rm hyper}$ are unknown a priori and must be estimated through a Bayesian inference procedure. The computational complexity of the GP procedure is dominated by the matrix inverse in Equations \eqref{eq:gp_mean} and \eqref{eq:gp_covar}. If there are $T$ observations, the asymptotic scaling goes as $\mathcal{O} \left(T^3\right)$ in general.

\subsection{Kalman filtering} \label{sec:gp_vs_kf}
A Kalman filter is related to, yet differs from, a GP as follows. First, rather than constructing a model of a physical process by postulating a mean $\boldsymbol{\mu}$ and covariance function, $\boldsymbol{K}(\boldsymbol{t}, \boldsymbol{t}')$, cf. Equation \eqref{eq:gp_prior}, one specifies a (parameterised) stochastic differential equation (SDE) for a sequence of hidden states which are related to the measurements through a (parameterised) measurement equation. That is to say, one uses a state-space representation. Specifying an SDE and measurement equation implicitly defines a mean and a kernel function in Equation \eqref{eq:gp_prior}. Indeed, an essential result from \cite{Hartikainen2010} and \cite{Sarkka2012} demonstrates that for a given GP and covariance function, it is always possible to find an equivalent SDE. Second, for GPs one wants to determine $\boldsymbol{\theta}_{\rm hyper}$. For the Kalman filter, one wants to determine the parameters of the SDE and the measurement equation, i.e. $\boldsymbol{\theta}$, cf. Section \ref{sec:summary_of_static_parameters}. Third, GPs often operate in batch mode on all discrete data points simultaneously. Accordingly, the computational cost is $\mathcal{O} \left(T^3\right)$. The Kalman filter operates recursively, iterating though the dataset, and has cost $\mathcal{O} \left(T \right)$ (see also Section \ref{sec:computation_costs}). Fourth, its nature as an iterative estimator means that the Kalman filter can be readily applied to non-stationary processes \citep{zarchan2000fundamentals}. In contrast, standard GP implementations are restricted to stationary processes as they rely on the conjugacy of Gaussians to obtain tractable, closed-form posteriors. Extensions to GPs to handle non-stationary processes have been developed \citep[e.g.][]{pmlr-v51-heinonen16,2023arXiv230519242S}. Finally, the Kalman filter can be applied readily to nongaussian systems. Indeed, the linear Kalman filter is an optimal estimator for a linear system, in the sense that it minimises the mean square error, without assuming Gaussianity \citep{2024arXiv240500058U}. For nongaussian likelihoods, standard GPs are often intractable, and approximation methods based on variational inference \citep[e.g.][]{2020arXiv200211451G} or MCMC sampling \citep[e.g.][]{2022arXiv220903117K} must be used.
\bsp	
\label{lastpage}
\end{document}